\documentclass[twocolumn, twocolappendix]{aastex631}

\usepackage{multirow}
\usepackage{mysymbol}
\usepackage{comment}
\usepackage{amsmath}
\usepackage{grffile}

\accepted{July, 2023}

\begin{document}

\title{Early Planet Formation in Embedded Disks (eDisk) V: Possible Annular Substructure in a Circumstellar Disk in the Ced110 IRS4 System}

\correspondingauthor{Jinshi Sai}
\email{jsai@asiaa.sinica.edu.tw}

\author[0000-0003-4361-5577]{Jinshi Sai (Insa Choi)}
\affil{Academia Sinica Institute of Astronomy and Astrophysics, 11F of Astronomy-Mathematics Building, AS/NTU, No.\ 1, Sec.\ 4, Roosevelt Rd, Taipei 10617, Taiwan}

\author[0000-0003-1412-893X]{Hsi-Wei Yen}
\affil{Academia Sinica Institute of Astronomy and Astrophysics, 11F of Astronomy-Mathematics Building, AS/NTU, No.\ 1, Sec.\ 4, Roosevelt Rd, Taipei 10617, Taiwan}

\author[0000-0003-0998-5064]{Nagayoshi Ohashi}
\affil{Academia Sinica Institute of Astronomy and Astrophysics, 11F of Astronomy-Mathematics Building, AS/NTU, No.\ 1, Sec.\ 4, Roosevelt Rd, Taipei 10617, Taiwan}

\author[0000-0002-6195-0152]{John J.\ Tobin}
\affil{National Radio Astronomy Observatory, 520 Edgemont Rd., Charlottesville, VA 22903 USA} 

\author[0000-0001-9133-8047]{Jes K.\ J{\o}rgensen}
\affil{Niels Bohr Institute, University of Copenhagen, {\O}ster Voldgade 5--7, DK~1350 Copenhagen K., Denmark}

\author[0000-0003-0845-128X]{Shigehisa Takakuwa}
\affil{Academia Sinica Institute of Astronomy and Astrophysics, 11F of Astronomy-Mathematics Building, AS/NTU, No.\ 1, Sec.\ 4, Roosevelt Rd, Taipei 10617, Taiwan}
\affiliation{Department of Physics and Astronomy, Graduate School of Science and Engineering, Kagoshima University, 1-21-35 Korimoto, Kagoshima,Kagoshima 890-0065, Japan}

\author[0000-0003-1549-6435]{Kazuya Saigo}
\affiliation{Department of Physics and Astronomy, Graduate School of Science and Engineering, Kagoshima University, 1-21-35 Korimoto, Kagoshima,Kagoshima 890-0065, Japan}

\author[0000-0002-8238-7709]{Yusuke Aso}
\affiliation{Korea Astronomy and Space Science Institute, 776 Daedeok-daero, Yuseong-gu, Daejeon 34055, Republic of Korea}

\author[0000-0001-7233-4171]{Zhe-Yu Daniel Lin}
\affiliation{University of Virginia, 530 McCormick Rd., Charlottesville, Virginia 22904, USA}

\author[0000-0003-2777-5861]{Patrick M.\ Koch}
\affil{Academia Sinica Institute of Astronomy and Astrophysics, 11F of Astronomy-Mathematics Building, AS/NTU, No.\ 1, Sec.\ 4, Roosevelt Rd, Taipei 10617, Taiwan}

\author[0000-0003-3283-6884]{Yuri Aikawa}
\affiliation{Department of Astronomy, Graduate School of Science, The University of Tokyo, 7-3-1 Hongo, Bunkyo-ku, Tokyo 113-0033, Japan}

\author[0000-0002-8591-472X]{Christian Flores}
\affil{Academia Sinica Institute of Astronomy and Astrophysics, 11F of Astronomy-Mathematics Building, AS/NTU, No.\ 1, Sec.\ 4, Roosevelt Rd, Taipei 10617, Taiwan}

\author[0000-0003-4518-407X]{Itziar de Gregorio-Monsalvo}
\affiliation{European Southern Observatory, Alonso de Cordova 3107, Casilla 19, Vitacura, Santiago, Chile}

\author[0000-0002-9143-1433]{Ilseung Han}
\affiliation{Korea Astronomy and Space Science Institute, 776 Daedeok-daero, Yuseong-gu, Daejeon 34055, Republic of Korea}
\affiliation{Division of Astronomy and Space Science, University of Science and Technology, 217 Gajeong-ro, Yuseong-gu, Daejeon 34113, Republic of Korea}

\author[0000-0002-2902-4239]{Miyu Kido}
\affiliation{Department of Physics and Astronomy, Graduate School of Science and Engineering, Kagoshima University, 1-21-35 Korimoto, Kagoshima,Kagoshima 890-0065, Japan}

\author[0000-0003-4022-4132]{Woojin Kwon}
\affiliation{Department of Earth Science Education, Seoul National University, 1 Gwanak-ro, Gwanak-gu, Seoul 08826, Republic of Korea}
\affiliation{SNU Astronomy Research Center, Seoul National University, 1 Gwanak-ro, Gwanak-gu, Seoul 08826, Republic of Korea}

\author[0000-0001-5522-486X]{Shih-Ping Lai}
\affil{Academia Sinica Institute of Astronomy and Astrophysics, 11F of Astronomy-Mathematics Building, AS/NTU, No.\ 1, Sec.\ 4, Roosevelt Rd, Taipei 10617, Taiwan}
\affiliation{Institute of Astronomy, National Tsing Hua University, No. 101, Section 2, Kuang-Fu Road, Hsinchu 30013, Taiwan}
\affiliation{Center for Informatics and Computation in Astronomy, National Tsing Hua University, No. 101, Section 2, Kuang-Fu Road, Hsinchu 30013, Taiwan}
\affiliation{Department of Physics, National Tsing Hua University, No. 101, Section 2, Kuang-Fu Road, Hsinchu 30013, Taiwan}

\author[0000-0002-3179-6334]{Chang Won Lee}
\affiliation{Korea Astronomy and Space Science Institute, 776 Daedeok-daero, Yuseong-gu, Daejeon 34055, Republic of Korea}
\affiliation{Division of Astronomy and Space Science, University of Science and Technology, 217 Gajeong-ro, Yuseong-gu, Daejeon 34113, Republic of Korea}

\author[0000-0003-3119-2087]{Jeong-Eun Lee}
\affiliation{Department of Physics and Astronomy, Seoul National University, 1 Gwanak-ro, Gwanak-gu, Seoul 08826, Korea}

\author[0000-0002-7402-6487]{Zhi-Yun Li}
\affiliation{University of Virginia, 530 McCormick Rd., Charlottesville, Virginia 22904, USA}

\author[0000-0002-4540-6587]{Leslie W.\ Looney}
\affiliation{Department of Astronomy, University of Illinois, 1002 West Green St, Urbana, IL 61801, USA}

\author[0000-0002-7002-939X]{Shoji Mori}
\affiliation{Astronomical Institute, Graduate School of Science, Tohoku University, Sendai 980-8578, Japan}

\author[0000-0002-4372-5509]{Nguyen Thi Phuong}
\affiliation{Korea Astronomy and Space Science Institute, 776 Daedeok-daero, Yuseong-gu, Daejeon 34055, Republic of Korea}
\affiliation{Department of Astrophysics, Vietnam National Space Center, Vietnam Academy of Science and Techonology, 18 Hoang Quoc Viet, Cau Giay, Hanoi, Vietnam}

\author[0000-0001-6267-2820]{Alejandro Santamaría-Miranda}
\affiliation{European Southern Observatory, Alonso de Cordova 3107, Casilla 19, Vitacura, Santiago, Chile}

\author[0000-0002-0549-544X]{Rajeeb Sharma}, 
\affiliation{Niels Bohr Institute, University of Copenhagen, \O ster Voldgade 5-7, 1350, Copenhagen K, Denmark}

\author[0000-0003-0334-1583]{Travis J.\ Thieme}
\affiliation{Institute of Astronomy, National Tsing Hua University, No. 101, Section 2, Kuang-Fu Road, Hsinchu 30013, Taiwan}
\affiliation{Center for Informatics and Computation in Astronomy, National Tsing Hua University, No. 101, Section 2, Kuang-Fu Road, Hsinchu 30013, Taiwan}
\affiliation{Department of Physics, National Tsing Hua University, No. 101, Section 2, Kuang-Fu Road, Hsinchu 30013, Taiwan}

\author[0000-0001-8105-8113]{Kengo Tomida}
\affiliation{Astronomical Institute, Graduate School of Science, Tohoku University, Sendai 980-8578, Japan}

\author[0000-0001-5058-695X]{Jonathan P.\ Williams}
\affiliation{Institute for Astronomy, University of Hawai‘i at Mānoa, 2680 Woodlawn Dr., Honolulu, HI 96822, USA}




\begin{abstract}

We have observed the Class 0/I protostellar system Ced110 IRS4 at an angular resolution of $0\farcs05$ ($\sim$10 au) as a part of the ALMA large program; {\it Early Planet Formation in the Embedded Disks (eDisk)}. The 1.3 mm dust continuum emission reveals that Ced110 IRS4 is a binary system with a projected separation of $\sim$250 au. The continuum emissions associated with the main source and its companion, named Ced110 IRS4A and IRS4B respectively, exhibit disk-like shapes and likely arise from dust disks around the protostars. The continuum emission of Ced110 IRS4A has a radius of $\sim$91.7 au ($\sim$0$\farcs485$), and shows bumps along its major axis with an asymmetry. The bumps can be interpreted as an shallow, ring-like structure at a radius of $\sim$40 au ($\sim$0$\farcs2$) in the continuum emission, as demonstrated from two-dimensional intensity distribution models. A rotation curve analysis on the C$^{18}$O and $^{13}$CO $J=2$--1 lines reveals the presence of a Keplerian disk within a radius of 120 au around Ced110 IRS4A, which supports the interpretation that the dust continuum emission arises from a disk. The ring-like structure in the dust continuum emission might indicate a possible, annular substructure in the surface density of the embedded disk, although the possibility that it is an apparent structure due to the optically thick continuum emission cannot be ruled out.

\end{abstract}



\section{Introduction} \label{sec:intro}

Protoplanetary disks ubiquitously form as part of the star formation process \citep[e.g.,][]{Terebey:1984aa}, and are the future sites of planet formation. Recent observations at an angular resolution of $\sim$5 au with the Atacama Millimeter/submillimeter Array (ALMA) have revealed that most of disks around Class \II~sources exhibit substructures such as rings and gaps \citep[e.g.,][]{ALMA-Partnership:2015aa, Andrews:2018aa, Long:2018ab, Bi:2020aa}. Several mechanisms have been proposed to explain these substructures \citep[][]{Flock:2015aa, Zhang:2015aa, Okuzumi:2016aa, Takahashi:2016aa}, including a scenario that these substructures are created by protoplanets through the disk-planet interaction \citep[][]{Lin:1986ab, Takeuchi:1996aa,Pinilla:2012aa, Zhu:2012aa}. Indeed, some observations have reported evidence of protoplanets present within disks \citep[][]{Keppler:2018aa, Haffert:2019aa, Benisty:2021aa, Currie:2022aa}, suggesting that the disk-planet interaction could commonly take place in substructured disks.

The progress in characterizing planet formation within disks raises the question of when planet formation within disks begins. It has been suggested that protoplanetary disks around Class \II~sources are not massive enough for forming gas giant planets \citep[e.g.,][]{Manara:2018aa}, although the masses of the protoplanetary disks could be underestimated by ignoring the effects of the dust scattering \citep{Liu:2019aa, Carrasco-Gonzalez:2019aa, Zhu:2019aa, Ueda:2020aa}. On the other hand, observations toward younger disks around Class 0 and I protostars, which are still embedded in surrounding envelopes, have shown that the embedded disks have sufficient mass to enable giant planets formation without the need for unrealistically high efficiency of planet formation \citep{Tychoniec:2018ab, Tobin:2020aa}. Moreover, several embedded disks exhibit substructures \citep[][]{Sheehan:2017ab, Sheehan:2018aa, Teague:2019aa, Sheehan:2020aa, Segura-Cox:2020aa, Ohashi:2022aa}, which could be more direct signature of planet formation. However, how common such substructures are among embedded disks is still unclear since few protostars have been imaged on scales of $\sim$5--10 au.

In this paper, we report observations of the protostellar system Ced110 IRS4 at an angular resolution of $0\farcs05$ ($\sim$10 au) in dust continuum and molecular lines, as part of the ALMA large program; \textit{Early Planet Formation in the Embedded Disks} \citep[eDisk;][]{Ohashi:2023aa}. Ced110 IRS4 is a Class 0/I protostellar system in the Cederblad (Ced) 110 region of the Chamaeleon (Cha) I dark cloud. Its bolometric temperature and luminosity are 68 K and 1.0 $\Lsun$, respectively \citep{Ohashi:2023aa}, classifying this source as a Class 0 protostar, while it has been considered as a Class I source in previous works \citep[][]{Zinnecker:1999aa, Pontoppidan:2005aa}. Thus, this source may be near the division of Class 0 and I phases, although it should be noted that the classification of Class 0 and I with a boundary of the bolometric temperature of 70 K \citep{Chen:1995aa, Evans:2009aa} is somewhat arbitrary and not based on any particular physical change in the system. The Ced110 region is clustered and possesses eight young stellar objects (YSOs) within $\sim$0.2 pc \citep{Prusti:1991aa, Lehtinen:2001aa, Persi:2001aa}. Molecular line observations with single-dish telescopes have revealed a high-velocity outflow, which is thought to be driven from Ced110 IRS4, although the launching point was not spatially resolved \citep{Prusti:1991aa, Hiramatsu:2007aa}. A reflection nebula extending in a north-south direction is associated with the protostar and the presence of an edge-on disk was suggested from shadow around the reflection nebula \citep{Zinnecker:1999aa, Pontoppidan:2005aa}.

The distance to Ced110 IRS4 is not well constrained, as is not directly detected by Gaia. While several works reported slightly different distances to the Cha dark cloud of 179--$210~\pc$ \citep{Voirin:2018aa, Roccatagliata:2018aa, Dzib:2018aa, Zucker:2019aa, Zucker:2020aa, Galli:2021aa}, the most precise measurement using Gaia DR2 was provided \cite{Galli:2021aa}, considering both statistical and systematic errors. They reported two different stellar populations with distances of 191.4 and $186.7~\pc$ in the Cha I dark cloud. We adopt their measurement of a mean distance to the entire Cha I of $189~\pc$ in this paper, as the two populations spatially overlap and which group Ced110 IRS4 belongs to is uncertain. We also adopt a systemic velocity of $4.67~\kmps$ derived in this work (Section \ref{sec:ana}) for the Ced110 IRS4 system throughout this paper.

The outline of this paper is as follows. The details of the observations are summarized in Section \ref{sec:obs} and the immediate results in Sect.~\ref{sec:results}. Analyses of the dust continuum and molecular line emissions are presented in Sect.~\ref{sec:ana}. The implications of the observational results and  analyses  are discussed in Sect.~\ref{sec:discuss}. Finally, we summarize the conclusions of the paper in Sect.~\ref{sec:summary}.

\section{ALMA Observations} \label{sec:obs}
Ced110 IRS4 was observed during ALMA's Cycle 7 and 8 at ALMA Band 6 with antenna configurations of C-5 and C-8, as a part of the ALMA large program eDisk (2019.1.00261.L: PI N.\ Ohashi). The observations consisted of four Execution Blocks (EBs), whose details are summarized in Table \ref{tab:obs}. CO isotopologues, SO and other lines were observed with Frequency Division Mode (FDM), and the 1.3 mm (225 GHz) continuum emission was obtained from the line-free channels of the spectral windows with a maximum window width of 1.875 GHz. The phase center was set to $(\mathrm{RA}, \mathrm{Dec})=(11^\mathrm{h}06^\mathrm{m}46\fs440, -77^\circ22\arcmin32\farcs20)$ for observations of the target source. The shortest projected baseline was $15.1~\mathrm{m}$, providing a sensitivity to extended structures such that 10\% of the total emission at 225~GHz extending to $\sim$11$''$ is recovered \citep{Wilner:1994aa}. Main target lines and final velocity resolutions of reduced maps are summarized in Table \ref{tab:maps}. Further details of the observations including full spectral setup are presented by \cite{Ohashi:2023aa}.

\begin{deluxetable*}{lccccccc}
\tablecaption{Summary of ALMA observations \label{tab:obs}}
\tabletypesize{\footnotesize}
\tablehead{
\colhead{Date} & \colhead{
\begin{tabular}{c}
     Antenna \\
     Configuration
\end{tabular}
} & \colhead{
\begin{tabular}{c}
     Number of \\
     Antennas
\end{tabular}
} & \colhead{
\begin{tabular}{c}
     Projected \\
     Baseline Length
\end{tabular}
} & \colhead{
\begin{tabular}{c}
     Total \\
     On-source Time
\end{tabular}
} & \multicolumn{2}{c}{Calibrators} & \colhead{Check Source} \\
\colhead{} & \colhead{} & \colhead{} & \colhead{} & \colhead{} & \colhead{Bandpass and Flux} & \colhead{Phase} & \colhead{}
}
\startdata
2021 Apr.\ 24 & C-5 & 45 & 15.1 m--1.4 km & 50 mins & J1107$-$4449 & J1058$-$8003 & - \\
2021 Oct.\ 4 & C-8 & 42 & 69.9 m--10.8 km & 41 mins & J0519$-$4546 & J1058$-$8003 & J1224-8313 \\
2021 Oct.\ 20 & C-8 & 47 & 46.8 m--9.0 km & 41 mins & J0519$-$4546 & J1058$-$8003 & J0942-7731 \\
2021 Oct.\ 20 & C-8 & 44 & 46.8 m--9.0 km & 41 mins & J1427$-$4206 & J1058$-$8003 & J0942$-$7731
\enddata
\end{deluxetable*}

All data were reduced with the Common Astronomy Software Applications package \citep[CASA;][]{McMullin:2007aa} 6.2.1. The details of the imaging procedure can be found in \cite{Ohashi:2023aa}\footnote{The scripts used for reduction can be found at \url{https://github.com/jjtobin/edisk} \citep{Tobin:2023ab}.}. In summary, we conducted ALMA standard calibration through the ALMA pipeline, and then performed phase and amplitude self-calibration for the 1.3 mm continuum emission until the signal-to-noise ratio (S/N) was not improved. The gain tables obtained by the self-calibration were also applied to line data. All images were produced using the \texttt{tclean} task with clean masks determined by the auto-masking algorithm. Clean components were drawn down to three sigma noise levels both for the continuum and line data. We present two maps with robust parameters of 0.0 and 2.0 for the 1.3 mm continuum emission to emphasize different details that are present on distinct spatial scales. The continuum map with a robust parameter of 0.0 has a higher angular resolution but poor sensitivity to extended structures, while the other map with a robust parameter of 2.0 has a lower angular resolution but better sensitivity to extended structures. For the subsequent analysis, we used the map with a robust parameter of 0.0, as it was optimal for the main focus of this paper. For line imaging, the $uv$ taper with a FWHM of 2000 k$\lambda$ was applied regardless of robust parameters. We mainly present CO isotopologues and SO lines, and the maps of other lines are also shown in Appendix \ref{sec:app_mommaps}. We calculated rms noise levels of the continuum map with a robust parameter of 0.0 and all the line maps from the weight information of visibility with the \texttt{apparentsens} task. Since the continuum map with a robust parameter of 2.0 exhibited stronger sidelobes, we measured the rms noise level in a emission-free region for this map. All the maps were primary-beam corrected. Details of the main maps are summarized in Table \ref{tab:maps}.

\begin{deluxetable*}{lccccc}
\tablecaption{Summary of ALMA maps \label{tab:maps}}
\tabletypesize{\footnotesize}
\tablehead{
\colhead{Continuum/Line} & \colhead{Frequency} & \colhead{Robust} & \colhead{Beam Size} & 
\colhead{Velocity Resolution} & \colhead{RMS} \\
\colhead{} & \colhead{(GHz)} & \colhead{} & \colhead{} & \colhead{($\kmps$)} & \colhead{($\mjypbm$)}
}
\startdata
1.3 mm continuum & 225 & 0 & $0\farcs 054 \times 0\farcs 035$ ($-12.5^\circ$) & - & 0.020 \\
1.3 mm continuum & 225 & 2 & $0\farcs 122 \times 0\farcs 086$ ($-18.9^\circ$) & - & 0.018 \\
$^{12}$CO $J=2$--1 & 230.538000 & 0.5 & $0\farcs 115 \times 0\farcs 083$ ($-12.5^\circ$) & 0.635 & 0.91 \\
$^{13}$CO $J=2$--1 & 220.398684 & 1 & $0\farcs 155 \times 0\farcs 107$ ($-22.2^\circ$) & 0.167 & 1.9 \\
C$^{18}$O $J=2$--1 & 219.560354 & 1 & $0\farcs 153 \times 0\farcs 107$ ($-19.4^\circ$) & 0.167 & 1.5 \\
SO $J_N=6_5$--$5_4$ & 219.949442 & 2 & $0\farcs 178 \times 0\farcs 127$ ($-20.9^\circ$) & 0.167 & 1.8
\enddata
\end{deluxetable*}

\section{Results} \label{sec:results}
\subsection{1.3 mm Continuum}
\begin{figure*}
    \centering
    \includegraphics[width=2\columnwidth]{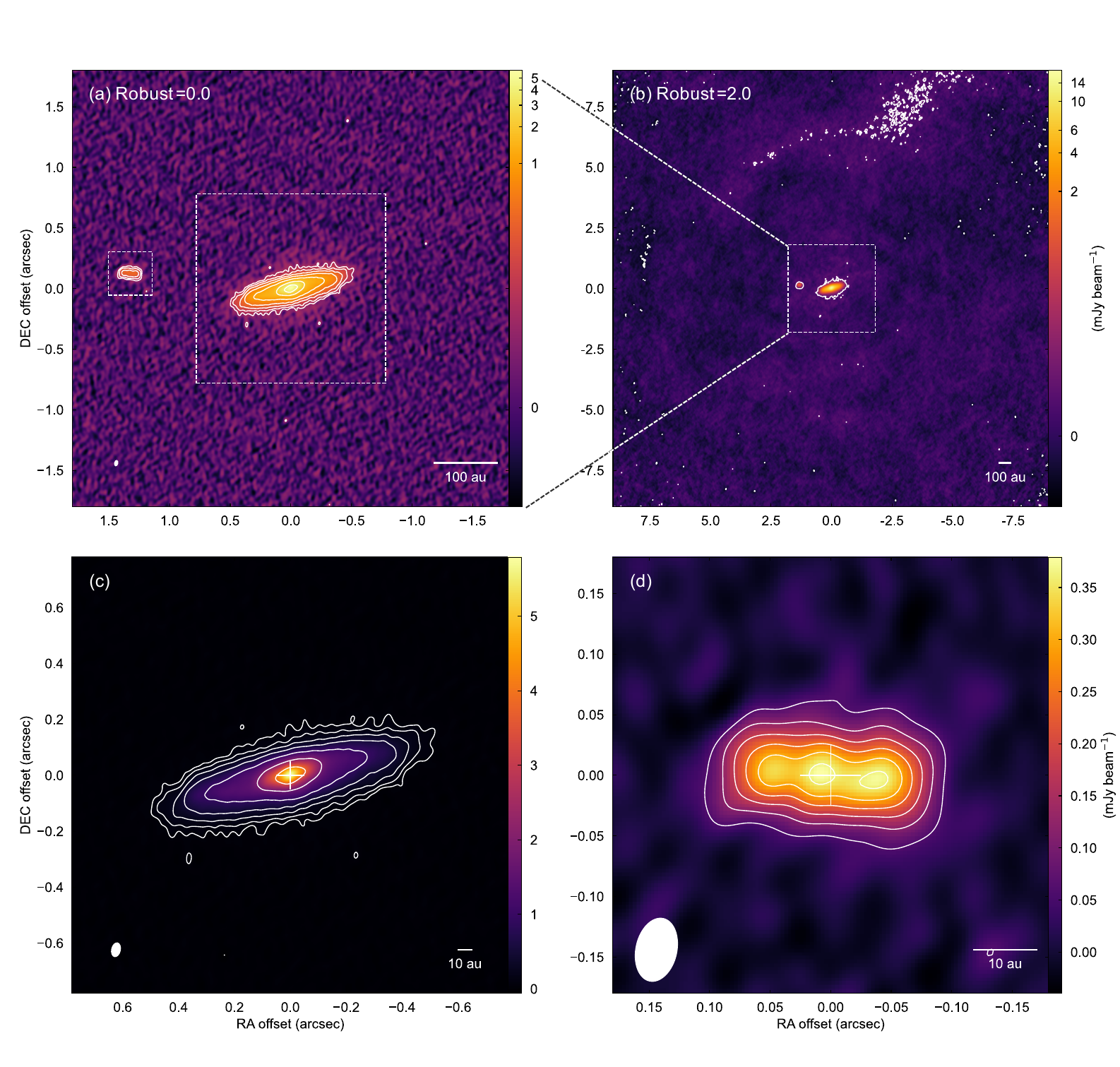}
    \caption{(a) The 1.3 mm continuum emission map with a robust parameter of 0.0. Contours levels are 3, 6, 12, 24, ... $\times \sigma$, where $\sigma=0.020~\mjypbm$. Dashed boxes show zoom-in areas, shown in panels (c) and (d). (b) The 1.3 mm continuum emission map with a robust parameter of 2.0. The contour level is 3$\sigma$, where $\sigma = 0.018~\mjypbm$. (c) Zoom-in view of the continuum emission associated with Ced110 IRS4A with a robust parameter of 0.0. Contour levels are the same as in panel (a). (d) Zoom-in view of the continuum emission associated with Ced110 IRS4B with a robust parameter of 0.0. Contour levels are 3, 6, 9, 12, 15, ... $\times \sigma$. Note that color scales are in an asinh stretch in panels (a) and (b) to cover a wide dynamic range, while they are in a linear scale in panels (c) and (d). In all panels, white ellipses at the bottom left corners denote the synthesized beam size.}
    \label{fig:cont_summary}
\end{figure*}

The 1.3 mm dust continuum emission maps are presented in Figure \ref{fig:cont_summary}. Figure \ref{fig:cont_summary}(a) shows two compact peaks that are associated with a main protostar, Ced110 IRS4A, and a fainter companion, Ced110 IRS4B, which has not been reported in previous lower-sensitivity observations. The projected distance between the two sources is $\sim$1$\farcs3$ ($\sim$250 au). In addition to the two compact emission around Ced110 IRS4A and IRS4B, an extended emission ($\sim$2$''$ in radius) is detected at $\sim$5$\sigma$ around ($\Delta \mathrm{RA},~\Delta \mathrm{Dec})=(-3'',7.5'')$ on the northern side of the protostellar system in a wider view of the 1.3 mm continuum map, presented in Figure \ref{fig:cont_summary}(b). An arc-like structure with a marginal detection at 3$\sigma$, extending from RA offset of $2\farcs5$ to $-2\farcs5$, is connecting to the extended emission.

The primary emission, that is associated with Ced110 IRS4A, exhibits a flattened, disk-like shape (Figure \ref{fig:cont_summary}(c)), and likely traces the circumstellar disk around the protostar. The peak brightness temperature of the emission is $\sim$79 K. The peak position is measured through the Gaussian fitting to be $(\mathrm{RA},~\mathrm{Dec})=(11^\mathrm{h}06^\mathrm{m}46\fs3682,~-77^\circ 22\arcmin 32\farcs 882)$. We adopt this position for the coordinates of Ced110 IRS4A and for the map center throughout this paper. The continuum emission does not show any clear substructures such as rings or gaps often found in Class \II~disks \citep[e.g.,][]{Andrews:2018aa}, despite the angular resolution of $\sim$0$\farcs05$ ($\sim$10 au).

The secondary emission, that is associated with Ced110 IRS4B, also shows a flattened, disk-like shape (Figure \ref{fig:cont_summary}(d)) and likely arises from the dust disk around Ced110 IRS4B, whereas it is not spatially well resolved along its minor axis. The emission around Ced110 IRS4B is less bright than the one around Ced110 IRS4A, and the peak brightness temperature is $\sim$9 K. It exhibits three intensity peaks separated by $\rtsim0\farcs05$ (or $\rtsim9~\au$) along its major axis, but the distinction of these peaks is marginal. The difference in the intensity between the three peaks and the dips between the peaks is $\sim$1--3$\sigma$. The position of Ced110 IRS4B is measured through the Gaussian fitting to be $(\mathrm{RA},~\mathrm{Dec})=(11^\mathrm{h}06^\mathrm{m}46\fs7718,~ -77^\circ 22\arcmin 32\farcs 757)$.

\begin{deluxetable*}{lcccc}
\tablecaption{Flux density and dust mass measured from the 1.3 mm continuum emission \label{tab:cont_flux}}
\tabletypesize{\footnotesize}
\tablehead{
\colhead{Source} & \colhead{Flux Density} & \colhead{Aperture} & \multicolumn{2}{c}{Dust Mass} \\
\colhead{} & \colhead{} & \colhead{} & \colhead{20 K$^{a}$} & \colhead{43 K$^{a}$} \\
\colhead{} & \colhead{(mJy)} & \colhead{} & \colhead{($\Mearth$)} & \colhead{($\Mearth$)}
}
\startdata
Ced110 IRS4A & $92.32\pm0.04$ & $1\farcs60 \times 0\farcs38$ ($104^\circ$) & $97.43\pm0.04$ & $38.85\pm0.02$ \\
Ced110 IRS4B & $1.96\pm0.01$ & $0\farcs31 \times 0\farcs09$ ($85^\circ$) & $2.07\pm0.01$ &  $0.825\pm0.004$
\enddata
\tablecomments{$^{a}$Assumed temperature.}
\end{deluxetable*}

We measured flux densities and dust disk radii to characterize the dust disks of Ced110 IRS4A and IRS4B. The flux densities were measured with elliptical apertures, whose semi-major axes were determined by the curve-of-growth method \citep{Ansdell:2016aa}. In this method, we applied larger apertures until the measured flux density converges. The aperture shapes were assumed based on inclination and position angles of the dust continuum emission, which are derived in Section \ref{subsec:ana_cont}. The adopted final apertures and measured flux densities are summarized in Table \ref{tab:cont_flux}. Uncertainties of the flux densities are statistical errors derived through the error propagation of the rms noise, and do not include the absolute flux calibration error, which is typically 10\% for ALMA Band 6 observations. The dust disk radii were also estimated with the curve-of-growth method. We adopted a definition of the dust disk radius that was used in the Lupus Class \II~ disk survey \citep{Ansdell:2016aa}, where the dust disk radius encloses 90\% of the flux density. Following the method of the Lupus survey, uncertainties of the dust disk radii are calculated by taking the range of radii within the uncertainties of the 90\% flux densities. The estimated dust disk radii of Ced110 IRS4A and IRS4B are $\rtsim 91.7 \pm 0.2~\au$ (or $\rtsim 0\farcs485 \pm 0\farcs001$) and $\rtsim 33.6 \pm 0.6~\au$ (or $\rtsim 0\farcs178 \pm 0\farcs003$), respectively.

Dust masses were calculated from the measured flux densities, assuming that the dust continuum emission is optically thin. We adopt the dust opacity of $2.3~\mathrm{cm}^2~\mathrm{g}^{-1}$ \citep[][]{Beckwith:1990aa} with the gas-to-dust mass ratio of 100. Two different dust temperatures are adopted: a temperature of 20 K which represents a characteristic temperature of Class \II~disks \citep{Andrews:2005aa}, and a temperature of 43 K which is derived from an empirical scaling relation of $T_\mathrm{dust} = 43 (\lbol/1 ~\Lsun)^{0.25}$ representing the averaged temperature of disks embedded in envelopes \citep[][]{Tobin:2020aa}. Note that the peak brightness temperature of the primary emission is much higher than these averaged temperatures of disks, suggesting that an inner part of the emission is optically thick, and thus the derived mass would be a lower limit. The estimated dust masses are presented in Table \ref{tab:cont_flux}. Uncertainties are calculated through the error propagation of the statistical errors of the flux densities. The dust mass of Ced110 IRS4A of $97.43~\Mearth$ assuming 20 K is much larger than the typical dust mass of Class \II~disks of $\sim$1 $\Mearth$ adopting the same temperature and a similar dust opacity \citep{Andrews:2013aa, Ansdell:2016aa, Ansdell:2018aa, Barenfeld:2017aa, Pascucci:2016aa, Long:2018aa, Ruiz-Rodriguez:2018aa, Cieza:2019aa, Williams:2019aa, Akeson:2019aa, Andrews:2020aa}. On the other hand, it agrees with the typical dust mass of Class 0 sources of $\sim$110 $\Mearth$ under similar assumptions on the temperature and dust opacity \citep[][]{Tobin:2020aa}. The flux density and dust mass of Ced110 IRS4B are smaller than those of Ced110 IRS4A approximately by two orders of magnitude and similar to the minimum value among the Orion Class 0 sources \citep{Tobin:2020aa}. As Ced110 IRS4B exhibits very small flux (and dust mass), we also compare its flux density with those measured in protostars with low masses or low luminosities. The flux density of Ced110 IRS4B is smaller than the typical 1.3 mm flux densities measured for disks around substellar-mass or low-mass protostellar objects \citep[$M_\ast \lesssim 0.1~\Msun$;][]{Lee:2018ac, Hsieh:2019aa, Riaz:2019aa, Aso:2023ab} by a factor of $\rtsim5$, but is comparable to those reported for disk components of very low-luminosity objects  \citep[$1.3\pm0.4~\mJy$ in L1521F and $3.6\pm1~\mJy$ in IRAM 04191;][]{Maury:2019aa}. Hence, the dust mass of Ced110 IRS4B is smaller than or comparable to the dust mass of these low-mass or low-luminosity protostellar systems, when the same dust temperature and opacity are adopted.

\begin{figure}
    \centering
    \includegraphics[width=\columnwidth]{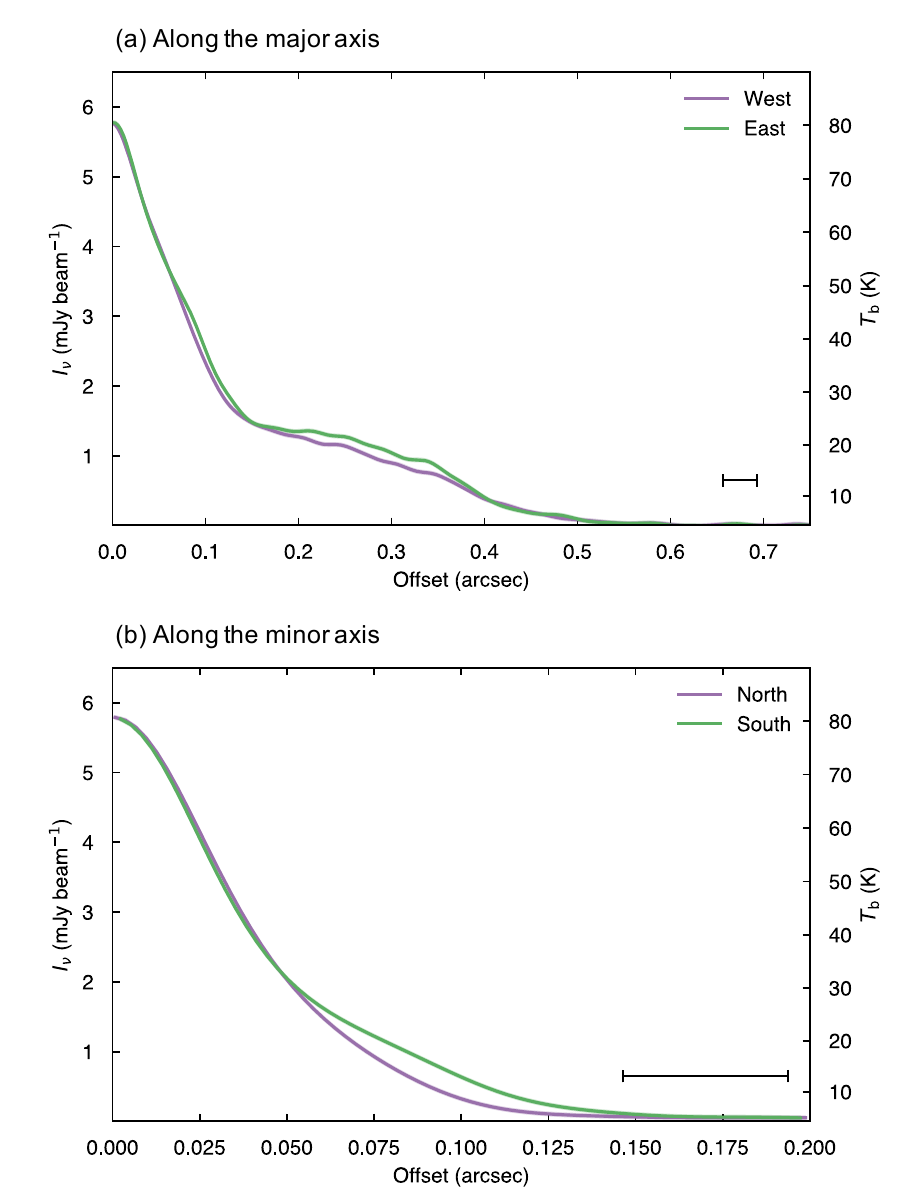}
    \caption{Intensity profiles of the 1.3 mm continuum emission associated with Ced110 IRS4A along the (a) major and (b) minor axes. One $\sigma$ uncertainty of the intensity is comparable to the width of the solid lines. Bars at the bottom right corners denote FWHMs of the synthesized beam along directions of the slices.}
    \label{fig:cont_1d-primary}
\end{figure}

\begin{figure}
    \centering
    \includegraphics[width=\columnwidth]{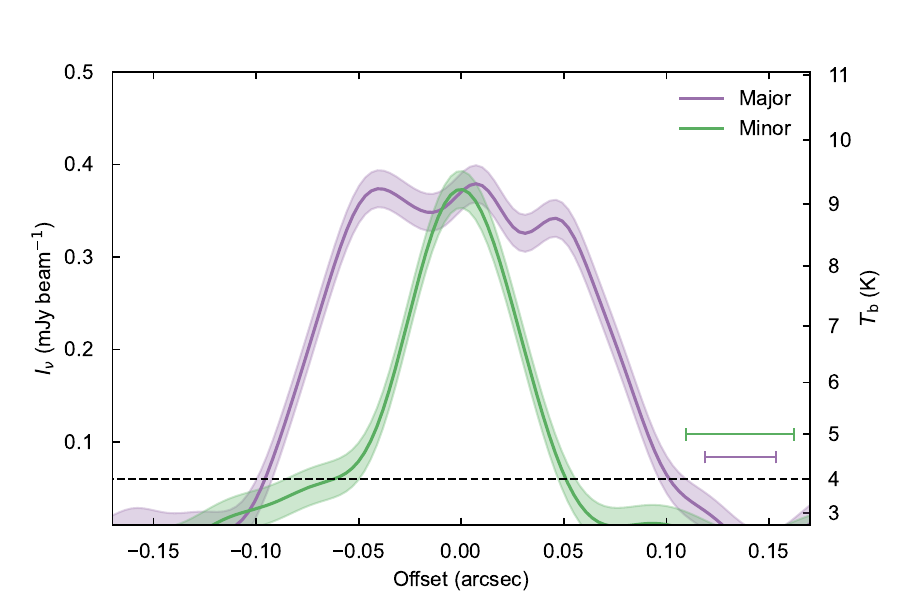}
    \caption{Intensity profiles of the 1.3 mm continuum emission associated with Ced110 IRS4B along the major and minor axes. Shaded regions indicate the 1$\sigma$ uncertainties. The horizontal dashed line indicates the 3$\sigma$ level. Purple and green bars at the bottom right corner denote FWHMs of the synthesized beam along the major and minor axes, respectively.}
    \label{fig:cont_1d-sec}
\end{figure}

In order to investigate structures of the continuum emission in more detail, we present one-dimensional intensity profiles along their major and minor axes with the cut width corresponding to the size of one pixel ($\sim$1$/10$ beam size) in Figures \ref{fig:cont_1d-primary} and \ref{fig:cont_1d-sec}. The position angles of $104^\circ$ and $85^\circ$ were adopted for the primary and secondary continuum emission, respectively, which were derived through fitting of intensity-distribution models in Section \ref{subsec:ana_cont}. The intensity profile along the major axis for Ced110 IRS4A, presented in Figure \ref{fig:cont_1d-primary}(a), interestingly, shows bumpy plateaus at a radius of $\sim$0$\farcs$2--0$\farcs$4, where the intensity distributions are close to flat. It also shows an asymmetry across a radial range of $0\farcs05$--$0\farcs4$, where the emission on the eastern side of the protostar is brighter than on the western side by $\sim$6$\sigma$ on average. An asymmetry is also clearly seen along the minor axis in Figure \ref{fig:cont_1d-primary}(b), where the emission on the southern side is significantly stronger than on the northern side. The intensity profiles for the continuum emission of Ced110 IRS4B are presented in Figure \ref{fig:cont_1d-sec}. The intensity profile along the continuum major axis for Ced110 IRS4B shows three intensity peaks, although these structures are not significant considering the rms noise level shown as a shaded region. The shape of the intensity profile along the minor axis is not well resolved.

\subsection{Lines} \label{subsec:res_lines}
Moment maps of the C$^{18}$O, $^{13}$CO, $^{12}$CO $J=2$--1 and SO $J_N=6_5$--$5_4$ lines are presented in Figure \ref{fig:moment01_c18o}--\ref{fig:moment01_so}. Moment maps of other lines are shown in Appendix \ref{sec:app_mommaps}. 
The moment zero and one maps are calculated by integrating emission detected above $3\sigma$.

The C$^{18}$O emission is detected around Ced110 IRS4A, as shown in Figure \ref{fig:moment01_c18o}(a). The emission exhibits a flattened shape with a size comparable to that of the continuum emission and an apparent deficit of emission at the protostellar position. The deficit of emission at the center is caused because the cold, foreground line emission absorbs the continuum emission around the line center and thus the continuum is over-subtracted. Indeed, in the velocity channel maps shown in Figure \ref{fig:channel_c18o_zoom}, negative emission is clearly seen at the position of Ced110 IRS4A at 3.85--4.52 $\kmps$, where most of the C$^{18}$O emission is resolved out. In the moment one map presented in Figure \ref{fig:moment01_c18o}(b), a clear velocity gradient is seen along the direction of the continuum elongation across the protostar, suggesting a rotational motion of its disk and/or an envelope. The C$^{18}$O emission is also detected around Ced110 IRS4B at $\ge3\sigma$. Figure \ref{fig:moment01_c18o}(c) and (d) show wider views of the moment zero and one maps, respectively. An extended emission at the northern side of the protostar and an arc-like structure connecting to the extended component are observed at $5\sigma$ in the C$^{18}$O emission. These structures are very similar to the north-extended and arc-like structures of the 1.3 mm continuum emission. The extended C$^{18}$O emission on the northern side has a similar extent to that of the continuum emission ($\sim$2$''$ in semi-major axis). The arc-like structure is more clearly seen in the C$^{18}$O emission over a velocity range of 4.85--5.86 $\kmps$ (see also Figure \ref{fig:channel_c18o_wide}), and appears to be one-fourth of a circle across east to north. A clear velocity gradient is seen along the arc-like structure and the northern extended component.

The $^{13}$CO emission is detected around Ced110 IRS4A as well as the C$^{18}$O emission, as seen in moment zero and one maps in Figure \ref{fig:moment01_13co}(a) and (b). The overall features of the $^{13}$CO emission associated with Ced110 IRS4A are similar to that of the C$^{18}$O emission: the $^{13}$CO emission shows an elongated shape and a velocity gradient in a direction similar to the direction of the continuum elongation, suggesting rotation of a disk and/or an envelope. The moment zero map of the $^{13}$CO emission also exhibits an apparent deficit of emission at the central, which is likely due to the continuum over-subtraction with the cold, foreground line emission around the line center (see Figure \ref{fig:channel_13co_zoom}). In addition to these structures, the $^{13}$CO emission exhibits four peaks at the edge of the continuum emission of Ced110 IRS4A, making a dark lane along the continuum major axis at radii of $0\farcs1$--$0\farcs5$. These features are not seen in the C$^{18}$O emission. The four peaks are slightly asymmetric and the emission on the northern side is brighter. These features suggest that the $^{13}$CO emission around the Ced110 IRS4A originates mainly from the disk surface. This point is discussed in more detail in Section \ref{subsec:diskasym}. The $^{13}$CO emission is also seen around Ced110 IRS4B, but it is much more extended than the continuum emission and connected to Ced110 IRS4A. Figure \ref{fig:moment01_13co}(c) and (d) show wider views of the moment zero and one maps of the $^{13}$CO emission. Extended structures of the $^{13}$CO emission are more complex than those of the C$^{18}$O emission. The $^{13}$CO emission does not show a component corresponding to the arc-like structure found in the C$^{18}$O and continuum emission. This could be because the arc-like structure is mostly obscured by optically-thick, extended foreground gas in the $^{13}$CO emission, as the emission at the velocity range of 3.85--5.52 $\kmps$ is resolved out in the velocity channel maps in Figure \ref{fig:channel_13co_wide}. The $^{13}$CO emission shows the arc-like structure only at a velocity of 5.86 $\kmps$ in the velocity channel maps.

The $^{12}$CO emission shows four intensity peaks around Ced110 IRS4A and a dark lane along the continuum major axis, as the $^{13}$CO emission does, but more clearly in Figure \ref{fig:moment01_12co}(a). The two intensity peaks on the northern side of the continuum emission are brighter than those on the southern side. In Figure \ref{fig:moment01_12co}(b), a clear velocity gradient appears along the continuum elongation around Ced110 IRS4A, as was seen in the C$^{18}$O and $^{13}$CO emission. The $^{12}$CO emission is also extended around Ced110 IRS4B. On larger scales, as presented in Figure \ref{fig:moment01_12co}(c) and (d), the $^{12}$CO emission is extended from east to west around Ced110 IRS4A with a clear velocity gradient. The $^{12}$CO emission is resolved out over a velocity range of 3.50--5.41 $\kmps$ (see Figure \ref{fig:channel_12co_wide}), and thus does not show any structures corresponding to the arc-like structures found in the continuum and C$^{18}$O emission. No clear outflow from Ced110 IRS4A and IRS4B is seen in the $^{12}$CO emission, though $^{12}$CO emission is a typical tracer of outflows around protostars.

The SO emission is not detected around Ced110 IRS4A or IRS4B, but it shows an arc-like structure with a velocity gradient along the structure in Figure \ref{fig:moment01_so}, as in the case of the C$^{18}$O and continuum emission. The SO arc-like structure appears within the same velocity range of 4.85--5.86 $\kmps$ as that of the C$^{18}$O emission. The SO emission also exhibits a blue-shifted component at a distance of $\rtsim6''$ ($\rtsim1000~\au$) on the southern side of the Ced110 IRS4 system.

\begin{figure*}[tbhp]
    \centering
    \includegraphics[width=2\columnwidth]{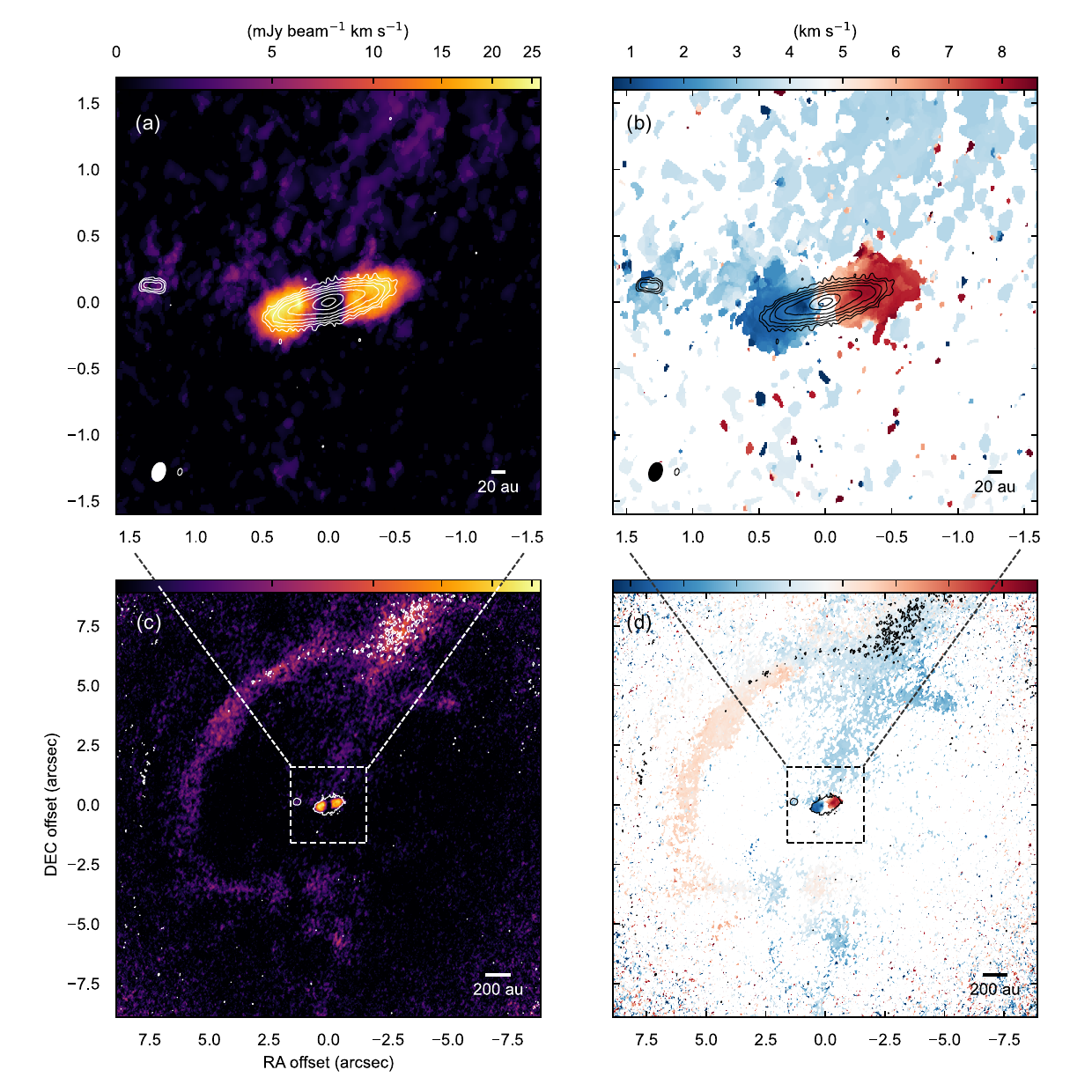}
    \caption{Moment zero and one maps of the C$^{18}$O $J=2$--1 emission (color) overlaid with the 1.3 mm continuum maps (contours). Panels (a) and (c) show zoom-in and wide views of the moment zero map, and panels (b) and (d) show zoom-in and wide views of the moment one map, respectively. The 1.3 mm continuum maps with robust parameters of 0.0 and 2.0 are overlaid on zoom-in and wide views of the moment maps, respectively. Contour levels are the same as those of Figure \ref{fig:cont_summary}(a) and (b). The filled and opened ellipses at the bottom left corners represent the synthesized beam sizes of the C$^{18}$O and continuum maps, respectively.}
    \label{fig:moment01_c18o}
\end{figure*}

\begin{figure*}[tbhp]
    \centering
    \includegraphics[width=2\columnwidth]{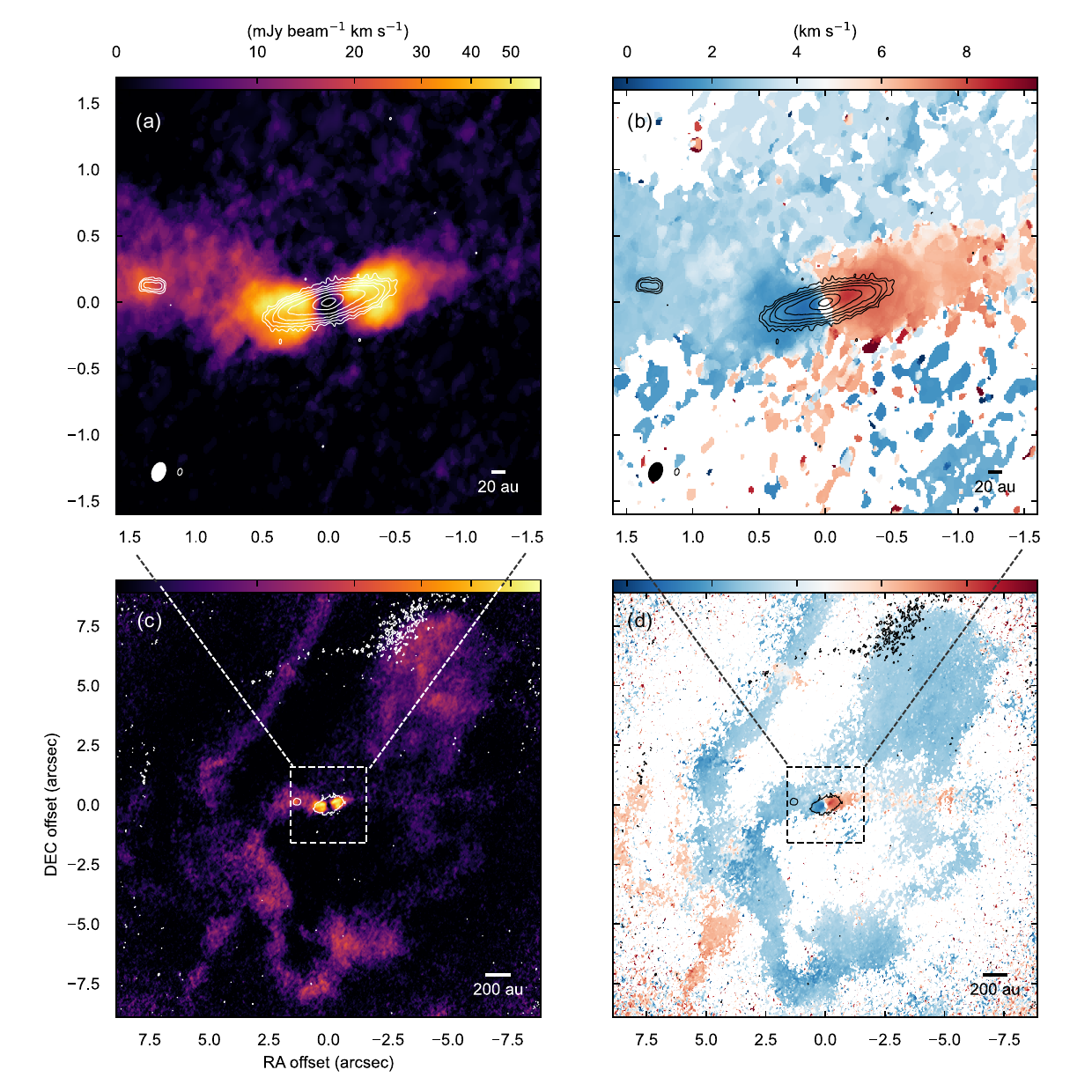}
    \caption{Same as Figure \ref{fig:moment01_c18o} but for the $^{13}$CO emission.}
    \label{fig:moment01_13co}
\end{figure*}

\begin{figure*}[tbhp]
    \centering
    \includegraphics[width=2\columnwidth]{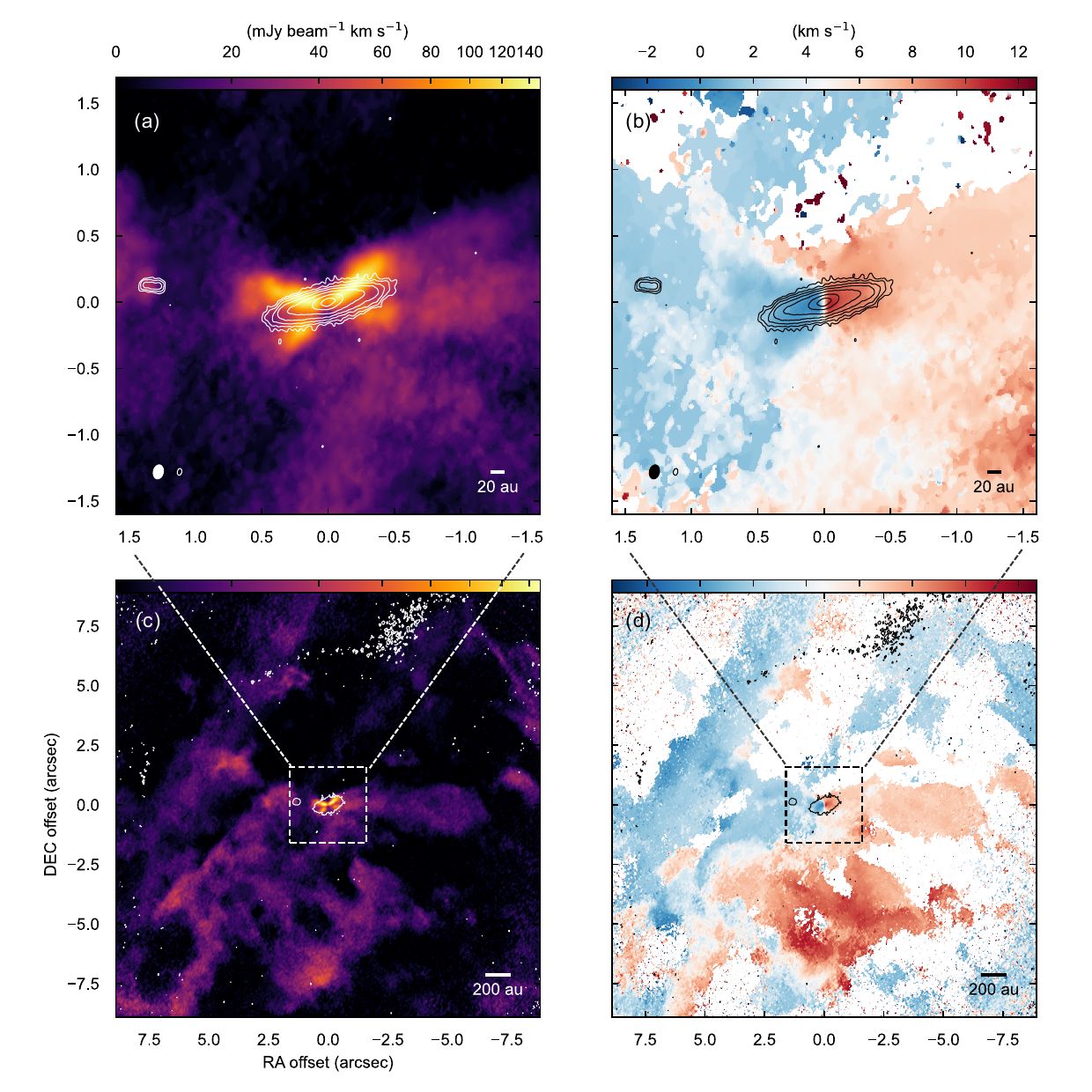}
    \caption{Same as Figure \ref{fig:moment01_c18o} but for the $^{12}$CO emission.}
    \label{fig:moment01_12co}
\end{figure*}

\begin{figure*}[tbhp]
    \centering
    \includegraphics[width=2\columnwidth]{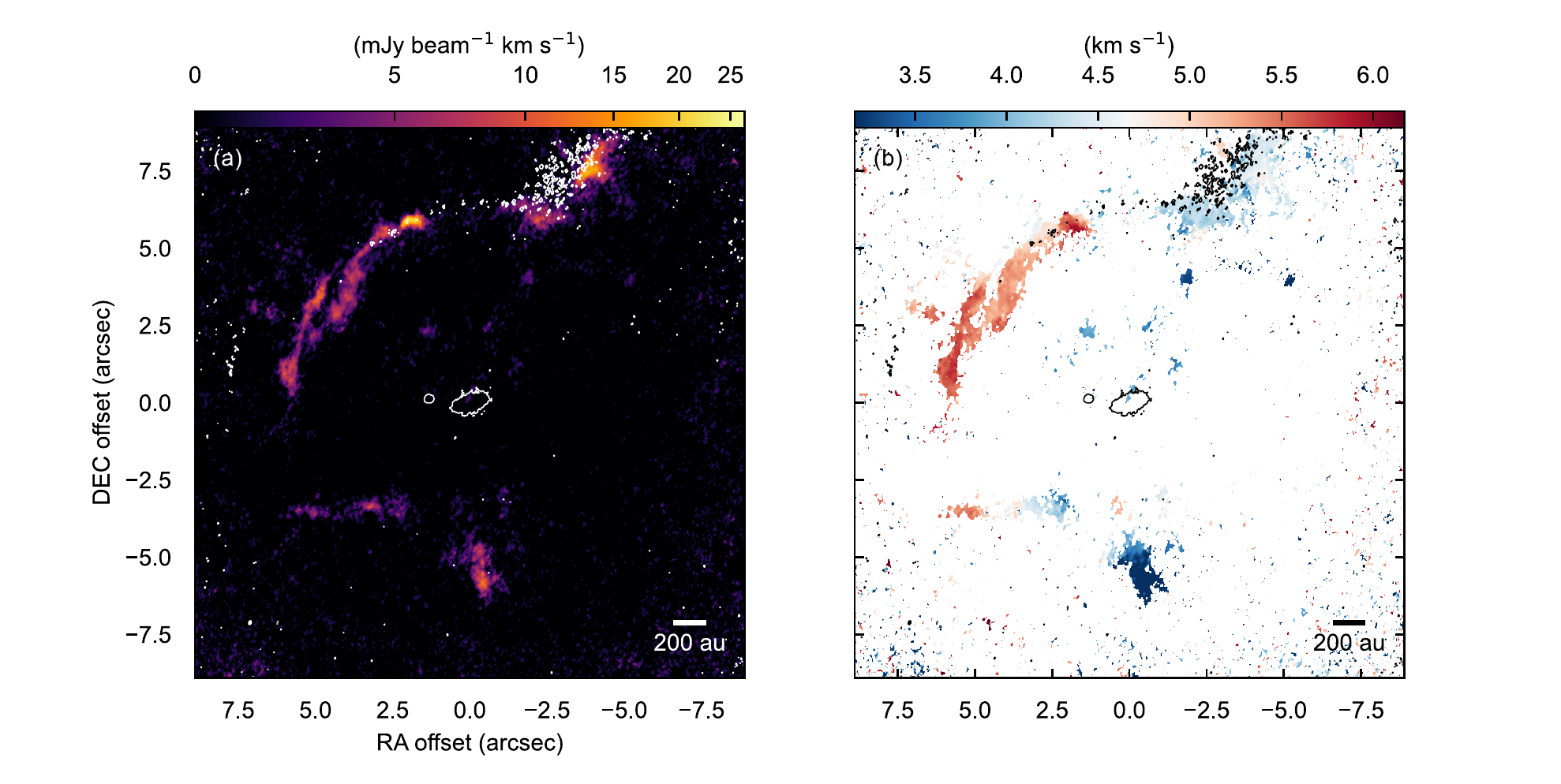}
    \caption{(a) Moment zero and (b) one maps of the SO emission (color) overlaid with the 1.3 mm continuum maps with a robust parameter of 2.0 (contours). Contour levels are as the same as those of Figure \ref{fig:cont_summary}(b). The filled and opened ellipses at the bottom left corners represent the synthesized beam sizes of the SO and continuum maps, respectively.}
    \label{fig:moment01_so}
\end{figure*}

\section{Analysis} \label{sec:ana}
\subsection{Intensity Distribution Model \label{subsec:ana_cont}}

\subsubsection{Ced110 IRS4A \label{subsubsec:ana_cont_irs4a}}

\begin{figure*}[tbhp]
    \centering
    \includegraphics[width=2.1\columnwidth]{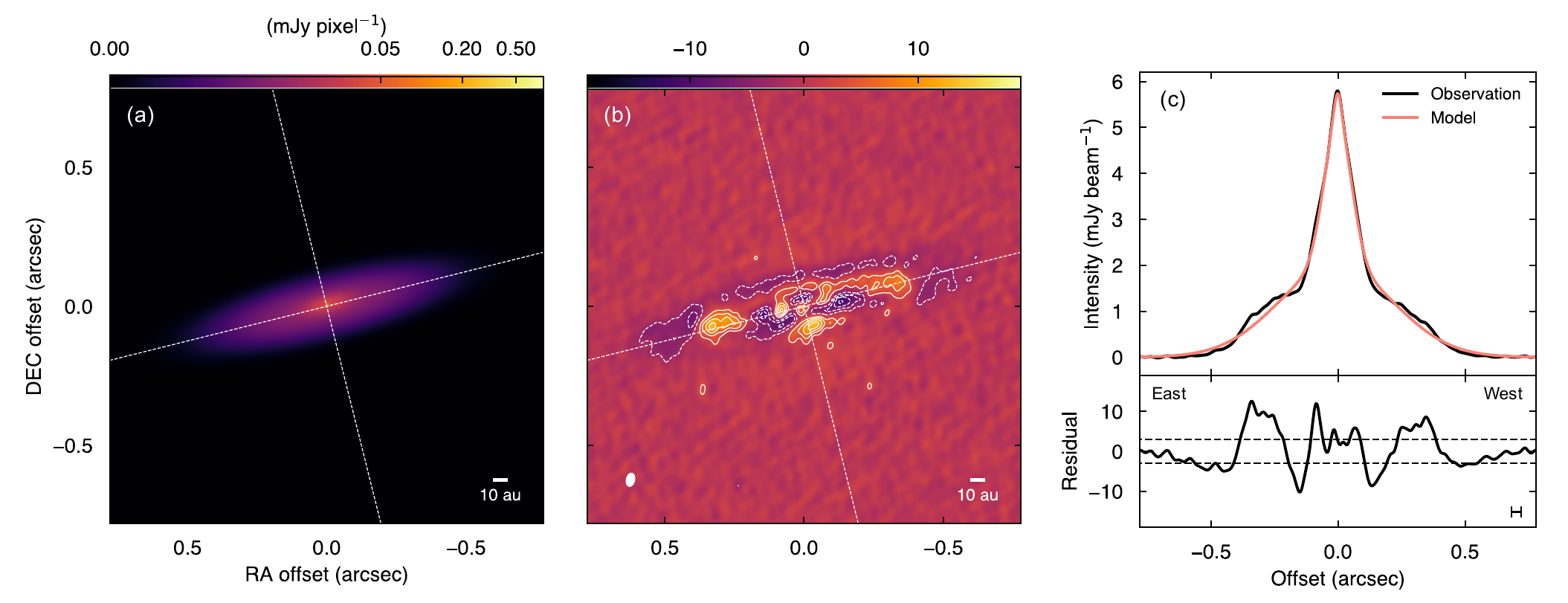}
    \caption{Comparisons between the smooth model and the observed continuum map of Ced110 IRS4A. (a) The model image before the beam convolution and (b) the residual image normalized by $\sigma$. Contours start from $3\sigma$ and increase in steps of $3\sigma$. Dashed contours indicate the negative residuals. Dashed lines across the maps show directions of the continuum major and minor axes. The white ellipse at the bottom left corner in panel (b) denotes the synthesized beam size. (c) The one-dimensional intensity and residual profiles along the continuum major axis with a cut width corresponding to the size of one pixel ($\sim$1$/10$ beam size). Horizontal dashed lines indicate $\pm 3 \sigma$. The bar in the bottom right corner represents the FWHM of the synthesized beam along the continuum major axis.}
    \label{fig:modelfit_dg}
\end{figure*}

The continuum emission associated with Ced110 IRS4A exhibits bumps along its major axis, suggesting a shallow, ring-like structure in the intensity distribution. In order to investigate these structures in more detail, we performed fitting of two-dimensional intensity models.

\begin{figure*}[tbh]
    \centering
    \includegraphics[width=2.1\columnwidth]{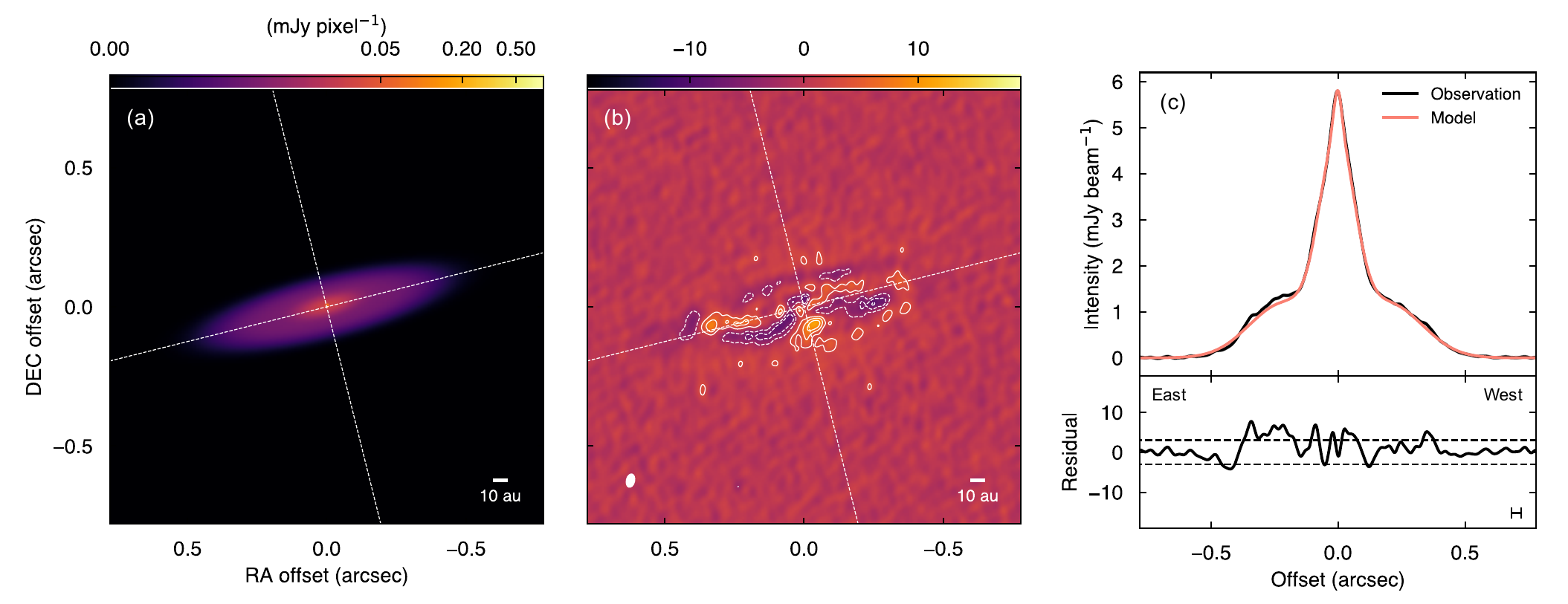}
    \caption{Same as Figure \ref{fig:modelfit_dg} but for the Gaussian ring model.}
    \label{fig:modelfit_gr}
\end{figure*}

We construct two types of intensity distributions: a smooth intensity distribution and an intensity distribution with a ring component at a certain radius. The smooth intensity distribution consists of a point source, a compact Gaussian, and a broad Gaussian, while the intensity distribution with a ring component is modeled with a point source, a compact Gaussian, and an azimuthally symmetric Gaussian ring, which is expressed by the following equation:
\begin{eqnarray}
I_\nu (r) = a_\mathrm{r} \exp \left\{ -\frac{(r - r_\mathrm{r})^2}{2 \sigma_\mathrm{r}^2} \right\},
\end{eqnarray}
where $r$ is the radius deprojected by the inclination angle $i_\mathrm{r}$ and the position angle $pa_\mathrm{r}$, $a_\mathrm{r}$ is the peak intensity of the ring component, $r_\mathrm{r}$ is the ring location in radius, and $\sigma_\mathrm{r}$ is the ring width.
A point source is included in both models to reproduce the observed high brightness temperature at the peak position, which cannot be reproduced by a compact Gaussian component only. The intensity models are compared to the observations after beam convolution. We fitted these two models to the observed continuum map with the MCMC code \texttt{emcee} \citep[][]{Foreman-Mackey:2013aa}. We performed the fitting in the image plane rather than in the uv-plane for simplicity, as Ced110 IRS4B is contained in the same field of view. All model parameters and their best-fit values are summarized in Table \ref{tab:param_intensitymodel}.

\begin{deluxetable*}{lclc}
\tablecaption{Parameters of the intensity distribution models \label{tab:param_intensitymodel}}
\tabletypesize{\footnotesize}
\tablehead{
\colhead{Parameter} & \colhead{Unit} & \colhead{Description} & \colhead{Best-fit value}
}

\startdata
\multicolumn{4}{c}{Smooth model ($k=13$, $\mathrm{BIC}=-43800$)$^{a}$} \\ \hline
$f_0$ & mJy & Flux density of the point source & $0.66 \pm 0.01$ \\
$a_0$ & $\times 10^{-3}$ mJy pixel$^{-1}$ & Amplitude of the 1st Gaussian & $21.79\pm0.11$ \\
$x_0$ & mas & RA offset coordinate of the center of the 1st Gaussian, and of the point source & $-2.03 \pm 0.05$ \\ 
$y_0$ & mas & Dec offset coordinate of the center of the 1st Gaussain, and of the point source & $-0.60 \pm 0.09 $ \\
$\sigma_\mathrm{maj,0}$ & mas &  Width (standard deviation) of the 1st Gaussian along the major axis & $51.80 \pm 0.14 $ \\
$\sigma_\mathrm{min,0}$ & mas & Width (standard deviation) of the 1st Gaussian along the minor axis & $17.72 \pm 0.09$ \\
$pa_0$ & $^\circ$ & Position angle of the 1st Gaussian & $105.06 \pm 0.08$ \\
$a_1$ & $\times 10^{-3}$ mJy pixel$^{-1}$ & Amplitude of the 2nd Gaussian & $9.26\pm0.02$ \\ 
$x_1$ & mas & RA offset coordinate of the center of the 2nd Gaussian & $4.04 \pm 0.05$ \\ 
$y_1$ & mas & Dec offset coordinate of the center of the 2nd Gaussian & $7.07 \pm 0.20$ \\
$\sigma_\mathrm{maj,1}$ & mas & Width (standard deviation) of the 2nd Gaussian along the major axis & $223.09 \pm 0.25$ \\
$\sigma_\mathrm{min,1}$ & mas & Width (standard deviation) of the 2nd Gaussian along the minor axis & $53.23 \pm 0.06$ \\
$pa_1$ & $^\circ$ & Position angle of the 2nd Gaussian & $103.75 \pm 0.02$ \\
\hline
\multicolumn{4}{c}{Ring model ($k=14$, $\mathrm{BIC}=-51500$)$^{a}$} \\ \hline
$f_0$ & mJy & Flux density of the point source & $0.92 \pm 0.01$ \\
$a_0$ & $\times 10^{-3}$ mJy pixel$^{-1}$ & Amplitude of the Gaussian & $27.43\pm0.09$ \\
$x_0$ & mas & RA offset coordinate of the center of the Gaussian and of the point source & $-0.05 \pm 0.04$ \\ 
$y_0$ & mas & Dec offset coordinate of the center of the Gaussian and of the point source & $1.41 \pm 0.08 $ \\
$\sigma_\mathrm{maj,0}$ & mas & Width (standard deviation) of the Gaussian along the major axis & $65.98 \pm 0.20 $ \\
$\sigma_\mathrm{min,0}$ & mas & Width (standard deviation) of the Gaussian along the minor axis & $19.61 \pm 0.07$ \\
$pa_0$ & $^\circ$ & Position angle of the Gaussian & $104.44 \pm 0.05$ \\
$a_\mathrm{r}$ & $\times 10^{-3}$ mJy pixel$^{-1}$ & Amplitude of the Gaussian ring & $5.23\pm0.01$ \\ 
$x_\mathrm{r}$ & mas & RA offset coordinate of the center of the Gaussian ring & $4.89 \pm 0.23$ \\ 
$y_\mathrm{r}$ & mas & Dec offset coordinate of the center of the Gaussian ring & $-6.77 \pm 0.08$ \\
$r_\mathrm{r}$ & mas & Radius of the Gaussian ring & $215.77 \pm 1.03$ \\
$\sigma_\mathrm{r}$ & mas & Width (standard deviation) of the Gaussian ring & $132.05 \pm 0.58$ \\
$i_\mathrm{r}$ & $^\circ$ & Inclination angle of the Gaussian ring &  $76.31 \pm 0.02$ \\
$pa_\mathrm{r}$ & $^\circ$ & Position angle of the Gaussian ring & $103.72 \pm 0.02$
\enddata
\tablecomments{$^a k$ is the number of free parameters, and BIC is the Bayesian information criterion calculated with the best-fit parameters.}
\end{deluxetable*}

The best-fit model images and residual images for the two models are presented in Figure \ref{fig:modelfit_dg}(a) and (b), and in Figure \ref{fig:modelfit_gr}(a) and (b). Both the smooth and ring models show large residuals on the southern side of the protostar along the continuum minor axis. This is because the intensity models are constructed to be axisymmetric, while the observed continuum emission is asymmetric along the minor axis. The smooth model exhibits significant residuals $\ge 6\sigma$ and $\le -6\sigma$ along the continuum major axis, while the ring model shows fewer residuals. The one-dimensional intensity and residual profiles along the major axis are presented in Figure \ref{fig:modelfit_dg}(c) and \ref{fig:modelfit_gr}(c) to show this feature more clearly. The residual profile of the smooth model in Figure \ref{fig:modelfit_dg}(c) shows large, systemic residuals $\ge 6\sigma$ and $\le -6\sigma$ around offsets of $-0\farcs3$ and $0\farcs3$. On the other hand, residuals for the ring model along the major axis are mostly less than $3\sigma$. Slightly large residuals of 4$\sigma$ on average, with a maximum residual of 8$\sigma$, are seen at offsets of $\sim-0\farcs4$ to $\sim-0\farcs05$. These residuals are comparable to the significance level of the asymmetry of the observed continuum emission along the major axis ($\sim$6$\sigma$ on average) found within the same radial range in Figure \ref{fig:cont_1d-primary}(a). Therefore, the slightly large differences between the observations and the ring model at these offsets are due to the asymmetric structure of the observed continuum emission, which is not taken into account in the intensity models.

In order to evaluate which model better describes the observed map quantitatively taking into account difference in the number of free parameters, we calculate the Bayesian information criterion (BIC), which is defined as follows, for each model:
\begin{equation}
\mathrm{BIC} = k \ln \left(n \right) - 2 \ln ( \hat{L} ),
\end{equation}
where $k$ is the number of free parameters, $n$ is the number of data points, and $\hat{L}$ is the maximum likelihood. The logarithm of the likelihood is defined as follows:
\begin{equation}
\ln \{ L(\Theta) \} = -\frac{1}{2} \sum_{i=1}^{n} \left\{ \frac{(I_{\mathrm{obs},i} - I_{\mathrm{model},i})^2}{\sigma^2} + \ln(2 \pi \sigma^2) \right\},
\label{eq:likelihood}
\end{equation}
where $I_\mathrm{obs}$ is the observed intensity, $I_\mathrm{model}$ is the model intensity with a parameter set $\Theta$, and $\sigma$ is the rms noise of the continuum map. The logarithm of the maximum likelihood ($\ln(\hat{L})$) is, thus, derived from Equation (\ref{eq:likelihood}) and the best-fit parameters of $\hat{\Theta}$. BICs for the smooth model and the ring model are calculated to be $-43800$ and $-51500$, respectively, where difference between BIC values larger than 10 indicates a very strong evidence in a favor of the model with the lower BIC value \citep{Kass:1995aa}. Hence, we conclude that an intensity distribution with a ring-component better explains the observed dust continuum emission. We have also tested intensity distributions with a power-law profile, that is often used to describe dust continuum emission from disks, and obtained similar results that including a ring component better explains the observations, as presented in Appendix \ref{sec:app_models}. We adopt the position angle of $104^\circ$ and inclination angle of $76^\circ$ of the ring model as those of the dust disk around Ced110 IRS4A. Note that the inclination angle is a lower limit, since the dust disk may not be geometry thin.

\subsubsection{Ced110 IRS4B \label{subsubsec:ana_cont_irs4b}}

\begin{figure*}[tbhp]
    \centering
    \includegraphics[width=2.1\columnwidth]{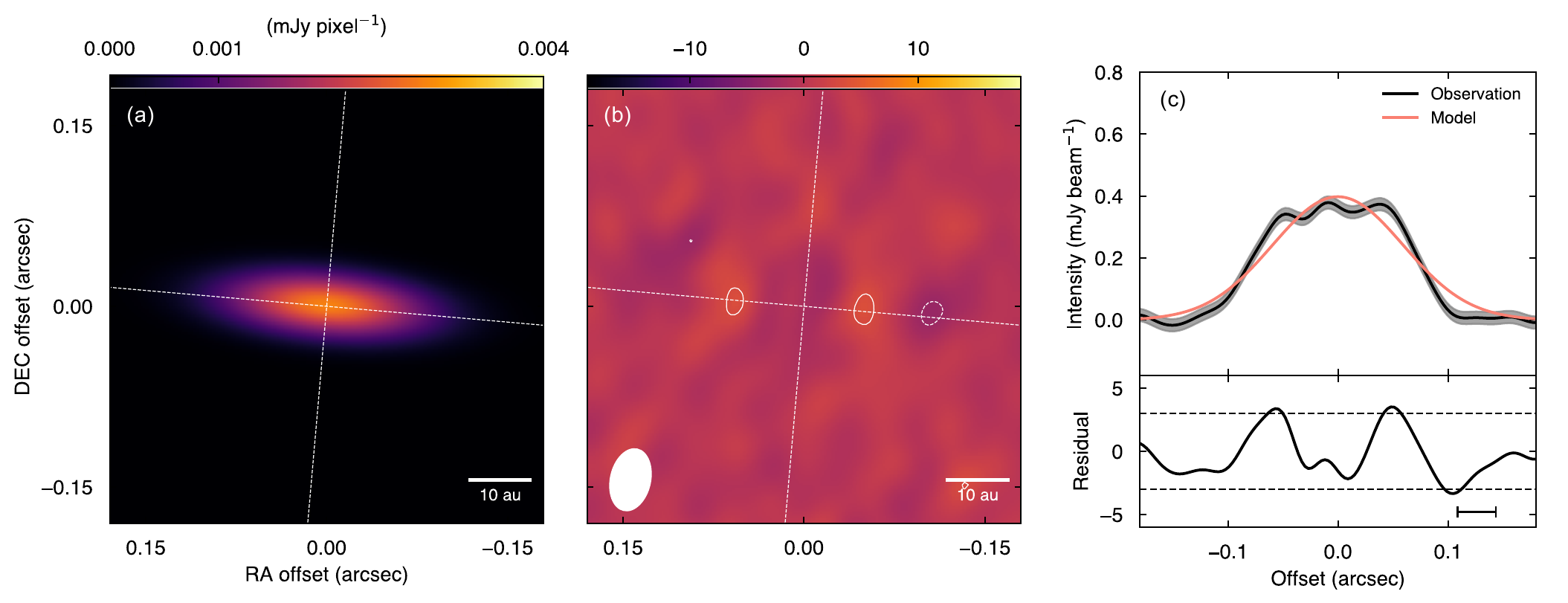}
    \caption{Results of the fitting of a two-dimensional Gaussian function to the dust continuum emission of Ced110 IRS4B. (a) The image of the deconvolved Gaussian and (b) The residual image normalized by $\sigma$. Solid and dashed contours indicate $3\sigma$ and $-3\sigma$, respectively. The white ellipse at the bottom left corner denotes the synthesized beam size. Dashed lines across the maps in panels (a) and (b) show directions of the continuum major and minor axes. (c) The one-dimensional intensity and residual profiles along the continuum major axis with a cut width corresponding to the size of one pixel ($\sim$1$/10$ beam size). Horizontal dashed lines indicate $\pm 3 \sigma$. The bar in the bottom right corner represents the FWHM of the synthesized beam along the continuum major axis.}
    \label{fig:modelfit_irs4b}
\end{figure*}

The structure of the dust continuum emission of Ced110 IRS4B is simpler than that of IRS4A. Thus, we have performed fitting of a two-dimensional Gaussian function to characterize its geometry in the same manner as the fitting for Ced110 IRS4A. The fitting was conducted in the image plane with the MCMC method including the beam convolution. The fitting results are summarized in Table \ref{tab:param_gaussfit_irs4b} and Figure \ref{fig:modelfit_irs4b}. The residual map in Figure \ref{fig:modelfit_irs4b}(b) shows that the dust continuum emission is mostly well reproduced by a two-dimensional Gaussian function. Only marginal residuals of $3\sigma$ appear at the locations of the two intensity peaks on the east and west sides of the protostar. The intensity and residual profiles presented in Figure \ref{fig:modelfit_irs4b}(c) show the correspondence of the residuals and intensity peaks more clearly. We adopt the estimated position angle of the deconvolved Gaussian of $85.0\pm0.6^\circ$ as that of the dust disk. The inclination angle of the dust disk is also estimated to be $72.5\pm0.8^\circ$ from the aspect ratio of the deconvolved Gaussian, assuming a geometry thin disk. Note that this inclination angle should be considered as a lower limit, as the assumption of the geometry thin disk may not be valid.
Combining the fitting results of the ring model for Ced110 IRS4A, the separation between centers of Ced110 IRS4A and IRS4B on the plane of the sky, and the position angle of the center of Ce110 IRS4B relative to that of IRS4A are calculated to be $1\farcs3291 \pm 0\farcs0003$ (or $251.21 \pm0.06~\au$ ) and $84.54 \pm 0.03^\circ$, respectively.

\begin{deluxetable*}{lclc}
\tablecaption{Summary of the fitting of a Gaussian function \label{tab:param_gaussfit_irs4b}}
\tabletypesize{\footnotesize}
\tablehead{
\colhead{Parameter} & \colhead{Unit} & \colhead{Description} & \colhead{Best-fit value}
}

\startdata
$a_0$ & $\times 10^{-3}$ mJy pixel$^{-1}$ & Amplitude of the Gaussian & $2.9\pm0.1$ \\
$x_0$ & & RA coordinate of the center & $11\mathrm{h}06\mathrm{m}46.7718\mathrm{s} \pm 0.0001\mathrm{s}$ \\ 
$y_0$ & & Dec coordinate of the center & $-77\mathrm{d}22\mathrm{m}32.7568\mathrm{s} \pm 0.0008\mathrm{s}$ \\
$\sigma_\mathrm{maj}$ & mas &  Width (standard deviation) along the major axis & $57.5 \pm 0.8 $ \\
$\sigma_\mathrm{min}$ & mas & Width (standard deviation) along the minor axis & $17.3 \pm 0.7$ \\
$pa$ & $^\circ$ & Position angle of the Gaussian & $85.0 \pm 0.6$
\enddata
\end{deluxetable*}

\subsection{Rotation Curve} \label{subsec:rotcurve}
\begin{figure*}[tbhp]
    \centering
    \includegraphics[width=2\columnwidth]{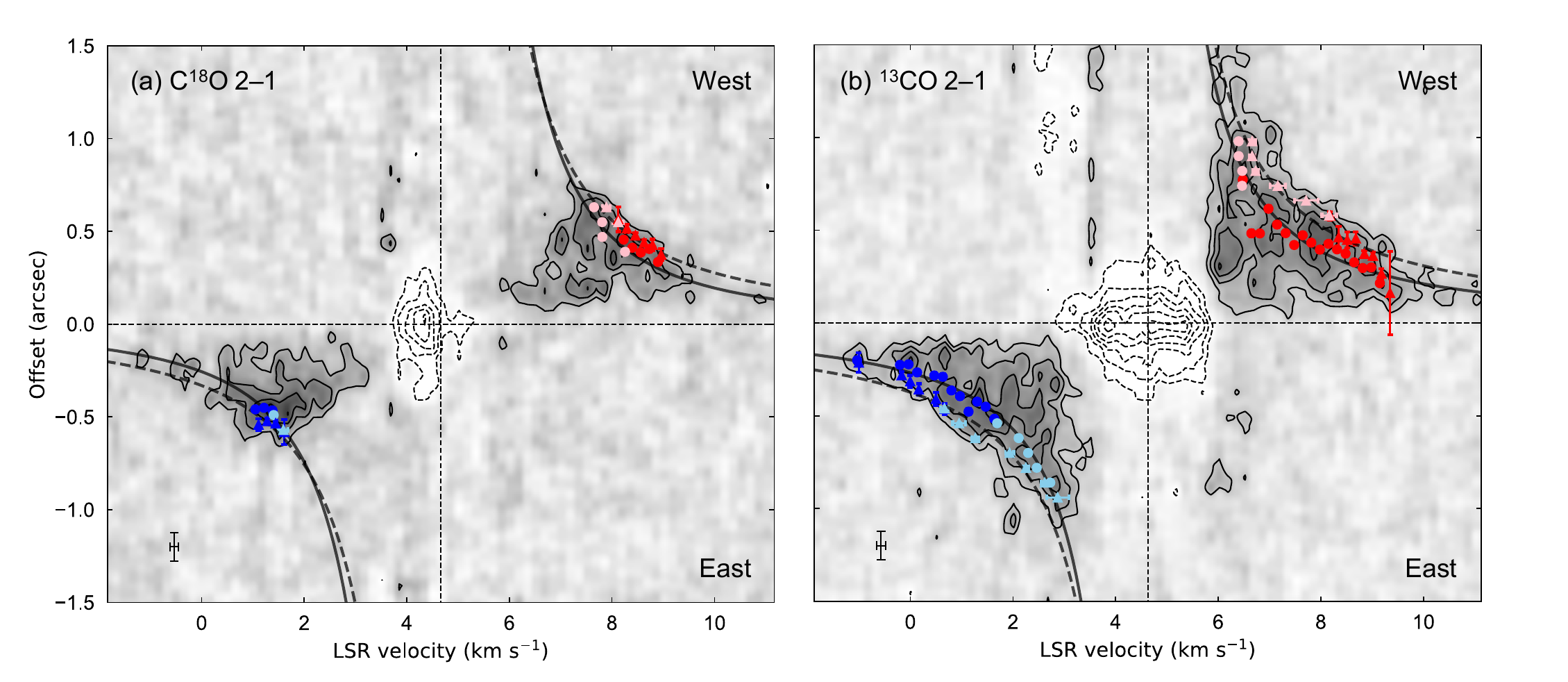}
    \caption{Position-velocity diagrams of (a) C$^{18}$O and (b) $^{13}$CO emission of Ced110 IRS4A cut along its continuum major axis. Contour levels are from 3$\sigma$ to 12$\sigma$ in steps of 3$\sigma$. Dashed contours represent negative intensities. Round and triangle markers denote the determined ridges and edges, respectively. The color of the marker indicates properties of the data points: blue and red markers show measurements of positions at a given velocity; cyan and pink markers show measurements of velocities at a given offset; blue and cyan markers indicate measurements at blue-shifted velocity; and red and pink markers indicate measurements at redshifted velocity. Solid and dashed curves show the power-law functions representing the best fits to the ridges and edges, respectively. The vertical and horizontal bars in the bottom left corners represent the FWHM of the synthesized beam and the velocity resolution, respectively.}
    \label{fig:pvds_curvefit}
\end{figure*}

Rotational motions are clearly seen in CO isotopologue lines around Ced110 IRS4A. These lines are also detected around Ced110 IRS4B, although they do not show clear velocity gradients around the protostar in the moment one maps. In this subsection, rotational motion around the protostars are investigated in more detail using the C$^{18}$O and $^{13}$CO lines, which are less affected by the absorption by the foreground cloud and imaged with higher velocity resolutions by a factor of four than the $^{12}$CO line.

\begin{figure*}[tbhp]
    \centering
    \includegraphics[width=2\columnwidth]{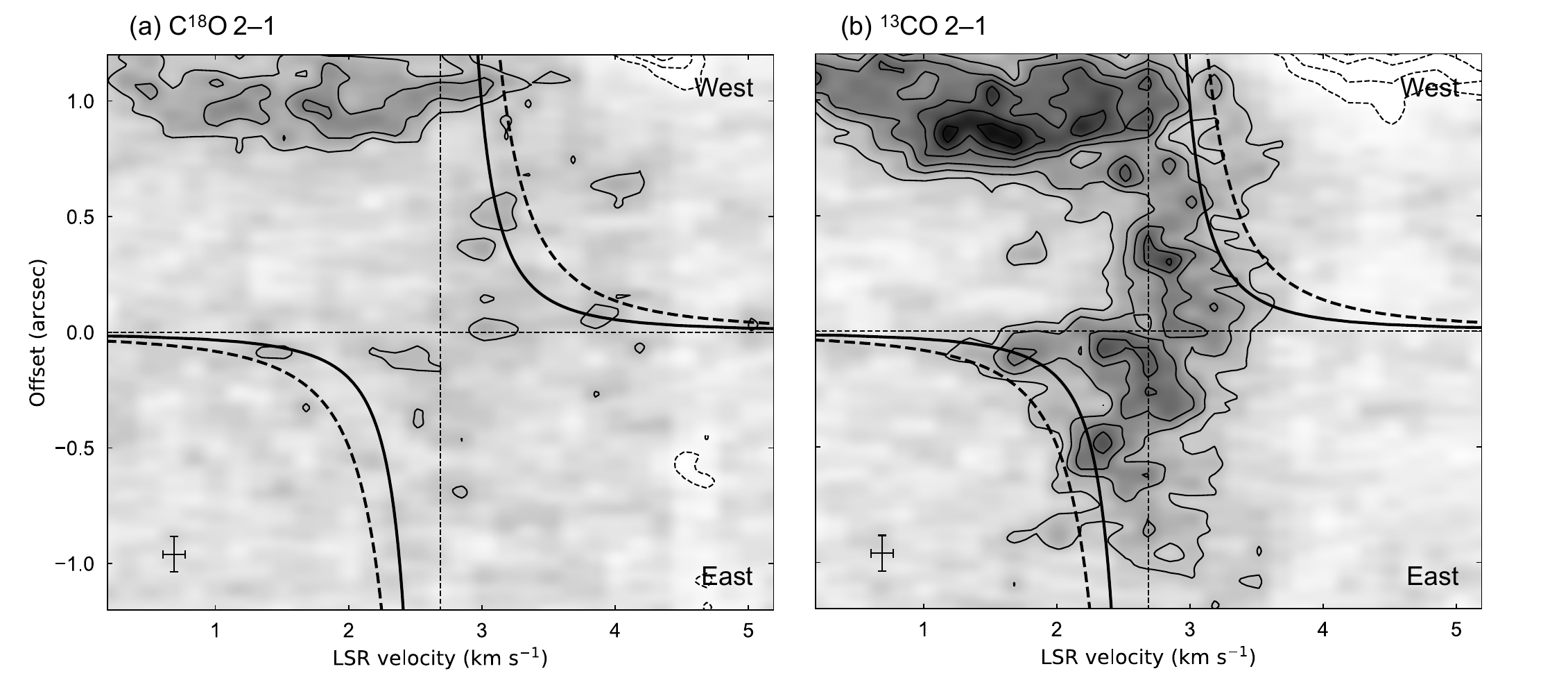}
    \caption{Position-velocity diagrams of (a) C$^{18}$O and (b) $^{13}$CO emission of Ced110 IRS4B cut along its continuum major axis. Contour levels are from 3$\sigma$ to 12$\sigma$ in steps of 3$\sigma$. Dashed contours represent negative intensities. Solid and dashed curves show Keplerian rotation with a central mass of 0.02 and 0.05 $\Msun$, respectively, without correction of the inclination angle. Vertical and horizontal bars in the bottom left corners represent the FWHM of the synthesized beam and the velocity resolution, respectively.}
    \label{fig:pvds_irs4b}
\end{figure*}

\begin{figure}
    \centering
    \includegraphics[width=\columnwidth]{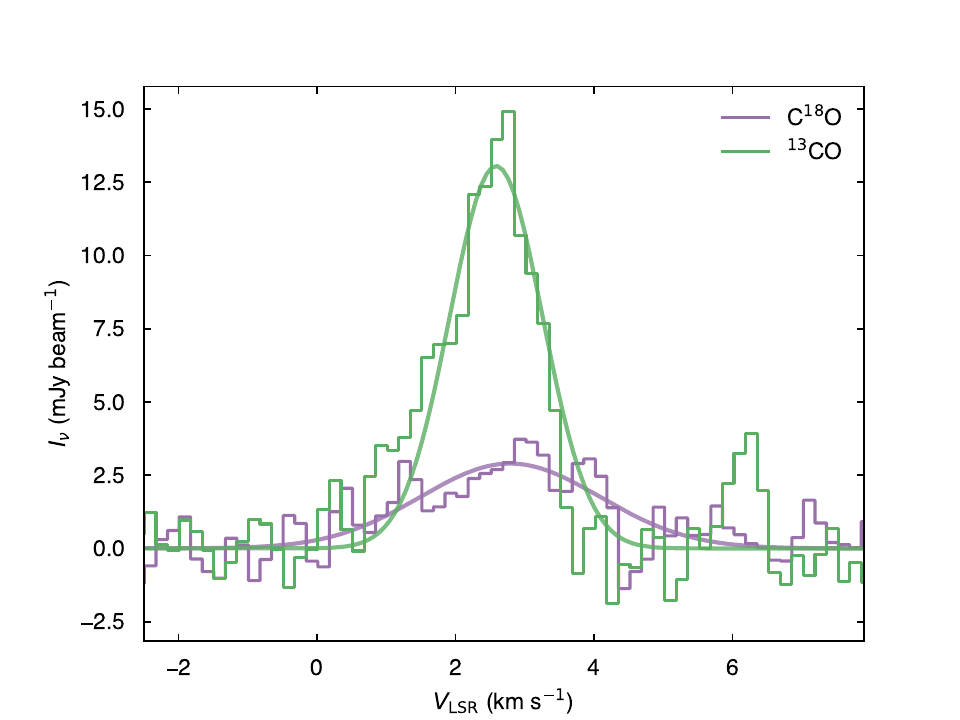}
    \caption{Spectra of the C$^{18}$O and $^{13}$CO $J=2$--1 lines measured within a radius of $0\farcs15$ around Ced110 IRS4B. Curves show the best-fit Gaussian functions.}
    \label{fig:spectra_irs4b}
\end{figure}

\subsubsection{Ced110 IRS4A}
Position-velocity (PV) diagrams of the C$^{18}$O and $^{13}$CO emission along the continuum major axis of Ced110 IRS4A are presented in Figure \ref{fig:pvds_curvefit}. The C$^{18}$O PV diagram clearly exhibits a feature of differential rotation, where velocity increases with decreasing radius. On the other hand, it shows no clear feature of symmetric infall, where both blue- and red-shifted velocity components appear at the same offset. These features imply that the differential motion traces Keplerian rotation of a disk rather than rotation of an infalling envelope. The $^{13}$CO PV diagram shows a similar feature of differential rotation, while it also exhibits faint, blue- and red-shifted emission on the west and east sides, respectively, which is not expected for a pure rotational motion.

To test whether the observed rotational motions originate from a Keplerian disk, we fit a power-law function to the spin-up feature using the \texttt{pvanalysis} tool of the \textit{Spectral Line Analysis/Modeling} \citep[\texttt{SLAM};][]{Aso:2023aa}\footnote{\url{https://github.com/jinshisai/SLAM}} code, and examine its radial dependence. The detail of the fitting method can be found in \cite{Ohashi:2023aa}. Here, we only briefly describe the fitting method. First, the representative ridge \citep[e.g.,][]{Yen:2013aa} and edge \citep[e.g.,][]{Seifried:2016aa} positions (or velocities) are derived from a one-dimensional cut of the PV diagram along the position (or velocity) axis. The ridge is defined as the intensity-weighted mean calculated using emission above a given threshold, while the edge is defined as the outermost contour with the given threshold, where the threshold of 6$\sigma$ is adopted to exclude noise in the analysis here. We use these two different definitions to discuss the lower and upper limit of the dynamical mass based on Keplerian rotation \citep{Maret:2020aa}. Then, we fit the following power-law function to the obtained data points to examine the radial dependence of the differential rotation:
\begin{eqnarray}
    \vrot = |\vlsr - \vsys| = V_0 \left( \frac{R}{R_0} \right)^{-p},
    \label{eq:spl}
\end{eqnarray}
where $\vrot$ is the rotational velocity, $\vlsr$ is the local standard of rest (LSR) velocity of the obtained data points, and $\vsys$ is the systemic velocity. We fix $V_0$ at the mean velocity of the data points, and set the remaining three parameters ($R_0$, $p$, $\vsys$) as free parameters in the fits.

The results from the fits are summarized in Table \ref{tab:rotcurve}. The best-fit power-law indices for edges and ridges for the C$^{18}$O emission are $0.669\pm0.070$ and $0.531\pm0.019$, respectively, which agree with Keplerian rotation ($\vrot \propto R^{-0.5}$) within fitting uncertainties of 2--$3\sigma$. The fitting to the $^{13}$CO emission, on the other hand, results in power-law indices of $0.761 \pm 0.009$ and $0.875 \pm 0.026$, which are larger than that for the Keplerian disk. This difference in the slopes obtained from the C$^{18}$O and $^{13}$CO emission is likely due to the different velocity ranges of the fitted data points. The data points obtained from the C$^{18}$O emission trace relatively high-velocity components of $\vrot \ge 3~\kmps$, while those for the $^{13}$CO emission include lower-velocity components of $\sim$1.5 $\kmps$. The lower-velocity components of the $^{13}$CO emission could trace a rotational motion of the infalling envelope \citep[$\vrot \propto r^{-1}$; e.g.,][]{Yen:2013aa} rather than the rotation of a disk. Moreover, the $^{13}$CO emission would be more affected by the optical depth and spatial filtering effects at lower velocities. We performed the fitting with high-velocity components of the $^{13}$CO emission within a velocity range similar to that for the C$^{18}$O emission ($\vrot \ge 3~\kmps$, corresponding to radii of $\lesssim$120 au), and obtained power-law indices of $0.518\pm0.024$ and $0.596\pm0.030$, which well agrees with Keplerian rotation. Hence, the inner region within a radius of $\sim$120 au is likely a part of a Keplerian disk.

The protostellar mass is estimated from the rotational motion measured with the ridges and edges for the C$^{18}$O emission to be $\sim$1.21 and $\sim$1.45 $\Msun$, respectively, assuming the inclination angle of $76^\circ$. The dynamical mass estimated using ridges and edges assuming Keplerian rotation tends to be under- and over-estimated, respectively \citep{Maret:2020aa}. Therefore, our analysis suggests the protostellar mass of Ced110 IRS4 is within a range of $\sim$1.21--1.45 $\Msun$. The systemic velocity of Ced110 IRS4A is measured to be $4.67 \pm 0.03~\kmps$ {by taking the average of the fitting results for C$^{18}$O ridges and edges, and calculating the error propagation of the fitting errors.

\subsubsection{Ced110 IRS4B}
Figure \ref{fig:pvds_irs4b} shows PV diagrams of the C$^{18}$O and $^{13}$CO emission cut along the continuum major axis of Ced110 IRS4B. To study gas motions relative to the protostellar system, we measure the systemic velocity of Ced110 IRS4B from the spectra of the two lines measured at the protostellar position, presented in Figure \ref{fig:spectra_irs4b}. Both lines exhibit Gaussian-shape spectra, although the $^{13}$CO spectrum would be affected by the absorption by the resolved-out, foreground gas at a velocity of $\sim$3.9 $\kmps$ (see also Figure \ref{fig:channel_13co_zoom}). The Gaussian fitting to the C$^{18}$O and $^{13}$CO spectra results in central velocities of $2.79\pm0.26~\kmps$ and $2.59\pm0.05~\kmps$, respectively. Thus, we adopted the mean value of $2.69 \pm 0.26~\kmps$ as the systemic velocity of Ced110 IRS4B with an uncertainty derived through the error propagation of the fitting errors. In Figure \ref{fig:pvds_irs4b}(a), the C$^{18}$O emission appears around the protostar across LSR velocities of $\sim$1.5--4.0 $\kmps$, although the emission is detected only at 3$\sigma$. The $^{13}$CO emission shows a marginal feature of a differential motion on the east side of the protostar: velocity slightly increases with decreasing radius at offsets of $-1''$ to $0''$ at the blue-shifted velocity. Such a feature of the differential rotation is less clear on the west side of the protostar. This could be because the relatively high-velocity components ($\sim$3.5 $\kmps$) are obscured by the optically thick foreground gas, which is resolved out in our map (see Figure \ref{fig:channel_13co_zoom}). Strong C$^{18}$O and $^{13}$CO emission found around an offset of $1''$ is associated with Ced110 IRS4A, as it is located along the PV cut direction.

While it is difficult to apply the same analysis as that for Ced110 IRS4A because of the less clear feature of the rotational motion, we overlay curves of Keplerian rotation to make a rough estimate of the mass of Ced110 IRS4B assuming the emission is associated with a Keplerian disk. Keplerian curves with the stellar mass of 0.02 and 0.05 $\Msun$ without correction of the inclination angle appear to agree with the intensity peaks and the outermost contours of the emission, respectively. The correction of the inclination angle of $\sim$73$^\circ$ increases the above mass estimate by 5\%. However, we note that the inclination angle of the disk of Ced110 IRS4B is quite uncertain, since the continuum emission is not well-resolved along its minor axis.

\begin{deluxetable*}{lccccccc}
\tablecaption{Results of the rotation curve fitting \label{tab:rotcurve}}
\tabletypesize{\small}
\colnumbers
\tablehead{
\colhead{Line} & \colhead{Velocity range} & \colhead{Ridge/edge} & \colhead{$V_0$} & \colhead{$R_0$} & \colhead{$p$} & \colhead{$\vsys$} & \colhead{$M_\ast$} \\
\colhead{} & \colhead{} & \colhead{} &\colhead{($\kmps$)} & \colhead{(au)} & \colhead{} & \colhead{($\kmps$)} & \colhead{($\Msun$)}
}
\startdata
\multirow{2}{*}{C$^{18}$O} & \multirow{2}{*}{Whole} & Ridge & $3.59$ & $78.6 \pm 0.7$ & $0.531\pm0.019$ & $4.64 \pm 0.01$ & $1.21 \pm 0.01$ \\
                           & & Edge & $3.63$ & $92.0 \pm 1.3$ & $0.669 \pm 0.070$ & $4.70 \pm 0.03$ & $1.45 \pm 0.02$ \\
\multirow{2}{*}{$^{13}$CO} & \multirow{2}{*}{Whole} & Ridge & $3.20$ & $80.6 \pm 0.5$ & $0.761\pm0.009$ & $4.55 \pm 0.01$ & $0.99 \pm 0.01$ \\
                           & & Edge & $3.10$ & $110.4 \pm 1.0$ & $0.876 \pm 0.026$ & $4.59 \pm 0.02$ & $1.27 \pm 0.01$ \\
\multirow{2}{*}{$^{13}$CO} & \multirow{2}{*}{$V_\mathrm{rot} \geq 3~\kmps$} & Ridge & $4.16$ & $57.3 \pm 0.8$ & $0.596\pm0.030$ & $4.61 \pm 0.03$ & $1.19 \pm 0.02$ \\
                           & & Edge & $4.16$ & $75.3 \pm 1.3$ & $0.518 \pm 0.024$ & $4.59 \pm 0.02$ & $1.56 \pm 0.03$ 
\enddata
\tablecomments{Column 1: Line used for the fitting. Column 2: Velocity range of the data used for the rotation curve fitting. Column 3: Definition of the representative points used for the fitting. Columns 4--7: Parameters in Equation \ref{eq:spl}. $V_0$ is the mean velocity of the data points, and the other three parameters are free parameters in the fitting. Column 8: Dynamical mass estimated from the mean velocity $V_0$ assuming Keplerian rotation and the inclination angle of 76$^\circ$.}
\end{deluxetable*}

\section{Discussion}\label{sec:discuss}
\subsection{Structures and Orientation of the Disk of Ced110 IRS4A}
The rotation curves derived in Section \ref{subsec:rotcurve} suggest that the inner region within a radius of $\sim$120 au around Ced110 IRS4A is part of a Keplerian disk. Thus, the dust continuum emission of Ced110 IRS4A with a radius of $\rtsim91.7~\au$ would also originate from a disk rather than a surrounding envelope. In the following subsections, we discuss the structures of the dust disk of Ced110 IRS4A based on the observed dust continuum structures and the orientation of the disk.

\subsubsection{Possible Annular Substructure}
One of the main focuses of this paper is to investigate whether the disk(s) exhibit substructures or not, which might tell us when planet formation begins. The dust continuum emission of Ced110 IRS4A does not show any clear gaps or rings, which are commonly reported in Class \II~disks \citep[e.g.,][]{Andrews:2018aa}. Interestingly, however, the continuum emission shows shallow bumps along its major axis, which can be interpreted as a shallow, ring-like structure at a radius of $\sim$40 au. This might indicate an early formation of an annular substructure in the dust disk. On the other hand, the continuum emission could be optically thick, as it shows a relatively high peak brightness temperature of 79 K, and thus the dust continuum emission might not exactly trace the actual surface density of the disk \citep[e.g.,][]{Ricci:2018aa}. In such a case, non-monotonic temperature profile in the disk could be contributing to the shallow, ring-like structure of the dust continuum emission \citep{Cleeves:2016aa, Facchini:2017aa}. The high inclination angle of the disk could also cause the cold disk mid-plane to obscure some of the emission from the disk itself, which may produce apparent inflections in the intensity profile, when the disk is not geometrically thin.

We can roughly estimate the optical thickness of the continuum emission at the bump radius ($\sim$0$\farcs2$ or $\sim$40 au) by comparing the dust brightness temperature to a typical disk temperature model. The observed brightness temperature is equal to the disk temperature when the dust emission is optically thick. The radial profile of the mid-plane temperature can be expressed as follows, assuming a passively heated, flared disk \citep[][]{Chiang:1997aa, Dalessio:1998aa, Dullemond:2001aa}:
\begin{equation}
    T_\mathrm{mid} (r) = \left( \frac{\varphi L_\ast}{8 \pi r^2 \sigma_\mathrm{SB}} \right)^{0.25},
\end{equation}
where $r$ is the radius, $L_\ast$ is the stellar luminosity, $\sigma_\mathrm{SB}$ is the Stefan--Boltzmann constant, and $\varphi$ is the disk flaring angle in radian. Although the flaring angle is uncertain, the flaring angle of 0.02 ($\sim$1$^\circ$) is often adopted for Class \II~ disks as a conservative assumption \citep[][]{Huang:2018ab}. This flaring angle yields a pressure scale height of $\sim$7 au at a radius of 100 au, assuming that the ratio of the photosphere height to the pressure scale height is about 2.5 as suggested by a simulation work \citep{Baillie:2014aa}. This is roughly consistent with a gas pressure scale height of $\sim$6 au at a radius of 100 au derived from the measured protostellar mass and a temperature calculated with the flaring angle of 0.02. Hence, the adopted flaring angle would be reasonable. Assuming $\varphi=0.02$ and $L_\ast = \lbol = 1~\Lsun$, the disk temperature is estimated to be about 20 K at the radius of 40 au, which is comparable to the observed brightness temperature of $\sim$20 K at 40 au derived with the full Planck function. The resultant temperature varies only by 20--30\%, when the flaring angle and the stellar luminosity changes by a factor of a few. Hence, we could not rule out the possibility that the continuum emission is optically thick around the shoulder radius even considering uncertainties of the assumed parameters by a factor of a few. We note that this is a very rough estimate based on a simple model of the temperature distribution. \cite{van-t-Hoff:2020aa} suggest that young, embedded disks could be warmer than Class \II~disks. A protostellar disk model of \cite{Harsono:2015aa}, where a total luminosity of the central source is $\sim$1 $\Lsun$ and an envelope mass of 1 $\Msun$, predicts a temperature $\sim$40 K at a radius of 40 au. The infalling envelope may also keep embedded disks warmer \citep{Whitney:2003aa, Tobin:2020aa}. Observations at longer wavelengths, where the optically thinner emission is expected, and/or the radiative transfer modeling are required to reveal the details and origin of the ring-like structure of the 1.3 mm continuum emission.

\subsubsection{Asymmetry and Orientation of the Disk \label{subsec:diskasym}}
The dust continuum emission shows asymmetry along both its major and minor axis. The asymmetry along the minor axis is also seen in several protostellar systems of the eDisk sample \citep{Ohashi:2023aa}. These structures can be explained by a geometric effect, which is a combination of the optical depth effect and disk flaring. When the dust continuum emission of an inclined disk is optically thick, we look at a cold disk edge on one side of the plane-of-sky but look at a warm disk surface on the other side along the minor axis \citep[e.g.,][]{Villenave:2020aa, Lin:2023aa}. If this is the case for Ced110 IRS4A, the southern side (brighter side) corresponds to the disk surface facing the observer. However, as the CO and $^{13}$CO emission suggests an opposite disk orientation, this would not be the case for Ced110 IRS4A and the asymmetry of the continuum emission along the minor axis seems to be associated with the temperature or surface-density structure of the disk. The CO and $^{13}$CO emission exhibit asymmetric intensity peaks in the northern and southern parts with respect to the dark lane along the  major axis of the continuum emission and likely arises from layers above the disk mid-plane. Such optically thick lines arising from the disk surface can tell the disk orientation with respect to the plane of the sky: the emitting layer in front of the disk mid-plane, as viewed from the observer, appears brighter than that behind the disk mid-plane \citep{Rosenfeld:2013aa, Pinte:2018ab}. The CO and $^{13}$CO emission of Ced110 IRS4A show stronger intensity peaks on the northern side, thus, suggesting that the northern side corresponds to the disk surface facing the observer. This is opposite to what is inferred from the continuum emission when assuming an optically thick and flared dust disk. Hence, if we assume that the orientations of the gas and dust disks indeed are consistent, the implication is that the dust disk is \emph{not} significantly flared. Also, as the continuum emission shows a clear asymmetry along the major axis, the ring-like structure could be asymmetric in the azimuthal direction and thus result in an asymmetry along the minor axis. The ring-like structure is not resolved along the minor axis because of the limited spatial resolution and its high optical depth at the center. Observations at higher angular resolution or longer wavelengths would reveal the asymmetry of the ring-like structure in more detail.

Note that previous studies suggest that an outflow is tilted to the near and far sides against the observer on the southern and northern sides of the protostar, respectively, on larger scales of $\sim$1000--10000 au \citep{Zinnecker:1999aa, Pontoppidan:2005aa, Hiramatsu:2007aa}, which is opposite to the tilt of the disk normal inferred from the CO and $^{13}$CO emission. However, the large-scale outflow, which should be ejected in past, would not necessarily be aligned orthogonal to the disk in the current epoch. A small tilt of $\sim$15$^\circ$ can change near and far sides of the outflow against the observer, as the disk is close to edge-on ($i\sim76^\circ$). As outflows typically have opening angles of $\gtrsim20^\circ$ \citep{Arce:2006aa, Velusamy:2014aa}, a part of the outflow cavity of Ced110 IRS4A comes to the near and far sides against the observer on the southern and northern sides of the protostar, respectively, even when the axis of the large-scale outflow is aligned with the current disk normal. Therefore, if one side of the outflow cavity is enhanced by interaction with an inhomogeneous envelope, it may cause the outflow orientation to appear opposite to that expected from the disk orientation. Our CO map does not show any clear bipolar outflow on smaller scales of $\sim$100--1000 au (see Figure \ref{fig:channel_12co_wide}). This could be because the outflow has a small projected velocity due to the large inclination angle of the disk (i.e., outflow axis nearly parallel to the plane-of-sky), and thus is obscured by the extended, optically-thick foreground gas. Observations of the outflow on scales of $\lesssim$1000 au would help to put a stronger constraint on the disk orientation.

\subsection{Extended Arc-like Structure}
The SO and C$^{18}$O emission show arc-like structures on the northern side of the protostellar system at a radius of $\sim$1100 au ($\sim$6$''$). A similar structure is also seen in the other lines, H$_2$CO and $c$-C$_3$H$_2$ of our data (see Appendix \ref{sec:app_mommaps}) and at near-infrared wavelengths \citep{Pontoppidan:2005aa}. A possible interpretation of this structure is a shocked shell caused by an outflow, as the SO line is often considered as a shock tracer \citep{Bachiller:1997aa, Wakelam:2005aa}. The arc-like structure is seen at LSR velocities of 4.69--6.02 $\kmps$ (see Figure \ref{fig:channel_c18o_wide} and \ref{fig:channel_so_wide}), which are red-shifted with respect to the systemic velocity of 4.67 $\kmps$. This is opposite to the disk orientation inferred from the CO emission but consistent with the orientation of the reflection nebula and the large-scale outflow on scales of $\ge 1000$ au \citep{Zinnecker:1999aa, Pontoppidan:2005aa, Hiramatsu:2007aa}. Thus, the arc-like structure could be formed by the large-scale outflow ejected in past. The LSR velocities of the arc-like structure correspond to velocities of about zero to 1.3 $\kmps$ relative to the systemic velocity. The highest velocity of the arc-like structure is comparable to the escape velocity of 1.37--1.50 $\kmps$ at the radius of $\sim$1130 au, assuming the protostellar mass of 1.21--$1.45~\Msun$. Although the inclination angle of the arc-like structure is uncertain as it is not necessarily aligned with the disk normal, the inclination angle of the system is estimated to be $70\pm5^\circ$ to explain the color gradient of the reflection nebula \citep{Pontoppidan:2005aa}. Assuming $i=70^\circ$, the deprojected velocity of the arc-like structure can be up to $\sim$3.8 $\kmps$ with respect to the systemic velocity and much higher than the escape velocity. Thus, this structure could be moving outward to escape from the system.

An infalling flow could be another possible origin of the arc-like structure, as non-axisymmetric, infalling flows are reported in a couple of protostellar systems \citep[e.g.,][]{Yen:2014aa, Yen:2019ac, Garufi:2022aa, Thieme:2022aa}. However, the trajectory of the arc-like structure is not connected to the disks associated with either Ced110 IRS4A or IRS4B. Moreover, the largest velocity of the arc-like structure is close to the escape velocity. Considering that the trajectory of the arc-like structure is not quite along the line of sight, the observed velocity would be too high for the arc-like structure to be accreted onto the protostellar system. The arc-like structure could also be a part of a filamentary structure of the foreground cloud. However, previous single-dish observations of the parental cloud in N$_2$H$^{+}$ show that it has an LSR velocity of around 4.3 $\kmps$, which is smaller than the highest velocity of the arc-like structure by 1.7 $\kmps$. Hence, the arc-like structure would not likely be a part of a filamentary structure in the foreground cloud.

\subsection{Binary System}
The binary system consisting of Ced110 IRS4A and IRS4B with a projected separation of $\sim$250 au is spatially resolved from the current observations. The properties of the two protostars are summarized in Table \ref{tab:binary}. The stellar mass ratio ($q =M_2/M_1$, where $M_1$ and $M_2$ are the primary and secondary stellar masses, respectively) of $\rtsim0.03$ is much smaller than typical values in Class \II~multiple systems \citep[$q\gtrsim0.1$;][]{Woitas:2001aa,Manara:2019ab} but similar to those of Class \II~binaries with planetary mass companions \citep[e.g., GQ Lup;][]{Wu:2017aa}. Given the relatively close sky separation, structures of the disks might be affected by the dynamical interaction between the two protostars, as it is hinted in Class \II~sources \citep[e.g.,][]{Manara:2019ab, Zurlo:2020aa, Zurlo:2021aa}. A survey observation of 32 Class \II~disks in the Taurus star-forming region has revealed that circumstellar dust disks in multiple systems of their samples ($q\gtrsim0.1$) tend to be smaller than those around single sources \citep{Manara:2019ab}, which may suggest that the disks in the multiple systems are truncated by the dynamical interaction \citep[e.g.,][]{Papaloizou:1977aa, Artymowicz:1994aa, Rosotti:2018aa}. The maximum dust disk radius in these multiple systems was $\rtsim80~\au$ and the ratio between the dust disk radius and the projected separation ($R/a$) was always less than $0.3$ \citep{Manara:2019ab}. Similar trends and maximum dust disk radii have been also observed in the surveys of Class \II~disks in the Lupus and Ophiuchus star-forming regions \citep{Zurlo:2020aa, Zurlo:2021aa}. In these studies of Class \II~disks, the disk radius is defined as the radius enclosing 90\% or 95\% of the flux density, similar to our study.

The dust disk radius of Ced110 IRS4A is larger than the maximum dust disk radius found in the Class \II~multiple systems mentioned above. The ratio between the dust disk radius and the projected separation for Ced110 IRS4A ($R/a \tsim 0.37$) is also larger than those reported in the Class \II~sources. On the other hand, the disk radius and $R/a$ of Ced110 IRS4B are in the range of those of the Class \II~multiple systems. These results could be due to the small stellar mass ratio of $q\tsim0.026$ observed in Ced110 IRS4. In a system with a smaller stellar mass ratio, dynamical interaction is expected to have a stronger effect on the disk around the secondary star but a less impact on the disk around the primary source \citep{Papaloizou:1977aa, Rosotti:2018aa}. It should be, however, noted that dust disk size can be also smaller than gas disk size due to radial drift of large dust grains \citep{Weidenschilling:1977ab}. This may also explain the larger dust disk of Ced110 IRS4A compared to those of the Class \II~multiple systems, since more radial drift would likely occur in more evolved disks. To estimate the expected truncation radius, we follow an analytic formula in \cite{Manara:2019ab}. Assuming the stellar mass ratio of 0.03 and a non-eccentric orbit, we obtain truncation radii for the disks of Ced110 IRS4A and IRS4B to be $\rtsim140~\au$ and $\rtsim30~\au$, respectively. The truncation radius calculated for Ced110 IRS4B is comparable with the observed dust disk radius. The expected truncation radius can be smaller approximately by a factor of two if we adopt an eccentricity of $\rtsim0.3$ \citep{Manara:2019ab}. Thus, the observed disk radius of Ced110 IRS4A could be also consistent with the tidal truncation if the binary orbit has an eccentricity between zero to 0.3. So far, no observations have constrained the binary orbit of Ced110 IRS4. More observations to reveal orbital motions of the protostars are required in order to evaluate how the disk is affected by the dynamical interaction of two protostars in more detail.

Although the binary orbit is unknown, the line-of-sight velocity difference of the protostars of $\sim$2.1 $\kmps$ agrees with the expected Keplerian velocity of 2.1--2.3 $\kmps$ at a radius of 250 au from Ced110 IRS4A, assuming the protostellar mass of Ced110 IRS4A of 1.21--1.45 $\Msun$ and that the two protostars are gravitationally bound. This may suggest that the orbital plane of the binary system is close to edge-on as well as the circumstellar disks, although more observational constraints on the binary orbit, such as proper motions, are necessary for confirming this. Ced110 IRS4B is on the eastern side of Ced110 IRS4A and has a blue-shifted velocity with respect to Ced110 IRS4A. Similarly, the CO isotopologue lines show blue-shifted velocities on the eastern sides both of Ced110 IRS4A and IRS4B with respect to their systemic velocities. This indicates that the orbital motion of Ced110 IRS4B and rotational motions of the disks/envelopes of Ced110 IRS4A and IRS4B are all in the same direction.

The relatively small projected separation of $\sim$250 au can be consistent with both disk fragmentation due to gravitational instability \citep{Adams:1989aa, Bonnell:1994aa} and turbulent fragmentation of dense cores \citep[][]{Offner:2010aa, Lee:2019ab}, as companions formed via turbulent fragmentation can migrate inward from thousands au to less than 100 au within a time scale of a few 10 kyr \citep[][]{Offner:2010aa, Lee:2019ab}. However, the ordered rotational motions and the large difference in protostellar and disk masses in the Ced110 IRS4 system would favor disk fragmentation over turbulent fragmentation. Although the disks around the two protostars are slightly misaligned by $\rtsim19 \pm 0.6 ^\circ$, which is naturally expected in turbulent fragmentation of dense cores \citep{Padoan:2002aa, Bate:2010aa, Offner:2010aa}, such small misalignment of $\lesssim20^\circ$ can also appear in multiple systems formed via disk fragmentation \citep{Stamatellos:2009aa, Bate:2018aa}.

\begin{deluxetable*}{lccccc}
\tablecaption{Summary of properties of the binary system \label{tab:binary}}
\tabletypesize{\small}
\tablehead{
\colhead{Source} & \colhead{$M_\ast$} & \colhead{$\vsys$} & \colhead{$M_\mathrm{disk}$~$^a$} & 
\colhead{$R_\mathrm{disk}$~$^b$} & \colhead{PA$_\mathrm{disk}$} \\
\colhead{} & \colhead{($\Msun$)} & \colhead{($\kmps$)} & \colhead{($\Msun$)} & \colhead{(au)} & \colhead{($^\circ$)}
}
\startdata
Ced110 IRS4A & 1.21--1.45 & $ 4.67 \pm 0.03$ & $(2.962 \pm 0.001)\times 10^{-2}$ & $91.7 \pm 0.2$ & $103.72\pm0.02$ \\
Ced110 IRS4B & 0.02--0.05 & $2.69\pm0.26$ & $(6.29\pm 0.03)\times 10^{-4}$ & $33.6 \pm 0.6$ & $85.0 \pm 0.6$ 
\enddata
\tablecomments{$^a$Disk gas mass, calculated from the dust mass derived with a disk temperature of 20 K (Table \ref{tab:cont_flux}) assuming the gas-to-dust mass ratio of 100. $^b$Radius of the dust disk.}
\end{deluxetable*}

\section{Conclusion} \label{sec:summary}
We have observed Ced110 IRS4, a Class 0/I protostellar system in the Chamaeleon I dark cloud, with ALMA at angular resolutions of $\sim$0$\farcs05$ ($\sim$10 au). The main results of our observations are as follows:

\begin{enumerate}
    \item The observations in the 1.3 mm dust continuum emission reveal that Ced110 IRS4 is a binary system with a projected separation of $\sim$250 au ($\sim$1$\farcs3$). In addition, a weak, extended emission and an arc-like structure connected to the extended emission are found at a radius of $\sim$6$''$ on the northern side of the protostellar system.
    \item The primary dust continuum emission associated with Ced110 IRS4A exhibits a disk-like shape with a radius of $\rtsim0\farcs 485$ ($\rtsim91.7~\au$), and likely traces a dust disk around the protostar, as also confirmed from gas kinematics with molecular lines. The continuum emission shows no clear gaps or rings that are commonly found in Class \II~sources, but exhibits small bumps along its major axis with an asymmetry. The bumps can be explained by a shallow, ring-like structure at a radius of $\rtsim0\farcs2$ ($\rtsim40~\au$) in the dust continuum emission, as demonstrated by the intensity distribution models. This might indicate a possible, annular substructure of the dust disk. However, the dust continuum emission does not necessarily trace the surface density structure of the disk, as it could be optically thick at the bump radii. More observations at optically thin wavelengths would confirm whether the ring-like structure originates from a substructure in the density distribution. 
    \item The CO isotopologue lines ($J=2$--1) are detected around Ced110 IRS4A. These lines show velocity gradients composed of blue- and red-shifted velocity components on the eastern and western sides of the protostar, respectively. The asymmetric intensity peaks of the $^{13}$CO and CO emission, tracing the disk emitting surfaces, suggest that the disk surface on the northern side is facing us. The PV diagrams of the C$^{18}$O and $^{13}$CO emission exhibit a signature of a differential rotation. The radial dependence of the rotational velocity is examined through fitting using the PV diagrams. The rotational velocity is proportional to $r^{\sim-0.5}$ within a radius of $\sim$120 au, which suggests the presence of a Keplerian disk. This confirms that the dust continuum emission of Ced110 IRS4A arises from a dusty disk around the protostar. The protostellar mass of Ced110 IRS4 is dynamically estimated to be 1.21--1.45 $\Msun$ with an inclination angle of $76^\circ$.
    \item The dust continuum emission associated with Ced110 IRS4B has a radius of $\sim$30 au, and exhibits three intensity peaks, though these continuum structures are marginal considering the rms noise. The PV diagram of the $^{13}$CO emission of Ced110 IRS4B shows a marginal feature of the differential rotation. Assuming Keplerian rotation, the protostellar mass is estimated to be $\sim$0.02--0.05 $\Msun$.
    \item The C$^{18}$O and SO emission show arc-like structures and weak extended components on the north side of the protostellar system as well as the dust continuum emission. As the velocity of the arc-like structures is close to the escape velocity expected for the protostellar mass, they could be associated with shocked gas caused by an outflow.
    \item The rotational motions of the disks/envelopes of Ced110 IRS4A and IRS4B, and possibly the orbital motion of Ced110 IRS4B if the two protostars are gravitationally bound, are all in the same direction, which seems to favor the binary formation scenarios of the disk fragmentation due to gravitational instability.
\end{enumerate}

\section*{Acknowledgments}
This paper makes use of the following ALMA data: ADS/JAO.ALMA\#2019.1.00261.L. ALMA is a partnership of ESO (representing its member states), NSF (USA) and NINS (Japan), together with NRC (Canada), MOST and ASIAA (Taiwan), and KASI (Republic of Korea), in cooperation with the Republic of Chile. The Joint ALMA Observatory is operated by ESO, AUI/NRAO and NAOJ. The National Radio Astronomy Observatory is a facility of the National Science Foundation operated under cooperative agreement by Associated Universities, Inc. H.-W.Y.\ acknowledges support from the National Science and Technology Council (NSTC) in Taiwan through the grant NSTC 110-2628-M-001-003-MY3 and from the Academia Sinica Career Development Award (AS-CDA-111- M03). N.O.~and C.F.~acknowledges support from National Science and Technology Council (NSTC) in Taiwan through the grants NSTC 109-2112-M-001-051 and 110-2112-M-001-031. J.J.T.\ acknowledges support from NASA XRP 80NSSC22K1159. J.K.J.\ and R.S.\ acknowledge support from the Independent Research Fund Denmark (grant No. 0135-00123B). S.T.\ is supported by JSPS KAKENHI Grant Numbers 21H00048 and 21H04495. K.S.~is supported by JSPS KAKENHI Grant No.~21H04495. This work was supported by NAOJ ALMA Scientific Research Grant Code 2022-20A. Z.Y.D.L.\ acknowledges support from NASA 80NSSC18K1095, the Jefferson Scholars Foundation, the NRAO ALMA Student Observing Support (SOS) SOSPA8-003, the Achievements Rewards for College Scientists (ARCS) Foundation Washington Chapter, the Virginia Space Grant Consortium (VSGC), and UVA research computing (RIVANNA). P.M.K.\ acknowledges support from NSTC 108-2112- M-001-012, NSTC 109-2112-M-001-022 and NSTC 110-2112-M-001-057. Y.A.\ acknowledges support by NAOJ ALMA Scientific Research Grant code 2019-13B, Grant-in-Aid for Scientific Research (S) 18H05222, and Grant-in-Aid for Transformative Research Areas (A) 20H05844 and 20H05847. I.d.G.\ acknowledges support from grant PID2020-114461GB-I00, funded by MCIN/AEI/10.13039/501100011033. W.K.\ was supported by the National Research Foundation of Korea (NRF) grant funded by the Korea government (MSIT) (NRF-2021R1F1A1061794). S.P.L.\ and T.J.T.\ acknowledge grants from the National Science and Technology Council of Taiwan 106-2119-M-007-021-MY3 and 109-2112-M-007-010-MY3. C.W.L.\ is supported by the Basic Science Research Program through the National Research Foundation of Korea (NRF) funded by the Ministry of Education, Science and Technology (NRF- 2019R1A2C1010851), and by the Korea Astronomy and Space Science Institute grant funded by the Korea government (MSIT; Project No. 2022-1-840-05). J.E.L.\ is supported by the National Research Foundation of Korea (NRF) grant funded by the Korean government (MSIT) (grant number 2021R1A2C1011718). Z.Y.L.\ is supported in part by NASA 80NSSC20K0533 and NSF AST-1910106. L.W.L.\ acknowledges support from NSF AST-2108794. S.M.\ is supported by JSPS KAKENHI Grant Numbers JP21J00086 and 22K14081. J.P.W.\ acknowledges support from NSF AST-2107841.

\vspace{5mm}
\facilities{ALMA}


\software{CASA \citep{McMullin:2007aa}, Numpy \citep{Oliphant:2006aa,van-der-Walt:2011aa}, Scipy \citep{Virtanen:2020aa}, Astropy \citep{Astropy-Collaboration:2013aa,Astropy-Collaboration:2018aa}, Matplotlib \citep{Hunter:2007aa}, \texttt{emcee} \citep[][]{Foreman-Mackey:2013aa}, SLAM (\url{https://github.com/jinshisai/SLAM})}



\restartappendixnumbering
\appendix

\section{Molecular Line Maps} \label{sec:app_mommaps}

The velocity channel maps of CO isotopologues and SO lines are presented in Figure \ref{fig:channel_c18o_zoom}--\ref{fig:channel_so_wide}, and moment zero maps of the other detected lines are shown in Figure \ref{fig:moments_gallery}. Moment zero maps are produced by integrating the emission detected above 3$\sigma$. Details of the maps of the other lines are summarized in Table \ref{tab:maps_app}.

\begin{figure*}[tbhp]
    \centering
    \includegraphics[width=2\columnwidth]{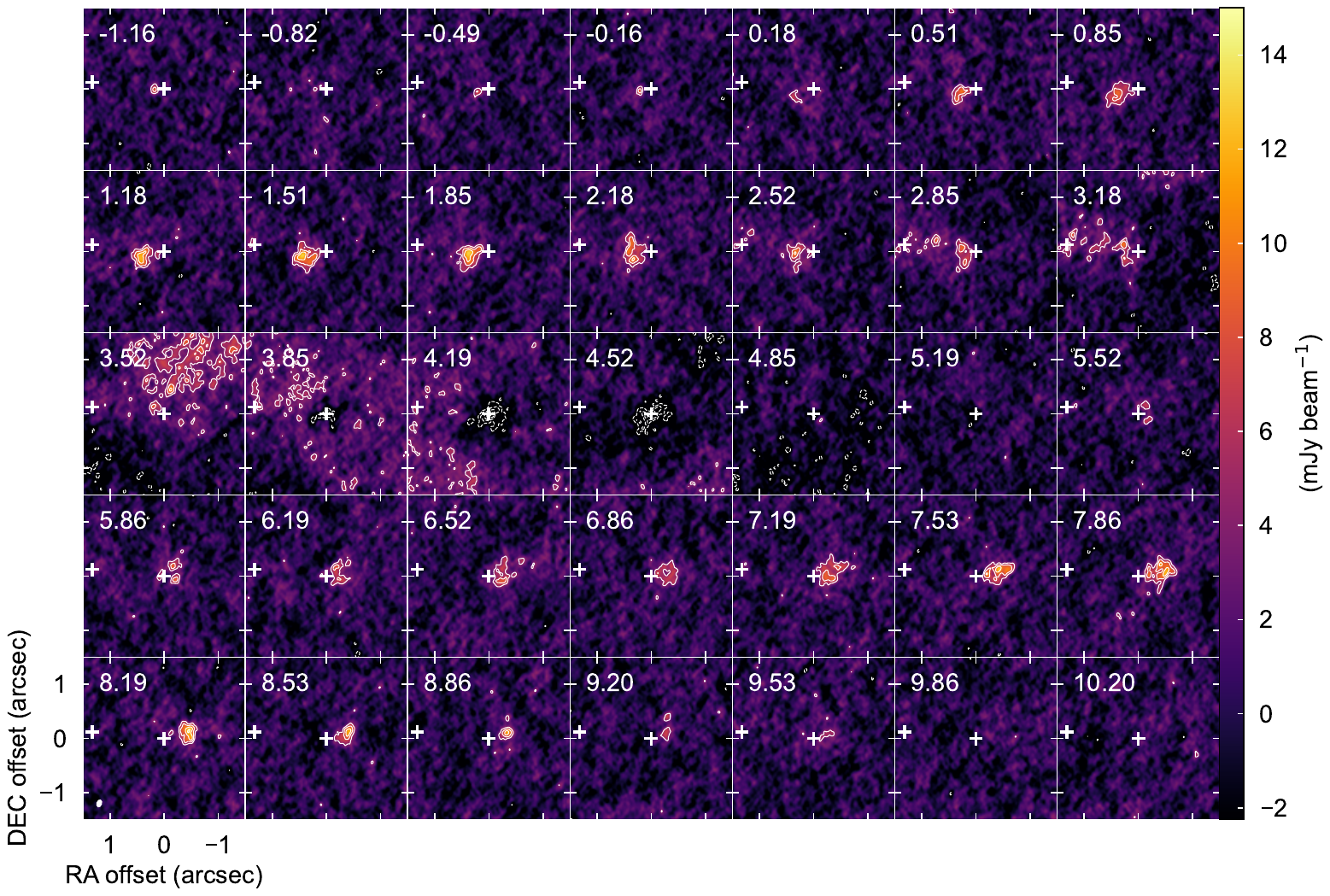}
    \caption{Zoom-in view of velocity channel maps of the C$^{18}$O $J=2$--1 emission. Contour levels are 3, 5, 7, 9... $\times \sigma$, where $\sigma$ is the rms noise listed in Table \ref{tab:maps}. Dashed contours show negative emission. Crosses indicate positions of Ced110 IRS4A and IRS4B. Labels at the top left corners denote the LSR velocities. The filled ellipse at the bottom left corner denotes the synthesized beam size.}
    \label{fig:channel_c18o_zoom}
\end{figure*}

\begin{figure*}[tbhp]
    \centering
    \includegraphics[width=2\columnwidth]{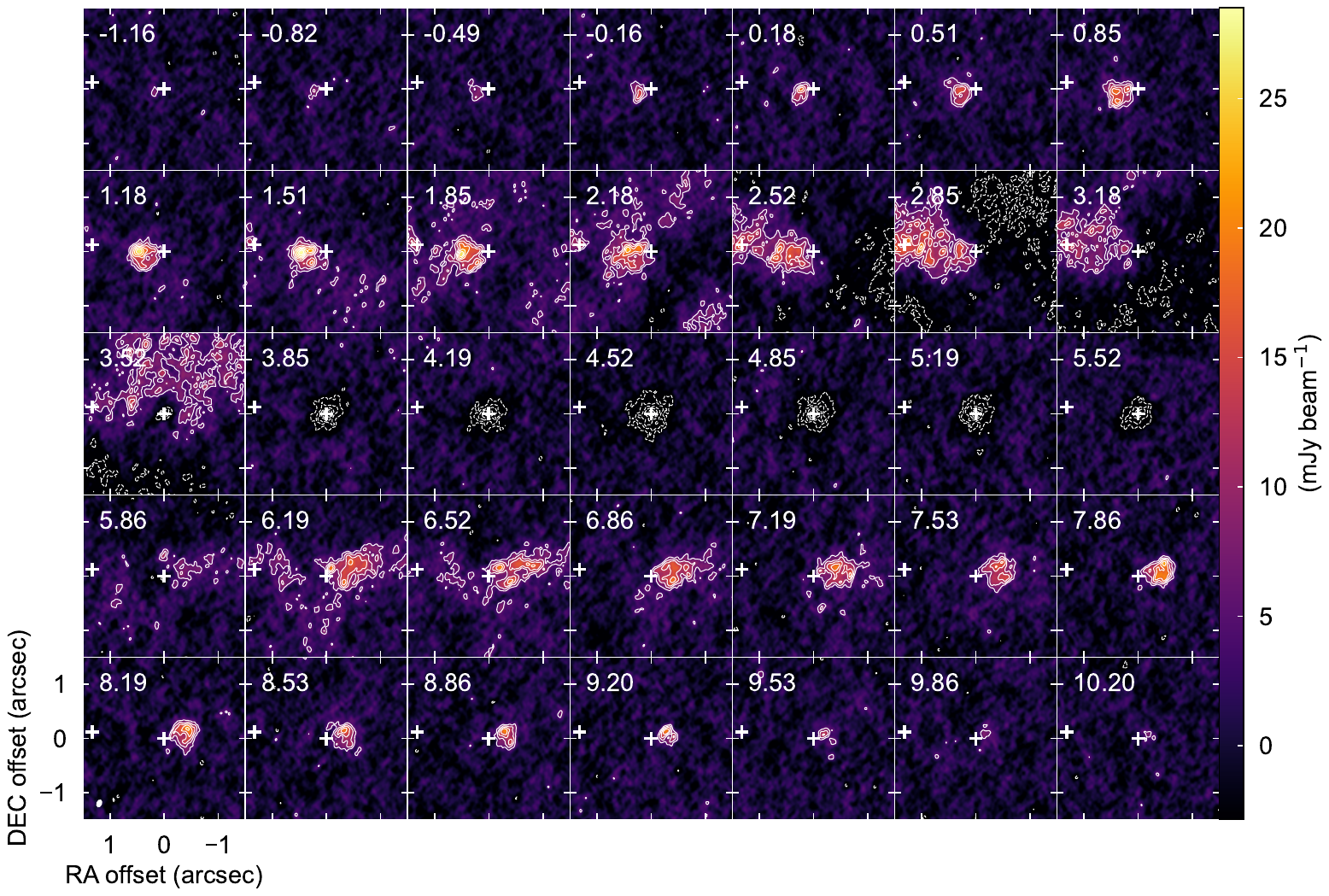}
    \caption{Same as Figure \ref{fig:channel_c18o_zoom} but for the $^{13}$CO $J=2$--1 emission.}
    \label{fig:channel_13co_zoom}
\end{figure*}

\begin{figure*}[tbhp]
    \centering
    \includegraphics[width=2\columnwidth]{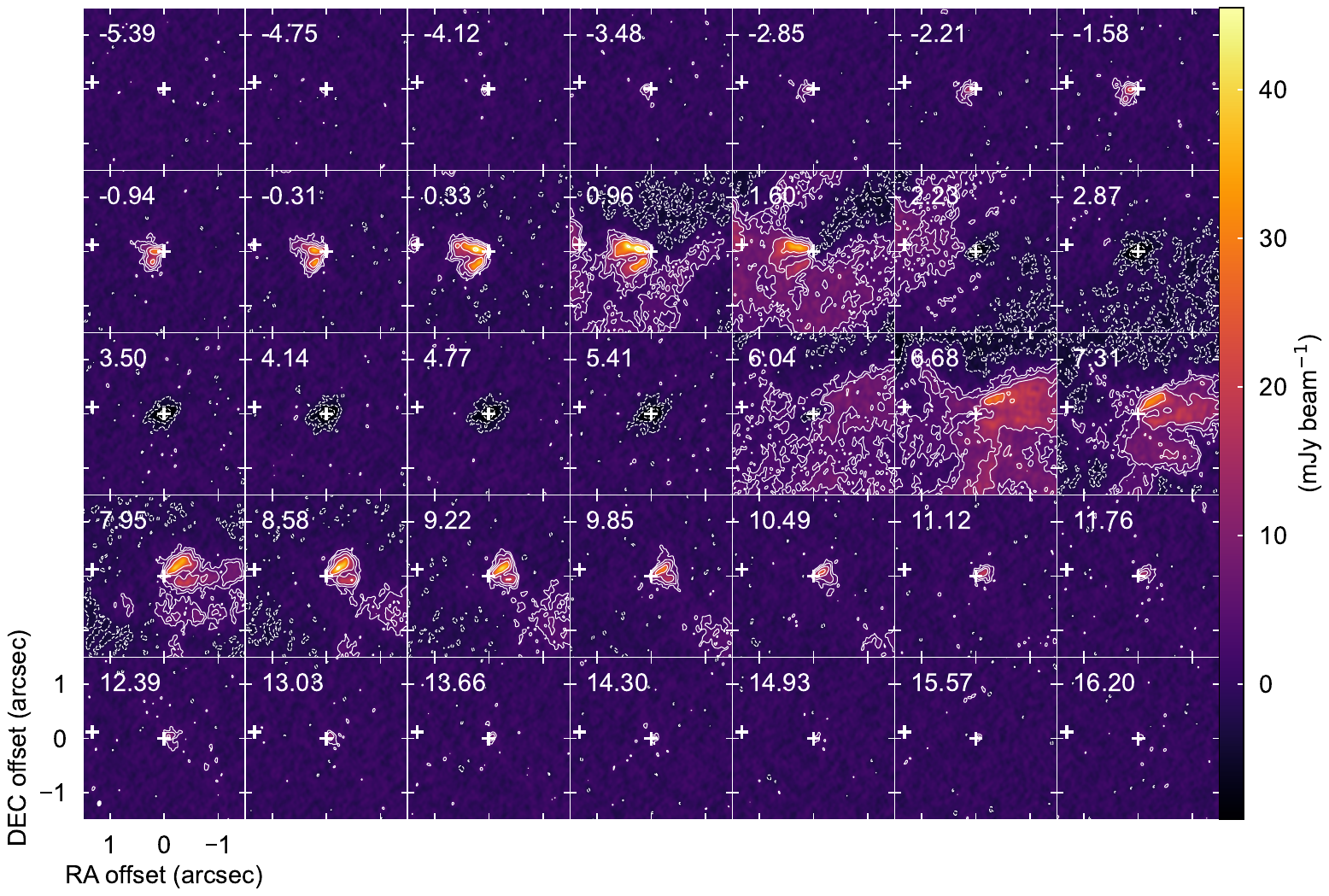}
    \caption{Same as Figure \ref{fig:channel_c18o_zoom} but for the $^{12}$CO $J=2$--1 emission with contour levels of 3, 6, 12, 24, ... $\times \sigma$.}
    \label{fig:channel_12co_zoom}
\end{figure*}

\begin{figure*}[tbhp]
    \centering
    \includegraphics[width=2\columnwidth]{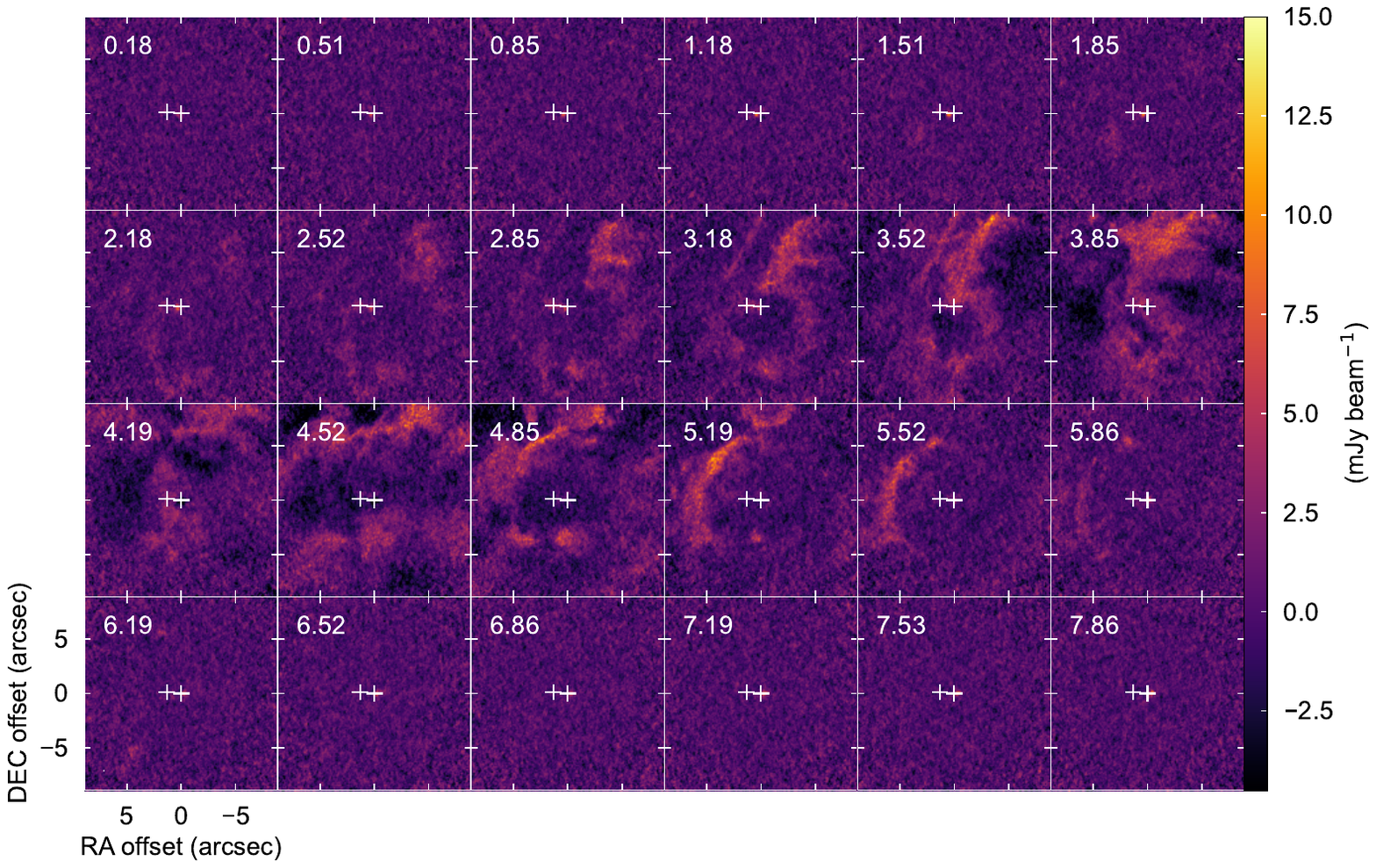}
    \caption{Wide view of velocity channel maps of the C$^{18}$O $J=2$--1 emission. Crosses indicate positions of Ced110 IRS4A and IRS4B. Labels at the top left corners denote the LSR velocities. The filled ellipse at the bottom left corner represents the synthesized beam size.}
    \label{fig:channel_c18o_wide}
\end{figure*}

\begin{figure*}[tbhp]
    \centering
    \includegraphics[width=2\columnwidth]{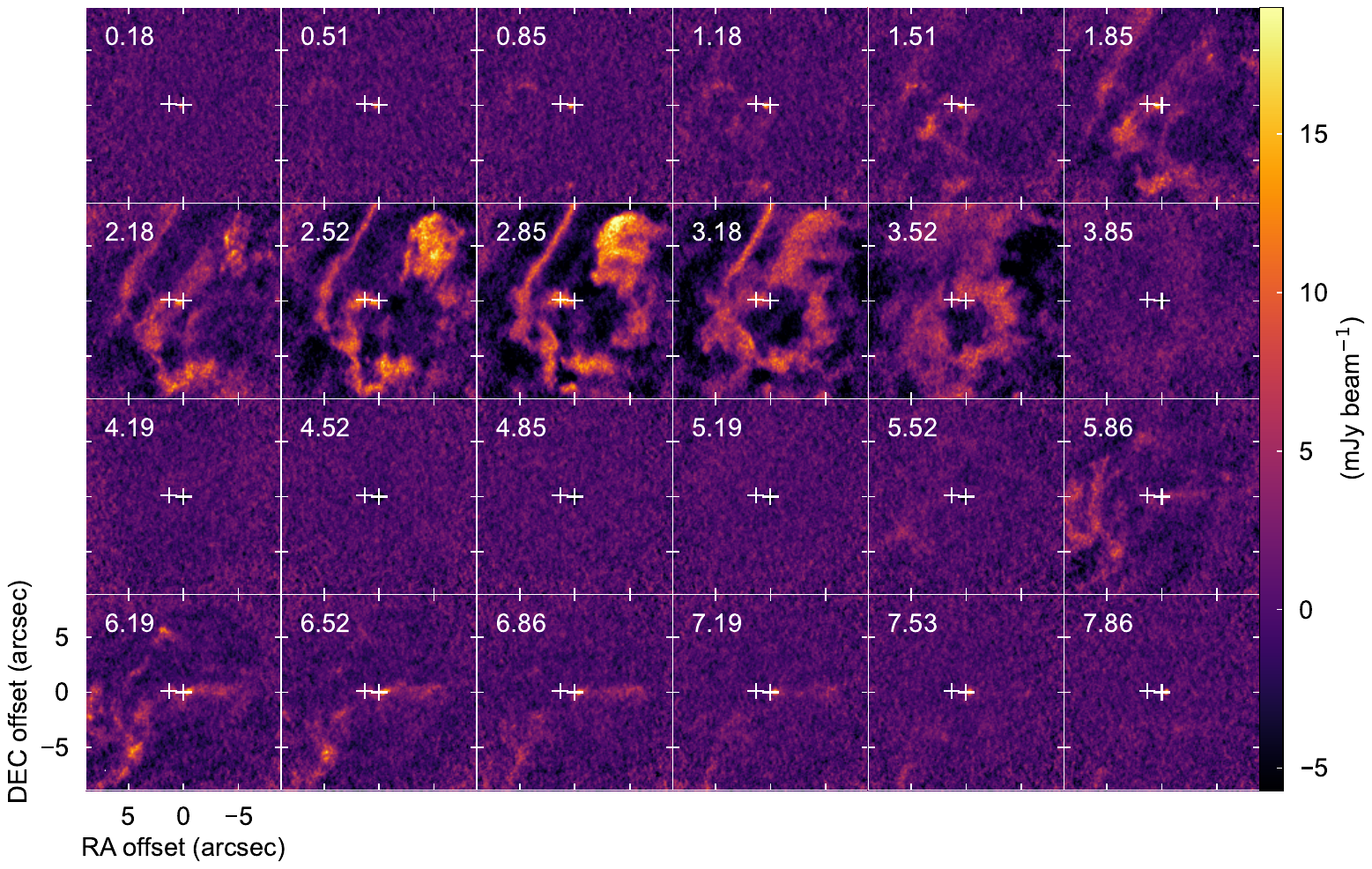}
    \caption{Same as Figure \ref{fig:channel_c18o_wide} but for the $^{13}$CO $J=2$--1 emission.}
    \label{fig:channel_13co_wide}
\end{figure*}

\begin{figure*}[tbhp]
    \centering
    \includegraphics[width=2\columnwidth]{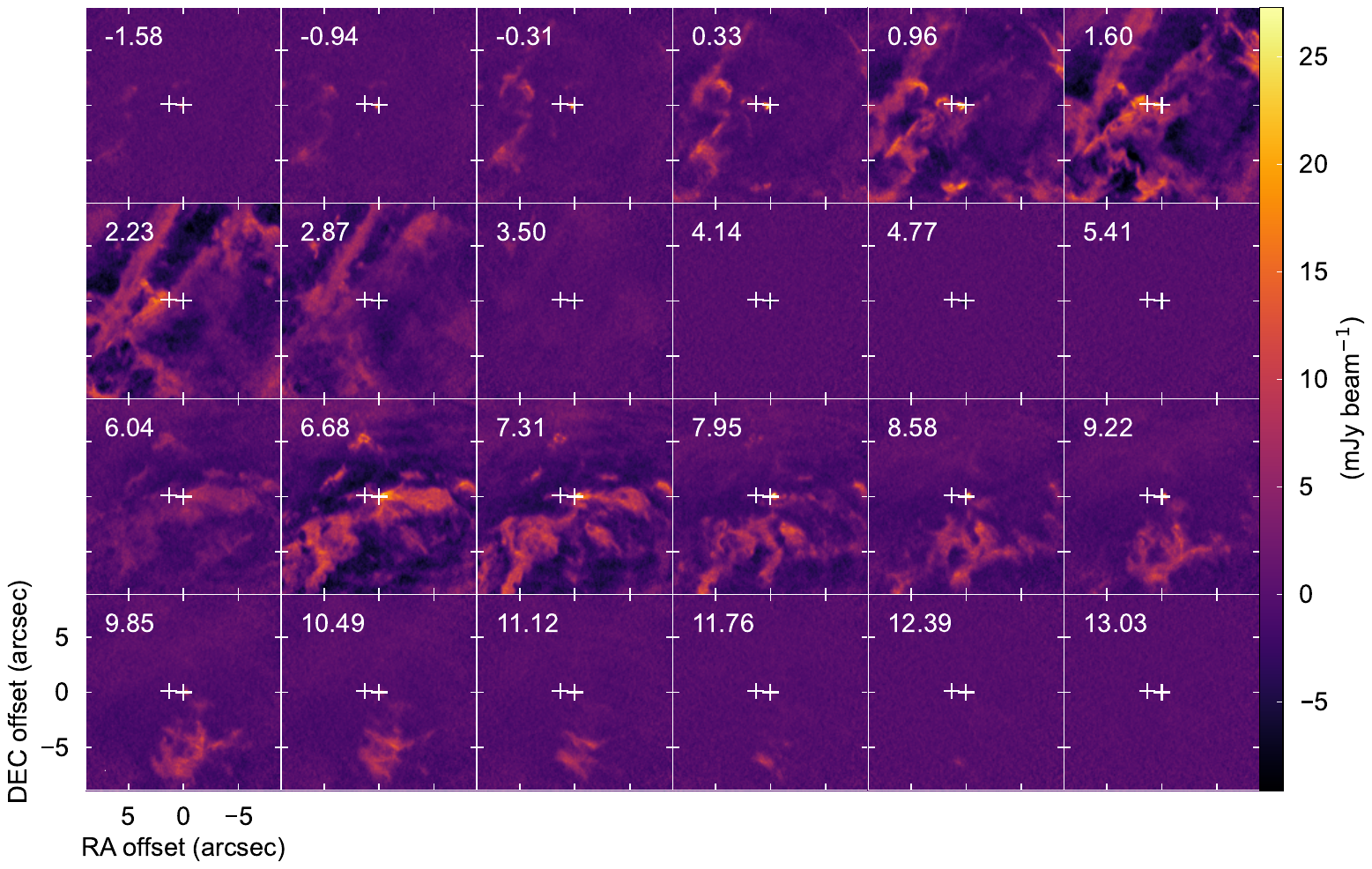}
    \caption{Same as Figure \ref{fig:channel_c18o_wide} but for the $^{12}$CO $J=2$--1 emission.}
    \label{fig:channel_12co_wide}
\end{figure*}

\begin{figure*}[tbhp]
    \centering
    \includegraphics[width=2\columnwidth]{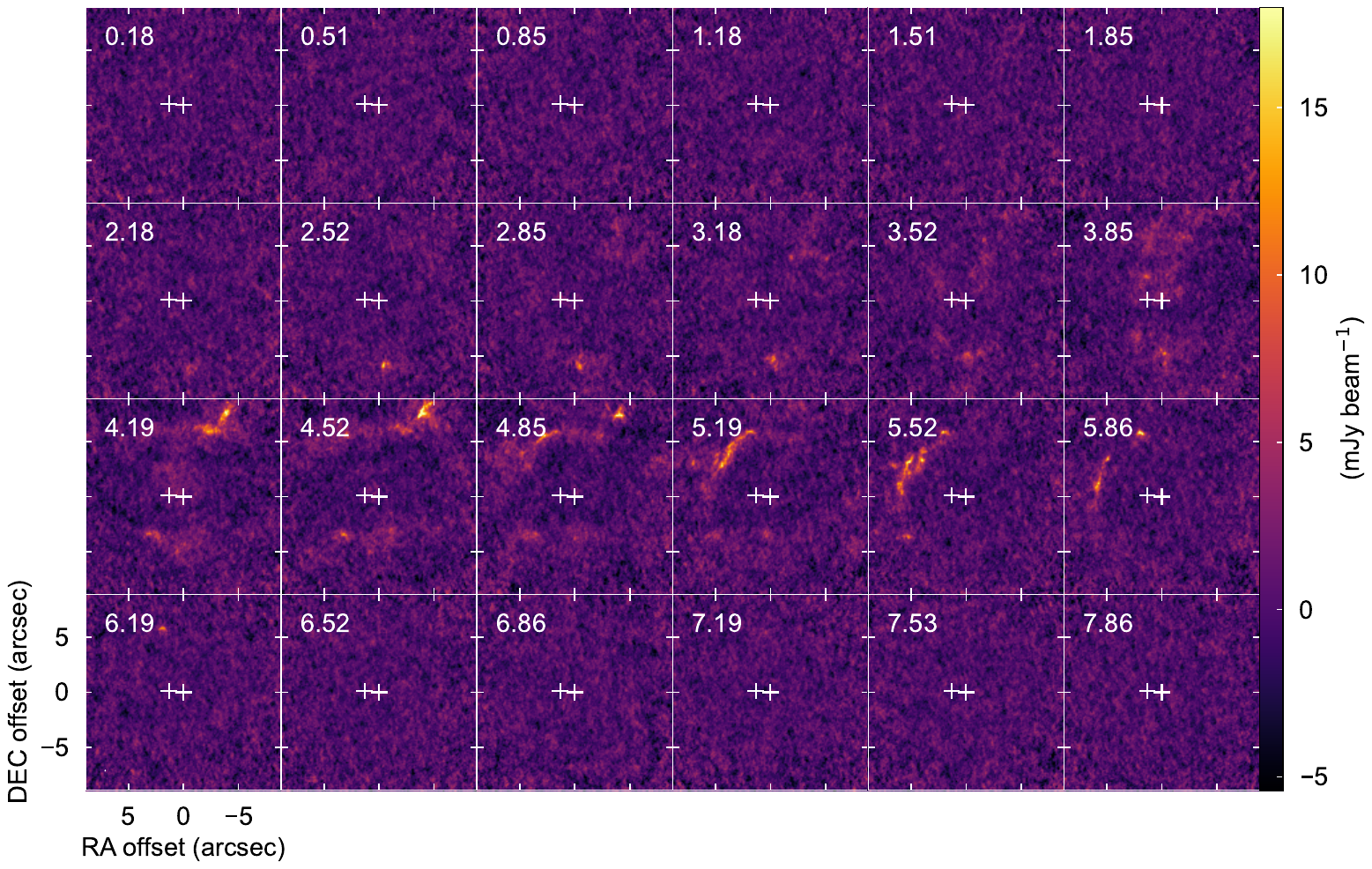}
    \caption{Same as Figure \ref{fig:channel_c18o_wide} but for the SO $J_N=6_5$--$5_4$ emission.}
    \label{fig:channel_so_wide}
\end{figure*}

\begin{deluxetable*}{ccccccc}
\tablecaption{Summary of ALMA maps of other detected lines \label{tab:maps_app}}
\tabletypesize{\footnotesize}
\tablehead{
\colhead{Molecule} & Transition & \colhead{Frequency} & \colhead{Robust} & \colhead{Beam Size} & 
\colhead{Velocity Resolution} & \colhead{RMS} \\
\colhead{} & \colhead{} & \colhead{(GHz)} & \colhead{} & \colhead{} & \colhead{($\kmps$)} & \colhead{($\mjypbm$)}
}
\startdata
H$_{2}$CO & 3$_{0,3}$--2$_{0,2}$ & 218.222192 & 0.5 & $0\farcs 113 \times 0\farcs 081$ ($-8.6^\circ$) & 1.340 & 0.52 \\
$c$-C$_{3}$H$_{2}$ & $6_{0,6}$--5$_{1,5}$~$^{a}$ & 217.822150 & 0.5 & $0\farcs 113 \times 0\farcs 081$ ($-8.5^\circ$) & 1.340 & 0.52 \\
$c$-C$_{3}$H$_{2}$ & $6_{1,6}$--5$_{0,5}$~$^{a}$ & 217.822150 & 0.5 & $0\farcs 113 \times 0\farcs 081$ ($-8.6^\circ$) & 1.340 & 0.52 \\
$c$-C$_{3}$H$_{2}$ & $5_{1,4}$--4$_{2,3}$ & 217.940050 & 0.5 & $0\farcs 113 \times 0\farcs 081$ ($-8.6^\circ$) & 1.340 & 0.52 \\
$c$-C$_{3}$H$_{2}$ & $5_{2,4}$--4$_{1,3}$ & 218.160440 & 0.5 & $0\farcs 113 \times 0\farcs 081$ ($-8.6^\circ$) & 1.340 & 0.52 \\
DCN & 3--2 & 217.238600 & 0.5 & $0\farcs 113 \times 0\farcs 081$ ($-8.5^\circ$) & 1.340 & 0.52 \\
\enddata
\tablecomments{$^a$These two transitions are blended.}
\end{deluxetable*}

\begin{figure*}[tbhp]
    \centering
    \includegraphics[width=2\columnwidth]{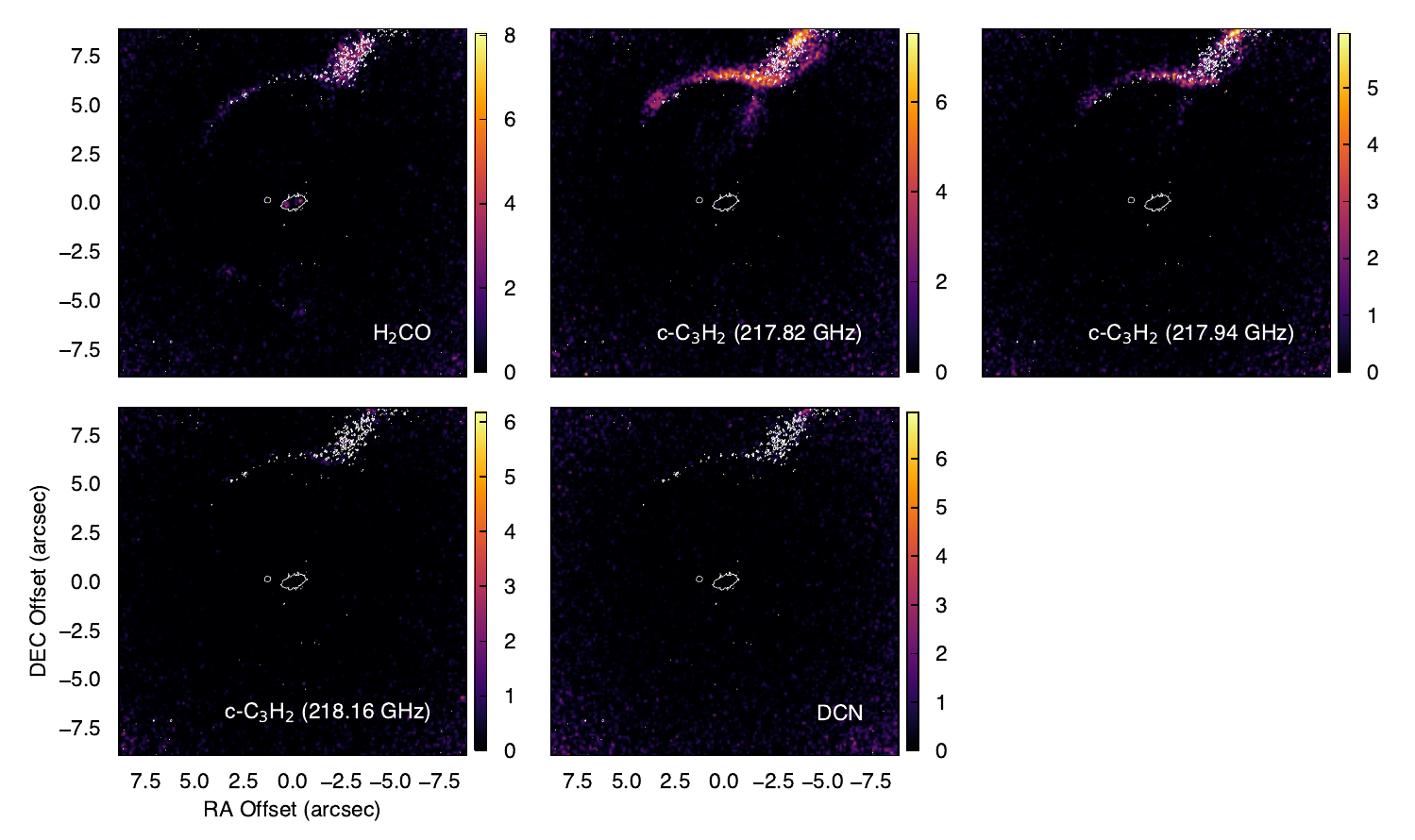}
    \caption{Moment zero maps of the other detected lines (color) overlaid with the 1.3 mm continuum map with a robust parameter of 2.0 (contours). Color scales are in units of $\mjypbm~\kmps$. The contour level is $3\sigma$, where $\sigma$ is 0.018 $\mjypbm$. The filled and opened ellipses at the bottom left corners represent the synthesized beam sizes of the line map and continuum map, respectively.}
    \label{fig:moments_gallery}
\end{figure*}

\section{Power-law Intensity Distribution Models} \label{sec:app_models}
Intensity distribution models that have a power-law intensity profile are examined in the same manner as described in Section \ref{subsec:ana_cont}.

Two types of intensity distributions, a broken power-law model and a power-law ring model, are tested. The broken power-law model is described as follows:
\begin{equation}
    I_\nu (r) = 
    \begin{cases}
     I_\mathrm{c} \left( \frac{r_\mathrm{b}}{r_\mathrm{c}} \right)^{-\beta} \exp \left\{- \left( \frac{r_\mathrm{b}}{r_\mathrm{c}} \right)^{2 - \gamma} \right\}
     \left( \frac{r}{r_\mathrm{b}} \right)^{-\alpha} & (r \leq r_\mathrm{b}), \\
     I_\mathrm{c} \left( \frac{r}{r_\mathrm{c}} \right)^{-\beta} \exp \left\{- \left( \frac{r}{r_\mathrm{c}} \right)^{2 - \gamma} \right\} & (r > r_\mathrm{b}).
    \end{cases}
\end{equation}
Here, $r$ is the radius deprojected by the inclination angle ($i$) and the position angle ($pa$), $r_\mathrm{c}$ is the characteristic radius where the intensity begins to decrease sharply, $r_\mathrm{b}$ is the break radius where the power-law index changes, $\alpha$ is the inner power-law index, $\beta$ is the outer power-law index, and $\gamma$ is the power-law index determining the sharpness of the intensity cutoff at the edge. The power-law ring model, on the other hand, consists of a power-law profile and a ring component as expressed by the following equation:
\begin{eqnarray}
    I_\nu (r) = 
     I_\mathrm{c} \left( \frac{r}{r_\mathrm{c}} \right)^{-\beta} \exp \left\{- \left( \frac{r}{r_\mathrm{c}} \right)^{2 - \gamma} \right\} \nonumber \\
     ~~~ + a_\mathrm{r} \exp \left\{ -\frac{(r - r_\mathrm{r})^2}{2 \sigma_\mathrm{r}^2} \right\},
\end{eqnarray}
where $a_\mathrm{r}$ is the peak intensity of the ring component, $r_\mathrm{r}$ is the ring location in radius, and $\sigma_\mathrm{r}$ is the ring width. For both the broken power-law and power-law ring models, we adopt the power-law intensity profile with an exponential cutoff, which is physically motivated by a Shakura--Sunyaev disk \citep{Shakura:1973aa} with a power-law temperature distribution and often used to describe the disk intensity distribution. Note that, however, the intensity distribution can be connected to the physical model only when the disk is optically and geometrically thin, which would not be the case for the disk of Ced110 IRS4A at 1.3 mm. Our focus here is therefore not to constrain the physical quantity of the disk but to demonstrate the significance of the observed bumps from a smooth intensity profile. The center of the model intensity distribution is fixed at the protostellar position in the both models. The two models are fitted to the observed continuum emission with the MCMC method. The beam convolution is calculated in every MCMC step. The best-fit parameters are summarized in Table \ref{tab:param_intensitymodel_pl}.

\begin{figure*}[tbhp]
    \centering
    \includegraphics[width=2.1\columnwidth]{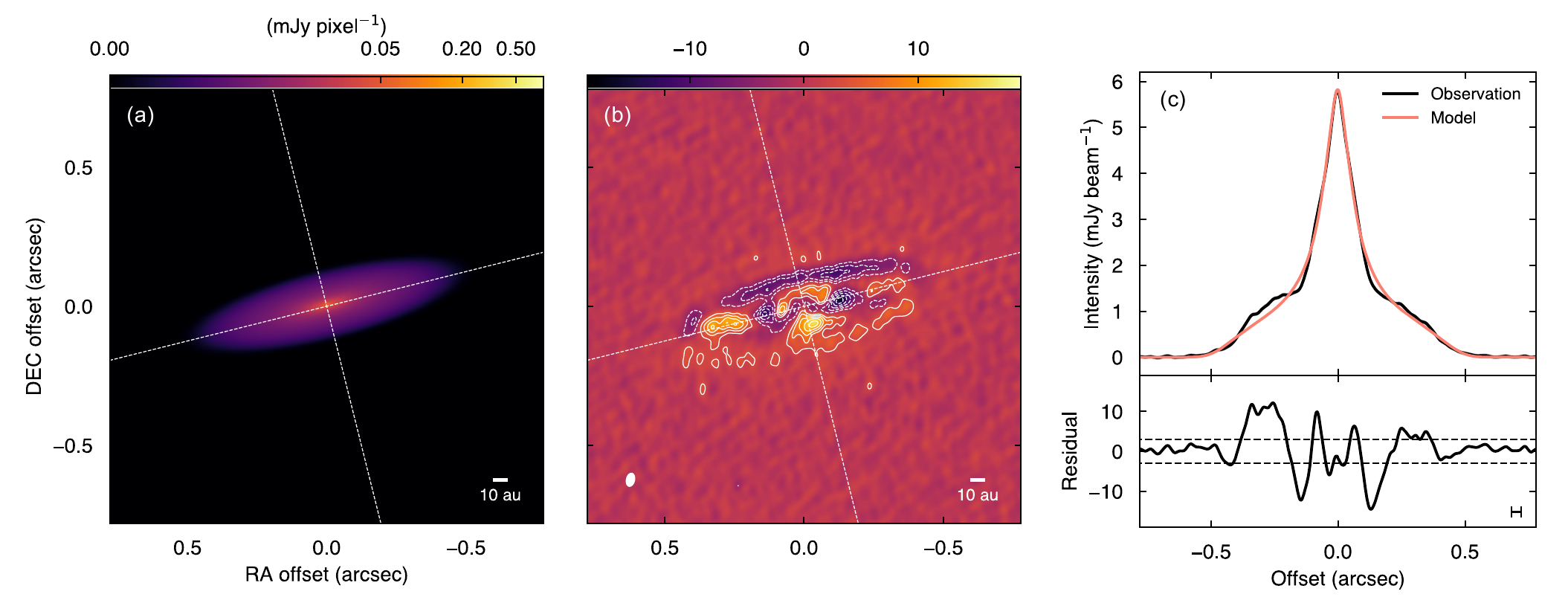}
    \caption{Same as Figure \ref{fig:modelfit_dg} but for the broken power-law model.}
    \label{fig:modelfit_dpl}
\end{figure*}

\begin{figure*}[tbhp]
    \centering
    \includegraphics[width=2.1\columnwidth]{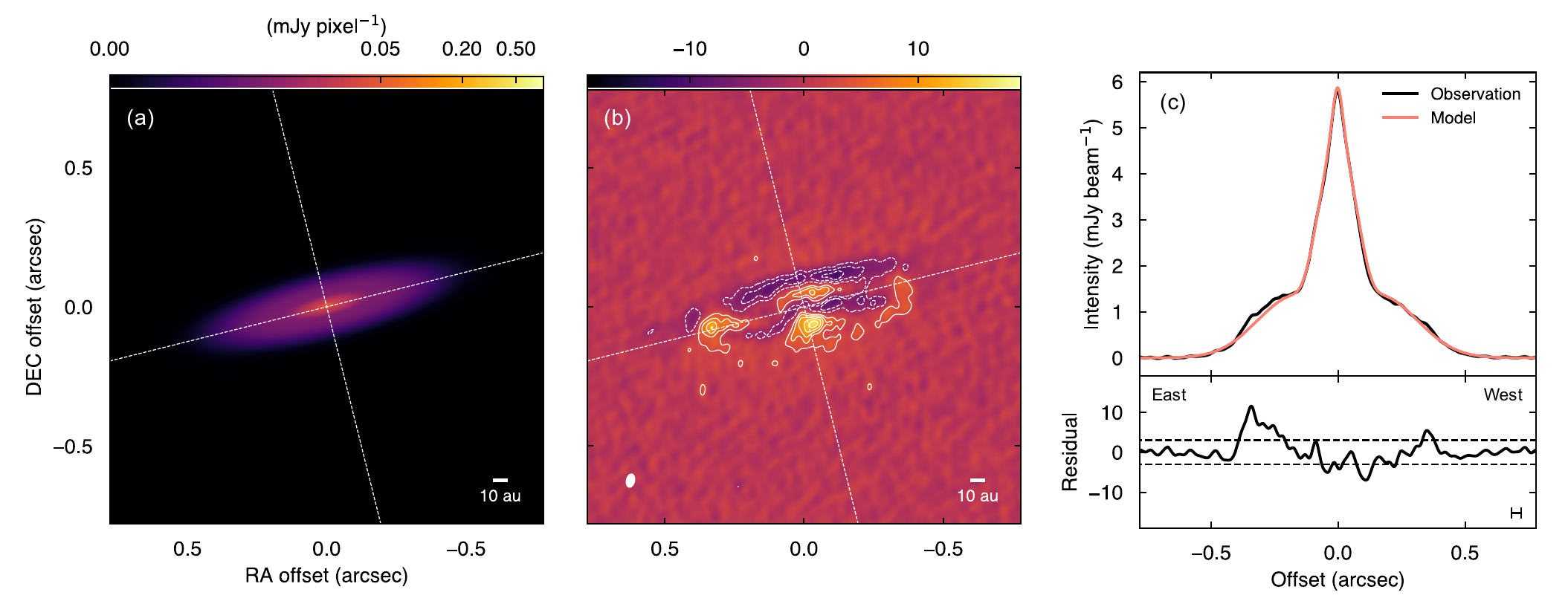}
    \caption{Same as Figure \ref{fig:modelfit_dg} but for the power-law ring model.}
    \label{fig:modelfit_plr}
\end{figure*}

\begin{deluxetable*}{lclc}
\tablecaption{Parameters of the power-law intensity distribution models \label{tab:param_intensitymodel_pl}}
\tabletypesize{\footnotesize}
\tablehead{
\colhead{Parameter} & \colhead{Unit} & \colhead{Description} & \colhead{Best-fit value}
}

\startdata
\multicolumn{4}{c}{Broken power-law model ($k=8$, $\mathrm{BIC}=-38400$)$^{a}$} \\ \hline
$I_\mathrm{c}$ & $\times 10^{-3}$ mJy pixel$^{-1}$ & Intensity at the characteristic radius & $2.47\pm0.01$ \\
$r_\mathrm{c}$ & mas & Characteristic radius & $459.53 \pm 0.46$ \\ 
$r_\mathrm{b}$ & mas & Break radius & $50.61 \pm 0.56$ \\ 
$\alpha$ & & Inner power-law index & $0.46 \pm 0.01 $ \\
$\beta$ & & Outer power-law index & $ 1.08 \pm 0.01 $ \\
$\gamma$ & & Power-law index for the exponential cutoff & $ -5.41 \pm 0.07 $ \\
$i$ & $^\circ$ & Inclination angle &  $75.58 \pm 0.02$ \\
$pa$ & $^\circ$ & Position angle & $103.89 \pm 0.02$ \\
\hline
\multicolumn{4}{c}{Power-law ring model ($k=9$, $\mathrm{BIC}=-43200$)$^{a}$} \\ \hline
$I_\mathrm{c}$ & $\times 10^{-3}$ mJy pixel$^{-1}$ & Intensity at the characteristic radius & $12.40\pm0.16$ \\
$r_\mathrm{c}$ & mas & Characteristic radius & $121.13 \pm 0.39$ \\ 
$\beta$ & & Power-law index & $ 0.60 \pm 0.01 $ \\
$\gamma$ & & Power-law index for the exponential cutoff & $ -2.41 \pm 0.11 $ \\
$a_\mathrm{r}$ & $\times 10^{-3}$ mJy pixel$^{-1}$ & Amplitude of the Gaussian ring & $5.99\pm0.03$ \\ 
$r_\mathrm{r}$ & mas & Radius of the Gaussian ring & $173.17 \pm 1.50$ \\
$\sigma_\mathrm{r}$ & mas & Width (standard deviation) of the Gaussian ring & $146.74 \pm 0.72$ \\
$i$ & $^\circ$ & Inclination angle &  $75.78 \pm 0.02$ \\
$pa$ & $^\circ$ & Position angle & $103.86 \pm 0.02$ \\
\enddata
\tablecomments{$^a k$ is the number of free parameters, and BIC is the Bayesian information criterion calculated with the best-fit parameters.}
\end{deluxetable*}

Figures \ref{fig:modelfit_dpl} and \ref{fig:modelfit_plr} show the fitting results. The broken power-law model shows systemic residuals $\geq6\sigma$ and $\leq-6\sigma$ around offsets of $\pm0\farcs3$ along the major axis (\ref{fig:modelfit_dpl}(c)). On the other hand, the power-law ring model exhibits fewer residuals at those offsets (Figure \ref{fig:modelfit_plr}(c)). Large residuals remaining around offsets of $-0\farcs3$ is due to the asymmetry of the observed continuum emission, as discussed in Section \ref{subsec:ana_cont}. BIC of the broken power-law and power-law ring models are calculated to be $-38400$ and $-43200$, respectively, indicating a very strong evidence that the power-law ring model better reproduces the observations. Hence, we conclude that a model including a ring component better explains the observations than the model with a smooth intensity profile. 


\bibliography{reference,reference_spl}{}

\begin{thebibliography}{}
\expandafter\ifx\csname natexlab\endcsname\relax\def\natexlab#1{#1}\fi
\providecommand{\url}[1]{\href{#1}{#1}}
\providecommand{\dodoi}[1]{doi:~\href{http://doi.org/#1}{\nolinkurl{#1}}}
\providecommand{\doeprint}[1]{\href{http://ascl.net/#1}{\nolinkurl{http://ascl.net/#1}}}
\providecommand{\doarXiv}[1]{\href{https://arxiv.org/abs/#1}{\nolinkurl{https://arxiv.org/abs/#1}}}

\bibitem[{{Adams} {et~al.}(1989){Adams}, {Ruden}, \& {Shu}}]{Adams:1989aa}
{Adams}, F.~C., {Ruden}, S.~P., \& {Shu}, F.~H. 1989, \apj, 347, 959,
  \dodoi{10.1086/168187}

\bibitem[{{Akeson} {et~al.}(2019){Akeson}, {Jensen}, {Carpenter}, {Ricci},
  {Laos}, {Nogueira}, \& {Suen-Lewis}}]{Akeson:2019aa}
{Akeson}, R.~L., {Jensen}, E. L.~N., {Carpenter}, J., {et~al.} 2019, \apj, 872,
  158, \dodoi{10.3847/1538-4357/aaff6a}

\bibitem[{{ALMA Partnership} {et~al.}(2015){ALMA Partnership}, {Brogan},
  {P{\'e}rez}, {Hunter}, {Dent}, {Hales}, {Hills}, {Corder}, {Fomalont},
  {Vlahakis}, {Asaki}, {Barkats}, {Hirota}, {Hodge}, {Impellizzeri}, {Kneissl},
  {Liuzzo}, {Lucas}, {Marcelino}, {Matsushita}, {Nakanishi}, {Phillips},
  {Richards}, {Toledo}, {Aladro}, {Broguiere}, {Cortes}, {Cortes}, {Espada},
  {Galarza}, {Garcia-Appadoo}, {Guzman-Ramirez}, {Humphreys}, {Jung}, {Kameno},
  {Laing}, {Leon}, {Marconi}, {Mignano}, {Nikolic}, {Nyman}, {Radiszcz},
  {Remijan}, {Rod{\'o}n}, {Sawada}, {Takahashi}, {Tilanus}, {Vila Vilaro},
  {Watson}, {Wiklind}, {Akiyama}, {Chapillon}, {de Gregorio-Monsalvo}, {Di
  Francesco}, {Gueth}, {Kawamura}, {Lee}, {Nguyen Luong}, {Mangum}, {Pietu},
  {Sanhueza}, {Saigo}, {Takakuwa}, {Ubach}, {van Kempen}, {Wootten},
  {Castro-Carrizo}, {Francke}, {Gallardo}, {Garcia}, {Gonzalez}, {Hill},
  {Kaminski}, {Kurono}, {Liu}, {Lopez}, {Morales}, {Plarre}, {Schieven},
  {Testi}, {Videla}, {Villard}, {Andreani}, {Hibbard}, \&
  {Tatematsu}}]{ALMA-Partnership:2015aa}
{ALMA Partnership}, {Brogan}, C.~L., {P{\'e}rez}, L.~M., {et~al.} 2015, \apjl,
  808, L3, \dodoi{10.1088/2041-8205/808/1/L3}

\bibitem[{{Andrews}(2020)}]{Andrews:2020aa}
{Andrews}, S.~M. 2020, \araa, 58, 483,
  \dodoi{10.1146/annurev-astro-031220-010302}

\bibitem[{{Andrews} {et~al.}(2013){Andrews}, {Rosenfeld}, {Kraus}, \&
  {Wilner}}]{Andrews:2013aa}
{Andrews}, S.~M., {Rosenfeld}, K.~A., {Kraus}, A.~L., \& {Wilner}, D.~J. 2013,
  \apj, 771, 129, \dodoi{10.1088/0004-637X/771/2/129}

\bibitem[{{Andrews} \& {Williams}(2005)}]{Andrews:2005aa}
{Andrews}, S.~M., \& {Williams}, J.~P. 2005, \apj, 631, 1134,
  \dodoi{10.1086/432712}

\bibitem[{{Andrews} {et~al.}(2018){Andrews}, {Huang}, {P{\'e}rez}, {Isella},
  {Dullemond}, {Kurtovic}, {Guzm{\'a}n}, {Carpenter}, {Wilner}, {Zhang}, {Zhu},
  {Birnstiel}, {Bai}, {Benisty}, {Hughes}, {{\"O}berg}, \&
  {Ricci}}]{Andrews:2018aa}
{Andrews}, S.~M., {Huang}, J., {P{\'e}rez}, L.~M., {et~al.} 2018, \apjl, 869,
  L41, \dodoi{10.3847/2041-8213/aaf741}

\bibitem[{{Ansdell} {et~al.}(2016){Ansdell}, {Williams}, {van der Marel},
  {Carpenter}, {Guidi}, {Hogerheijde}, {Mathews}, {Manara}, {Miotello},
  {Natta}, {Oliveira}, {Tazzari}, {Testi}, {van Dishoeck}, \& {van
  Terwisga}}]{Ansdell:2016aa}
{Ansdell}, M., {Williams}, J.~P., {van der Marel}, N., {et~al.} 2016, \apj,
  828, 46, \dodoi{10.3847/0004-637X/828/1/46}

\bibitem[{{Ansdell} {et~al.}(2018){Ansdell}, {Williams}, {Trapman}, {van
  Terwisga}, {Facchini}, {Manara}, {van der Marel}, {Miotello}, {Tazzari},
  {Hogerheijde}, {Guidi}, {Testi}, \& {van Dishoeck}}]{Ansdell:2018aa}
{Ansdell}, M., {Williams}, J.~P., {Trapman}, L., {et~al.} 2018, \apj, 859, 21,
  \dodoi{10.3847/1538-4357/aab890}

\bibitem[{{Arce} \& {Sargent}(2006)}]{Arce:2006aa}
{Arce}, H.~G., \& {Sargent}, A.~I. 2006, \apj, 646, 1070,
  \dodoi{10.1086/505104}

\bibitem[{{Artymowicz} \& {Lubow}(1994)}]{Artymowicz:1994aa}
{Artymowicz}, P., \& {Lubow}, S.~H. 1994, \apj, 421, 651,
  \dodoi{10.1086/173679}

\bibitem[{Aso \& Sai(2023)}]{Aso:2023aa}
Aso, Y., \& Sai, J. 2023, jinshisai/SLAM: First Release of SLAM, v1.0.0,
  Zenodo, \dodoi{10.5281/zenodo.7783868}

\bibitem[{{Aso} {et~al.}(2023){Aso}, {Kwon}, {Ohashi}, {Jørgensen}, {Tobin},
  {Aikawa}, {Gregorio-Monsalvo}, {Han}, {Kido}, {Koch}, {Lai}, {Lee}, {Lee},
  {Li}, {Lin}, {Looney}, {Narayanan}, {Phuong}, {Sai}, {Saigo},
  {Santamaría-Miranda}, {Sharma}, {Takakuwa}, {Thieme}, {Tomida}, {Williams},
  \& {Yen}}]{Aso:2023ab}
{Aso}, Y., {Kwon}, W., {Ohashi}, N., {et~al.} 2023, \apj, 954, 101,
  \dodoi{10.3847/1538-4357/ace624}

\bibitem[{{Astropy Collaboration} {et~al.}(2013){Astropy Collaboration},
  {Robitaille}, {Tollerud}, {Greenfield}, {Droettboom}, {Bray}, {Aldcroft},
  {Davis}, {Ginsburg}, {Price-Whelan}, {Kerzendorf}, {Conley}, {Crighton},
  {Barbary}, {Muna}, {Ferguson}, {Grollier}, {Parikh}, {Nair}, {Unther},
  {Deil}, {Woillez}, {Conseil}, {Kramer}, {Turner}, {Singer}, {Fox}, {Weaver},
  {Zabalza}, {Edwards}, {Azalee Bostroem}, {Burke}, {Casey}, {Crawford},
  {Dencheva}, {Ely}, {Jenness}, {Labrie}, {Lim}, {Pierfederici}, {Pontzen},
  {Ptak}, {Refsdal}, {Servillat}, \&
  {Streicher}}]{Astropy-Collaboration:2013aa}
{Astropy Collaboration}, {Robitaille}, T.~P., {Tollerud}, E.~J., {et~al.} 2013,
  \aap, 558, A33, \dodoi{10.1051/0004-6361/201322068}

\bibitem[{{Astropy Collaboration} {et~al.}(2018){Astropy Collaboration},
  {Price-Whelan}, {Sip{\H o}cz}, {G{\"u}nther}, {Lim}, {Crawford}, {Conseil},
  {Shupe}, {Craig}, {Dencheva}, {Ginsburg}, {VanderPlas}, {Bradley},
  {P{\'e}rez-Su{\'a}rez}, {de Val-Borro}, {Aldcroft}, {Cruz}, {Robitaille},
  {Tollerud}, {Ardelean}, {Babej}, {Bach}, {Bachetti}, {Bakanov}, {Bamford},
  {Barentsen}, {Barmby}, {Baumbach}, {Berry}, {Biscani}, {Boquien}, {Bostroem},
  {Bouma}, {Brammer}, {Bray}, {Breytenbach}, {Buddelmeijer}, {Burke},
  {Calderone}, {Cano Rodr{\'{\i}}guez}, {Cara}, {Cardoso}, {Cheedella},
  {Copin}, {Corrales}, {Crichton}, {D'Avella}, {Deil}, {Depagne}, {Dietrich},
  {Donath}, {Droettboom}, {Earl}, {Erben}, {Fabbro}, {Ferreira}, {Finethy},
  {Fox}, {Garrison}, {Gibbons}, {Goldstein}, {Gommers}, {Greco}, {Greenfield},
  {Groener}, {Grollier}, {Hagen}, {Hirst}, {Homeier}, {Horton}, {Hosseinzadeh},
  {Hu}, {Hunkeler}, {Ivezi{\'c}}, {Jain}, {Jenness}, {Kanarek}, {Kendrew},
  {Kern}, {Kerzendorf}, {Khvalko}, {King}, {Kirkby}, {Kulkarni}, {Kumar},
  {Lee}, {Lenz}, {Littlefair}, {Ma}, {Macleod}, {Mastropietro}, {McCully},
  {Montagnac}, {Morris}, {Mueller}, {Mumford}, {Muna}, {Murphy}, {Nelson},
  {Nguyen}, {Ninan}, {N{\"o}the}, {Ogaz}, {Oh}, {Parejko}, {Parley}, {Pascual},
  {Patil}, {Patil}, {Plunkett}, {Prochaska}, {Rastogi}, {Reddy Janga},
  {Sabater}, {Sakurikar}, {Seifert}, {Sherbert}, {Sherwood-Taylor}, {Shih},
  {Sick}, {Silbiger}, {Singanamalla}, {Singer}, {Sladen}, {Sooley},
  {Sornarajah}, {Streicher}, {Teuben}, {Thomas}, {Tremblay}, {Turner},
  {Terr{\'o}n}, {van Kerkwijk}, {de la Vega}, {Watkins}, {Weaver}, {Whitmore},
  {Woillez}, {Zabalza}, \& {Astropy
  Contributors}}]{Astropy-Collaboration:2018aa}
{Astropy Collaboration}, {Price-Whelan}, A.~M., {Sip{\H o}cz}, B.~M., {et~al.}
  2018, \aj, 156, 123, \dodoi{10.3847/1538-3881/aabc4f}

\bibitem[{{Bachiller} \& {P{\'e}rez Guti{\'e}rrez}(1997)}]{Bachiller:1997aa}
{Bachiller}, R., \& {P{\'e}rez Guti{\'e}rrez}, M. 1997, \apjl, 487, L93,
  \dodoi{10.1086/310877}

\bibitem[{{Bailli{\'e}} \& {Charnoz}(2014)}]{Baillie:2014aa}
{Bailli{\'e}}, K., \& {Charnoz}, S. 2014, \apj, 786, 35,
  \dodoi{10.1088/0004-637X/786/1/35}

\bibitem[{{Barenfeld} {et~al.}(2017){Barenfeld}, {Carpenter}, {Sargent},
  {Isella}, \& {Ricci}}]{Barenfeld:2017aa}
{Barenfeld}, S.~A., {Carpenter}, J.~M., {Sargent}, A.~I., {Isella}, A., \&
  {Ricci}, L. 2017, \apj, 851, 85, \dodoi{10.3847/1538-4357/aa989d}

\bibitem[{{Bate}(2018)}]{Bate:2018aa}
{Bate}, M.~R. 2018, \mnras, 475, 5618, \dodoi{10.1093/mnras/sty169}

\bibitem[{{Bate} {et~al.}(2010){Bate}, {Lodato}, \& {Pringle}}]{Bate:2010aa}
{Bate}, M.~R., {Lodato}, G., \& {Pringle}, J.~E. 2010, \mnras, 401, 1505,
  \dodoi{10.1111/j.1365-2966.2009.15773.x}

\bibitem[{{Beckwith} {et~al.}(1990){Beckwith}, {Sargent}, {Chini}, \&
  {Guesten}}]{Beckwith:1990aa}
{Beckwith}, S.~V.~W., {Sargent}, A.~I., {Chini}, R.~S., \& {Guesten}, R. 1990,
  \aj, 99, 924, \dodoi{10.1086/115385}

\bibitem[{{Benisty} {et~al.}(2021){Benisty}, {Bae}, {Facchini}, {Keppler},
  {Teague}, {Isella}, {Kurtovic}, {P{\'e}rez}, {Sierra}, {Andrews},
  {Carpenter}, {Czekala}, {Dominik}, {Henning}, {Menard}, {Pinilla}, \&
  {Zurlo}}]{Benisty:2021aa}
{Benisty}, M., {Bae}, J., {Facchini}, S., {et~al.} 2021, \apjl, 916, L2,
  \dodoi{10.3847/2041-8213/ac0f83}

\bibitem[{{Bi} {et~al.}(2020){Bi}, {van der Marel}, {Dong}, {Muto}, {Martin},
  {Smallwood}, {Hashimoto}, {Liu}, {Nomura}, {Hasegawa}, {Takami}, {Konishi},
  {Momose}, {Kanagawa}, {Kataoka}, {Ono}, {Sitko}, {Takahashi}, {Tomida}, \&
  {Tsukagoshi}}]{Bi:2020aa}
{Bi}, J., {van der Marel}, N., {Dong}, R., {et~al.} 2020, \apjl, 895, L18,
  \dodoi{10.3847/2041-8213/ab8eb4}

\bibitem[{{Bonnell} \& {Bate}(1994)}]{Bonnell:1994aa}
{Bonnell}, I.~A., \& {Bate}, M.~R. 1994, \mnras, 271, 999,
  \dodoi{10.1093/mnras/271.4.999}

\bibitem[{{Carrasco-Gonz{\'a}lez} {et~al.}(2019){Carrasco-Gonz{\'a}lez},
  {Sierra}, {Flock}, {Zhu}, {Henning}, {Chandler}, {Galv{\'a}n-Madrid},
  {Mac{\'\i}as}, {Anglada}, {Linz}, {Osorio}, {Rodr{\'\i}guez}, {Testi},
  {Torrelles}, {P{\'e}rez}, \& {Liu}}]{Carrasco-Gonzalez:2019aa}
{Carrasco-Gonz{\'a}lez}, C., {Sierra}, A., {Flock}, M., {et~al.} 2019, \apj,
  883, 71, \dodoi{10.3847/1538-4357/ab3d33}

\bibitem[{{Chen} {et~al.}(1995){Chen}, {Myers}, {Ladd}, \&
  {Wood}}]{Chen:1995aa}
{Chen}, H., {Myers}, P.~C., {Ladd}, E.~F., \& {Wood}, D.~O.~S. 1995, \apj, 445,
  377, \dodoi{10.1086/175703}

\bibitem[{{Chiang} \& {Goldreich}(1997)}]{Chiang:1997aa}
{Chiang}, E.~I., \& {Goldreich}, P. 1997, \apj, 490, 368,
  \dodoi{10.1086/304869}

\bibitem[{{Cieza} {et~al.}(2019){Cieza}, {Ru{\'\i}z-Rodr{\'\i}guez}, {Hales},
  {Casassus}, {P{\'e}rez}, {Gonzalez-Ruilova}, {C{\'a}novas}, {Williams},
  {Zurlo}, {Ansdell}, {Avenhaus}, {Bayo}, {Bertrang}, {Christiaens}, {Dent},
  {Ferrero}, {Gamen}, {Olofsson}, {Orcajo}, {Pe{\~n}a Ram{\'\i}rez},
  {Principe}, {Schreiber}, \& {van der Plas}}]{Cieza:2019aa}
{Cieza}, L.~A., {Ru{\'\i}z-Rodr{\'\i}guez}, D., {Hales}, A., {et~al.} 2019,
  \mnras, 482, 698, \dodoi{10.1093/mnras/sty2653}

\bibitem[{{Cleeves}(2016)}]{Cleeves:2016aa}
{Cleeves}, L.~I. 2016, \apjl, 816, L21, \dodoi{10.3847/2041-8205/816/2/L21}

\bibitem[{{Currie} {et~al.}(2022){Currie}, {Lawson}, {Schneider}, {Lyra},
  {Wisniewski}, {Grady}, {Guyon}, {Tamura}, {Kotani}, {Kawahara}, {Brandt},
  {Uyama}, {Muto}, {Dong}, {Kudo}, {Hashimoto}, {Fukagawa}, {Wagner}, {Lozi},
  {Chilcote}, {Tobin}, {Groff}, {Ward-Duong}, {Januszewski}, {Norris},
  {Tuthill}, {van der Marel}, {Sitko}, {Deo}, {Vievard}, {Jovanovic},
  {Martinache}, \& {Skaf}}]{Currie:2022aa}
{Currie}, T., {Lawson}, K., {Schneider}, G., {et~al.} 2022, Nature Astronomy,
  6, 751, \dodoi{10.1038/s41550-022-01634-x}

\bibitem[{{D'Alessio} {et~al.}(1998){D'Alessio}, {Cant{\"o}}, {Calvet}, \&
  {Lizano}}]{Dalessio:1998aa}
{D'Alessio}, P., {Cant{\"o}}, J., {Calvet}, N., \& {Lizano}, S. 1998, \apj,
  500, 411, \dodoi{10.1086/305702}

\bibitem[{{Dullemond} {et~al.}(2001){Dullemond}, {Dominik}, \&
  {Natta}}]{Dullemond:2001aa}
{Dullemond}, C.~P., {Dominik}, C., \& {Natta}, A. 2001, \apj, 560, 957,
  \dodoi{10.1086/323057}

\bibitem[{{Dzib} {et~al.}(2018){Dzib}, {Loinard}, {Ortiz-Le{\'o}n},
  {Rodr{\'\i}guez}, \& {Galli}}]{Dzib:2018aa}
{Dzib}, S.~A., {Loinard}, L., {Ortiz-Le{\'o}n}, G.~N., {Rodr{\'\i}guez}, L.~F.,
  \& {Galli}, P. A.~B. 2018, \apj, 867, 151, \dodoi{10.3847/1538-4357/aae687}

\bibitem[{{Evans} {et~al.}(2009){Evans}, {Dunham}, {J{\o}rgensen}, {Enoch},
  {Mer{\'{\i}}n}, {van Dishoeck}, {Alcal{\'a}}, {Myers}, {Stapelfeldt},
  {Huard}, {Allen}, {Harvey}, {van Kempen}, {Blake}, {Koerner}, {Mundy},
  {Padgett}, \& {Sargent}}]{Evans:2009aa}
{Evans}, II, N.~J., {Dunham}, M.~M., {J{\o}rgensen}, J.~K., {et~al.} 2009,
  \apjs, 181, 321, \dodoi{10.1088/0067-0049/181/2/321}

\bibitem[{{Facchini} {et~al.}(2017){Facchini}, {Birnstiel}, {Bruderer}, \& {van
  Dishoeck}}]{Facchini:2017aa}
{Facchini}, S., {Birnstiel}, T., {Bruderer}, S., \& {van Dishoeck}, E.~F. 2017,
  \aap, 605, A16, \dodoi{10.1051/0004-6361/201630329}

\bibitem[{{Flock} {et~al.}(2015){Flock}, {Ruge}, {Dzyurkevich}, {Henning},
  {Klahr}, \& {Wolf}}]{Flock:2015aa}
{Flock}, M., {Ruge}, J.~P., {Dzyurkevich}, N., {et~al.} 2015, \aap, 574, A68,
  \dodoi{10.1051/0004-6361/201424693}

\bibitem[{{Foreman-Mackey} {et~al.}(2013){Foreman-Mackey}, {Hogg}, {Lang}, \&
  {Goodman}}]{Foreman-Mackey:2013aa}
{Foreman-Mackey}, D., {Hogg}, D.~W., {Lang}, D., \& {Goodman}, J. 2013, \pasp,
  125, 306, \dodoi{10.1086/670067}

\bibitem[{{Galli} {et~al.}(2021){Galli}, {Bouy}, {Olivares}, {Miret-Roig},
  {Sarro}, {Barrado}, {Berihuete}, {Bertin}, \& {Cuillandre}}]{Galli:2021aa}
{Galli}, P.~A.~B., {Bouy}, H., {Olivares}, J., {et~al.} 2021, \aap, 646, A46,
  \dodoi{10.1051/0004-6361/202039395}

\bibitem[{{Garufi} {et~al.}(2022){Garufi}, {Podio}, {Codella}, {Segura-Cox},
  {Vander Donckt}, {Mercimek}, {Bacciotti}, {Fedele}, {Kasper}, {Pineda},
  {Humphreys}, \& {Testi}}]{Garufi:2022aa}
{Garufi}, A., {Podio}, L., {Codella}, C., {et~al.} 2022, \aap, 658, A104,
  \dodoi{10.1051/0004-6361/202141264}

\bibitem[{{Haffert} {et~al.}(2019){Haffert}, {Bohn}, {de Boer}, {Snellen},
  {Brinchmann}, {Girard}, {Keller}, \& {Bacon}}]{Haffert:2019aa}
{Haffert}, S.~Y., {Bohn}, A.~J., {de Boer}, J., {et~al.} 2019, Nature
  Astronomy, 3, 749, \dodoi{10.1038/s41550-019-0780-5}

\bibitem[{{Harsono} {et~al.}(2015){Harsono}, {Bruderer}, \& {van
  Dishoeck}}]{Harsono:2015aa}
{Harsono}, D., {Bruderer}, S., \& {van Dishoeck}, E.~F. 2015, \aap, 582, A41,
  \dodoi{10.1051/0004-6361/201525966}

\bibitem[{{Hiramatsu} {et~al.}(2007){Hiramatsu}, {Hayakawa}, {Tatematsu},
  {Kamegai}, {Onishi}, {Mizuno}, {Yamaguchi}, \& {Hasegawa}}]{Hiramatsu:2007aa}
{Hiramatsu}, M., {Hayakawa}, T., {Tatematsu}, K., {et~al.} 2007, \apj, 664,
  964, \dodoi{10.1086/519269}

\bibitem[{{Hsieh} {et~al.}(2019){Hsieh}, {Hirano}, {Belloche}, {Lee}, {Aso}, \&
  {Lai}}]{Hsieh:2019aa}
{Hsieh}, T.-H., {Hirano}, N., {Belloche}, A., {et~al.} 2019, \apj, 871, 100,
  \dodoi{10.3847/1538-4357/aaf4fe}

\bibitem[{{Huang} {et~al.}(2018){Huang}, {Andrews}, {Dullemond}, {Isella},
  {P{\'e}rez}, {Guzm{\'a}n}, {{\"O}berg}, {Zhu}, {Zhang}, {Bai}, {Benisty},
  {Birnstiel}, {Carpenter}, {Hughes}, {Ricci}, {Weaver}, \&
  {Wilner}}]{Huang:2018ab}
{Huang}, J., {Andrews}, S.~M., {Dullemond}, C.~P., {et~al.} 2018, \apjl, 869,
  L42, \dodoi{10.3847/2041-8213/aaf740}

\bibitem[{Hunter(2007)}]{Hunter:2007aa}
Hunter, J.~D. 2007, Computing in Science \& Engineering, 9, 90,
  \dodoi{10.1109/MCSE.2007.55}

\bibitem[{{Kass} \& {Raftery}(1995)}]{Kass:1995aa}
{Kass}, R.~E., \& {Raftery}, A.~E. 1995, Journal of the American Statistical
  Association, 90, 773, \dodoi{10.2307/2291091}

\bibitem[{{Keppler} {et~al.}(2018){Keppler}, {Benisty}, {M{\"u}ller},
  {Henning}, {van Boekel}, {Cantalloube}, {Ginski}, {van Holstein}, {Maire},
  {Pohl}, {Samland}, {Avenhaus}, {Baudino}, {Boccaletti}, {de Boer},
  {Bonnefoy}, {Chauvin}, {Desidera}, {Langlois}, {Lazzoni}, {Marleau},
  {Mordasini}, {Pawellek}, {Stolker}, {Vigan}, {Zurlo}, {Birnstiel},
  {Brandner}, {Feldt}, {Flock}, {Girard}, {Gratton}, {Hagelberg}, {Isella},
  {Janson}, {Juhasz}, {Kemmer}, {Kral}, {Lagrange}, {Launhardt}, {Matter},
  {M{\'e}nard}, {Milli}, {Molli{\`e}re}, {Olofsson}, {P{\'e}rez}, {Pinilla},
  {Pinte}, {Quanz}, {Schmidt}, {Udry}, {Wahhaj}, {Williams}, {Buenzli},
  {Cudel}, {Dominik}, {Galicher}, {Kasper}, {Lannier}, {Mesa}, {Mouillet},
  {Peretti}, {Perrot}, {Salter}, {Sissa}, {Wildi}, {Abe}, {Antichi},
  {Augereau}, {Baruffolo}, {Baudoz}, {Bazzon}, {Beuzit}, {Blanchard}, {Brems},
  {Buey}, {De Caprio}, {Carbillet}, {Carle}, {Cascone}, {Cheetham}, {Claudi},
  {Costille}, {Delboulb{\'e}}, {Dohlen}, {Fantinel}, {Feautrier}, {Fusco},
  {Giro}, {Gluck}, {Gry}, {Hubin}, {Hugot}, {Jaquet}, {Le Mignant}, {Llored},
  {Madec}, {Magnard}, {Martinez}, {Maurel}, {Meyer}, {M{\"o}ller-Nilsson},
  {Moulin}, {Mugnier}, {Orign{\'e}}, {Pavlov}, {Perret}, {Petit}, {Pragt},
  {Puget}, {Rabou}, {Ramos}, {Rigal}, {Rochat}, {Roelfsema}, {Rousset}, {Roux},
  {Salasnich}, {Sauvage}, {Sevin}, {Soenke}, {Stadler}, {Suarez}, {Turatto}, \&
  {Weber}}]{Keppler:2018aa}
{Keppler}, M., {Benisty}, M., {M{\"u}ller}, A., {et~al.} 2018, \aap, 617, A44,
  \dodoi{10.1051/0004-6361/201832957}

\bibitem[{{Lee} {et~al.}(2019){Lee}, {Offner}, {Kratter}, {Smullen}, \&
  {Li}}]{Lee:2019ab}
{Lee}, A.~T., {Offner}, S. S.~R., {Kratter}, K.~M., {Smullen}, R.~A., \& {Li},
  P.~S. 2019, \apj, 887, 232, \dodoi{10.3847/1538-4357/ab584b}

\bibitem[{{Lee} {et~al.}(2018){Lee}, {Kim}, {Myers}, {Saito}, {Kim}, {Kwon},
  {Lyo}, {Soam}, \& {Kim}}]{Lee:2018ac}
{Lee}, C.~W., {Kim}, G., {Myers}, P.~C., {et~al.} 2018, \apj, 865, 131,
  \dodoi{10.3847/1538-4357/aadcf6}

\bibitem[{{Lehtinen} {et~al.}(2001){Lehtinen}, {Haikala}, {Mattila}, \&
  {Lemke}}]{Lehtinen:2001aa}
{Lehtinen}, K., {Haikala}, L.~K., {Mattila}, K., \& {Lemke}, D. 2001, \aap,
  367, 311, \dodoi{10.1051/0004-6361:20000401}

\bibitem[{{Lin} \& {Papaloizou}(1986)}]{Lin:1986ab}
{Lin}, D.~N.~C., \& {Papaloizou}, J. 1986, \apj, 307, 395,
  \dodoi{10.1086/164426}

\bibitem[{{Lin} {et~al.}(2023){Lin}, {Li}, {Tobin}, {Ohashi}, {J{\o}rgensen},
  {Looney}, {Aso}, {Takakuwa}, {Aikawa}, {van't Hoff}, {de Gregorio-Monsalvo},
  {Encalada}, {Flores}, {Gavino}, {Han}, {Kido}, {Koch}, {Kwon}, {Lai}, {Lee},
  {Lee}, {Phuong}, {Sai}, {Sharma}, {Sheehan}, {Thieme}, {Williams}, {Yamato},
  \& {Yen}}]{Lin:2023aa}
{Lin}, Z.-Y.~D., {Li}, Z.-Y., {Tobin}, J.~J., {et~al.} 2023, \apj, 951, 9,
  \dodoi{10.3847/1538-4357/acd5c9}

\bibitem[{{Liu}(2019)}]{Liu:2019aa}
{Liu}, H.~B. 2019, \apjl, 877, L22, \dodoi{10.3847/2041-8213/ab1f8e}

\bibitem[{{Long} {et~al.}(2018{\natexlab{a}}){Long}, {Pinilla}, {Herczeg},
  {Harsono}, {Dipierro}, {Pascucci}, {Hendler}, {Tazzari}, {Ragusa}, \&
  {Salyk}}]{Long:2018ab}
{Long}, F., {Pinilla}, P., {Herczeg}, G.~J., {et~al.} 2018{\natexlab{a}}, \apj,
  869, 17, \dodoi{10.3847/1538-4357/aae8e1}

\bibitem[{{Long} {et~al.}(2018{\natexlab{b}}){Long}, {Herczeg}, {Pascucci},
  {Apai}, {Henning}, {Manara}, {Mulders}, {Sz{\H{u}}cs}, \&
  {Hendler}}]{Long:2018aa}
{Long}, F., {Herczeg}, G.~J., {Pascucci}, I., {et~al.} 2018{\natexlab{b}},
  \apj, 863, 61, \dodoi{10.3847/1538-4357/aacce9}

\bibitem[{{Manara} {et~al.}(2018){Manara}, {Morbidelli}, \&
  {Guillot}}]{Manara:2018aa}
{Manara}, C.~F., {Morbidelli}, A., \& {Guillot}, T. 2018, \aap, 618, L3,
  \dodoi{10.1051/0004-6361/201834076}

\bibitem[{{Manara} {et~al.}(2019){Manara}, {Tazzari}, {Long}, {Herczeg},
  {Lodato}, {Rota}, {Cazzoletti}, {van der Plas}, {Pinilla}, {Dipierro},
  {Edwards}, {Harsono}, {Johnstone}, {Liu}, {Menard}, {Nisini}, {Ragusa},
  {Boehler}, \& {Cabrit}}]{Manara:2019ab}
{Manara}, C.~F., {Tazzari}, M., {Long}, F., {et~al.} 2019, \aap, 628, A95,
  \dodoi{10.1051/0004-6361/201935964}

\bibitem[{{Maret} {et~al.}(2020){Maret}, {Maury}, {Belloche}, {Gaudel},
  {Andr{\'e}}, {Cabrit}, {Codella}, {Lef{\'e}vre}, {Podio}, {Anderl}, {Gueth},
  \& {Hennebelle}}]{Maret:2020aa}
{Maret}, S., {Maury}, A.~J., {Belloche}, A., {et~al.} 2020, \aap, 635, A15,
  \dodoi{10.1051/0004-6361/201936798}

\bibitem[{{Maury} {et~al.}(2019){Maury}, {Andr{\'e}}, {Testi}, {Maret},
  {Belloche}, {Hennebelle}, {Cabrit}, {Codella}, {Gueth}, {Podio}, {Anderl},
  {Bacmann}, {Bontemps}, {Gaudel}, {Ladjelate}, {Lef{\`e}vre}, {Tabone}, \&
  {Lefloch}}]{Maury:2019aa}
{Maury}, A.~J., {Andr{\'e}}, P., {Testi}, L., {et~al.} 2019, \aap, 621, A76

\bibitem[{{McMullin} {et~al.}(2007){McMullin}, {Waters}, {Schiebel}, {Young},
  \& {Golap}}]{McMullin:2007aa}
{McMullin}, J.~P., {Waters}, B., {Schiebel}, D., {Young}, W., \& {Golap}, K.
  2007, in Astronomical Society of the Pacific Conference Series, Vol. 376,
  Astronomical Data Analysis Software and Systems XVI, ed. R.~A. {Shaw},
  F.~{Hill}, \& D.~J. {Bell}, 127

\bibitem[{{Offner} {et~al.}(2010){Offner}, {Kratter}, {Matzner}, {Krumholz}, \&
  {Klein}}]{Offner:2010aa}
{Offner}, S. S.~R., {Kratter}, K.~M., {Matzner}, C.~D., {Krumholz}, M.~R., \&
  {Klein}, R.~I. 2010, \apj, 725, 1485, \dodoi{10.1088/0004-637X/725/2/1485}

\bibitem[{{Ohashi} {et~al.}(2023){Ohashi}, {Tobin}, {J{\o}rgensen}, {Takakuwa},
  {Sheehan}, {Aikawa}, {Li}, {Looney}, {Williams}, {Aso}, {Sharma}, {Choi},
  {Yamato}, {Lee}, {Tomida}, {Yen}, {Encalada}, {Flores}, {Gavino}, {Kido},
  {Han}, {Lin}, {Narayanan}, {Phuong}, {Santamar{\'\i}a-Miranda}, {Thieme},
  {van't Hoff}, {de Gregorio-Monsalvo}, {Koch}, {Kwon}, {Lai}, {Lee},
  {Plunkett}, {Saigo}, {Hirano}, {Lam}, \& {Mori}}]{Ohashi:2023aa}
{Ohashi}, N., {Tobin}, J.~J., {J{\o}rgensen}, J.~K., {et~al.} 2023, \apj, 951,
  8, \dodoi{10.3847/1538-4357/acd384}

\bibitem[{{Ohashi} {et~al.}(2022){Ohashi}, {Kobayashi}, {Sai}, \&
  {Sakai}}]{Ohashi:2022aa}
{Ohashi}, S., {Kobayashi}, H., {Sai}, J., \& {Sakai}, N. 2022, \apj, 933, 23,
  \dodoi{10.3847/1538-4357/ac6fcf}

\bibitem[{{Okuzumi} {et~al.}(2016){Okuzumi}, {Momose}, {Sirono}, {Kobayashi},
  \& {Tanaka}}]{Okuzumi:2016aa}
{Okuzumi}, S., {Momose}, M., {Sirono}, S.-i., {Kobayashi}, H., \& {Tanaka}, H.
  2016, \apj, 821, 82, \dodoi{10.3847/0004-637X/821/2/82}

\bibitem[{Oliphant(2006)}]{Oliphant:2006aa}
Oliphant, T.~E. 2006, A guide to {NumPy}, USA: Trelgol Publishing

\bibitem[{{Padoan} \& {Nordlund}(2002)}]{Padoan:2002aa}
{Padoan}, P., \& {Nordlund}, {\r{A}}. 2002, \apj, 576, 870,
  \dodoi{10.1086/341790}

\bibitem[{{Papaloizou} \& {Pringle}(1977)}]{Papaloizou:1977aa}
{Papaloizou}, J., \& {Pringle}, J.~E. 1977, \mnras, 181, 441,
  \dodoi{10.1093/mnras/181.3.441}

\bibitem[{{Pascucci} {et~al.}(2016){Pascucci}, {Testi}, {Herczeg}, {Long},
  {Manara}, {Hendler}, {Mulders}, {Krijt}, {Ciesla}, {Henning}, {Mohanty},
  {Drabek-Maunder}, {Apai}, {Sz{\H u}cs}, {Sacco}, \&
  {Olofsson}}]{Pascucci:2016aa}
{Pascucci}, I., {Testi}, L., {Herczeg}, G.~J., {et~al.} 2016, \apj, 831, 125,
  \dodoi{10.3847/0004-637X/831/2/125}

\bibitem[{{Persi} {et~al.}(2001){Persi}, {Marenzi}, {G{\'o}mez}, \&
  {Olofsson}}]{Persi:2001aa}
{Persi}, P., {Marenzi}, A.~R., {G{\'o}mez}, M., \& {Olofsson}, G. 2001, \aap,
  376, 907, \dodoi{10.1051/0004-6361:20010962}

\bibitem[{{Pinilla} {et~al.}(2012){Pinilla}, {Benisty}, \&
  {Birnstiel}}]{Pinilla:2012aa}
{Pinilla}, P., {Benisty}, M., \& {Birnstiel}, T. 2012, \aap, 545, A81,
  \dodoi{10.1051/0004-6361/201219315}

\bibitem[{{Pinte} {et~al.}(2018){Pinte}, {M{\'e}nard}, {Duch{\^e}ne}, {Hill},
  {Dent}, {Woitke}, {Maret}, {van der Plas}, {Hales}, {Kamp}, {Thi}, {de
  Gregorio-Monsalvo}, {Rab}, {Quanz}, {Avenhaus}, {Carmona}, \&
  {Casassus}}]{Pinte:2018ab}
{Pinte}, C., {M{\'e}nard}, F., {Duch{\^e}ne}, G., {et~al.} 2018, \aap, 609,
  A47, \dodoi{10.1051/0004-6361/201731377}

\bibitem[{{Pontoppidan} \& {Dullemond}(2005)}]{Pontoppidan:2005aa}
{Pontoppidan}, K.~M., \& {Dullemond}, C.~P. 2005, \aap, 435, 595,
  \dodoi{10.1051/0004-6361:20042059}

\bibitem[{{Prusti} {et~al.}(1991){Prusti}, {Clark}, {Whittet}, {Laureijs}, \&
  {Zhang}}]{Prusti:1991aa}
{Prusti}, T., {Clark}, F.~O., {Whittet}, D.~C.~B., {Laureijs}, R.~J., \&
  {Zhang}, C.~Y. 1991, \mnras, 251, 303, \dodoi{10.1093/mnras/251.2.303}

\bibitem[{{Riaz} {et~al.}(2019){Riaz}, {Machida}, \&
  {Stamatellos}}]{Riaz:2019aa}
{Riaz}, B., {Machida}, M.~N., \& {Stamatellos}, D. 2019, \mnras, 486, 4114,
  \dodoi{10.1093/mnras/stz1032}

\bibitem[{{Ricci} {et~al.}(2018){Ricci}, {Liu}, {Isella}, \&
  {Li}}]{Ricci:2018aa}
{Ricci}, L., {Liu}, S.-F., {Isella}, A., \& {Li}, H. 2018, \apj, 853, 110,
  \dodoi{10.3847/1538-4357/aaa546}

\bibitem[{{Roccatagliata} {et~al.}(2018){Roccatagliata}, {Sacco},
  {Franciosini}, \& {Randich}}]{Roccatagliata:2018aa}
{Roccatagliata}, V., {Sacco}, G.~G., {Franciosini}, E., \& {Randich}, S. 2018,
  \aap, 617, L4, \dodoi{10.1051/0004-6361/201833890}

\bibitem[{{Rosenfeld} {et~al.}(2013){Rosenfeld}, {Andrews}, {Hughes}, {Wilner},
  \& {Qi}}]{Rosenfeld:2013aa}
{Rosenfeld}, K.~A., {Andrews}, S.~M., {Hughes}, A.~M., {Wilner}, D.~J., \&
  {Qi}, C. 2013, \apj, 774, 16, \dodoi{10.1088/0004-637X/774/1/16}

\bibitem[{{Rosotti} \& {Clarke}(2018)}]{Rosotti:2018aa}
{Rosotti}, G.~P., \& {Clarke}, C.~J. 2018, \mnras, 473, 5630,
  \dodoi{10.1093/mnras/stx2769}

\bibitem[{{Ru{\'\i}z-Rodr{\'\i}guez} {et~al.}(2018){Ru{\'\i}z-Rodr{\'\i}guez},
  {Cieza}, {Williams}, {Andrews}, {Principe}, {Caceres}, {Canovas}, {Casassus},
  {Schreiber}, \& {Kastner}}]{Ruiz-Rodriguez:2018aa}
{Ru{\'\i}z-Rodr{\'\i}guez}, D., {Cieza}, L.~A., {Williams}, J.~P., {et~al.}
  2018, \mnras, 478, 3674, \dodoi{10.1093/mnras/sty1351}

\bibitem[{{Segura-Cox} {et~al.}(2020){Segura-Cox}, {Schmiedeke}, {Pineda},
  {Stephens}, {Fern{\'a}ndez-L{\'o}pez}, {Looney}, {Caselli}, {Li}, {Mundy},
  {Kwon}, \& {Harris}}]{Segura-Cox:2020aa}
{Segura-Cox}, D.~M., {Schmiedeke}, A., {Pineda}, J.~E., {et~al.} 2020, \nat,
  586, 228, \dodoi{10.1038/s41586-020-2779-6}

\bibitem[{{Seifried} {et~al.}(2016){Seifried}, {S{\'a}nchez-Monge}, {Walch}, \&
  {Banerjee}}]{Seifried:2016aa}
{Seifried}, D., {S{\'a}nchez-Monge}, {\'A}., {Walch}, S., \& {Banerjee}, R.
  2016, \mnras, 459, 1892, \dodoi{10.1093/mnras/stw785}

\bibitem[{{Shakura} \& {Sunyaev}(1973)}]{Shakura:1973aa}
{Shakura}, N.~I., \& {Sunyaev}, R.~A. 1973, \aap, 500, 33

\bibitem[{{Sheehan} \& {Eisner}(2017)}]{Sheehan:2017ab}
{Sheehan}, P.~D., \& {Eisner}, J.~A. 2017, \apjl, 840, L12,
  \dodoi{10.3847/2041-8213/aa6df8}

\bibitem[{{Sheehan} \& {Eisner}(2018)}]{Sheehan:2018aa}
---. 2018, \apj, 857, 18, \dodoi{10.3847/1538-4357/aaae65}

\bibitem[{{Sheehan} {et~al.}(2020){Sheehan}, {Tobin}, {Federman}, {Megeath}, \&
  {Looney}}]{Sheehan:2020aa}
{Sheehan}, P.~D., {Tobin}, J.~J., {Federman}, S., {Megeath}, S.~T., \&
  {Looney}, L.~W. 2020, \apj, 902, 141, \dodoi{10.3847/1538-4357/abbad5}

\bibitem[{{Stamatellos} \& {Whitworth}(2009)}]{Stamatellos:2009aa}
{Stamatellos}, D., \& {Whitworth}, A.~P. 2009, \mnras, 392, 413,
  \dodoi{10.1111/j.1365-2966.2008.14069.x10.48550/arXiv.0810.1687}

\bibitem[{{Takahashi} {et~al.}(2016){Takahashi}, {Tomida}, {Machida}, \&
  {Inutsuka}}]{Takahashi:2016aa}
{Takahashi}, S.~Z., {Tomida}, K., {Machida}, M.~N., \& {Inutsuka}, S.-i. 2016,
  \mnras, 463, 1390, \dodoi{10.1093/mnras/stw1994}

\bibitem[{{Takeuchi} {et~al.}(1996){Takeuchi}, {Miyama}, \&
  {Lin}}]{Takeuchi:1996aa}
{Takeuchi}, T., {Miyama}, S.~M., \& {Lin}, D.~N.~C. 1996, \apj, 460, 832,
  \dodoi{10.1086/177013}

\bibitem[{{Teague} {et~al.}(2019){Teague}, {Bae}, \& {Bergin}}]{Teague:2019aa}
{Teague}, R., {Bae}, J., \& {Bergin}, E.~A. 2019, \nat, 574, 378,
  \dodoi{10.1038/s41586-019-1642-0}

\bibitem[{{Terebey} {et~al.}(1984){Terebey}, {Shu}, \&
  {Cassen}}]{Terebey:1984aa}
{Terebey}, S., {Shu}, F.~H., \& {Cassen}, P. 1984, \apj, 286, 529,
  \dodoi{10.1086/162628}

\bibitem[{{Thieme} {et~al.}(2022){Thieme}, {Lai}, {Lin}, {Cheong}, {Lee},
  {Yen}, {Li}, {Lam}, \& {Zhao}}]{Thieme:2022aa}
{Thieme}, T.~J., {Lai}, S.-P., {Lin}, S.-J., {et~al.} 2022, \apj, 925, 32,
  \dodoi{10.3847/1538-4357/ac382b}

\bibitem[{Tobin(2023)}]{Tobin:2023ab}
Tobin, J. 2023, eDisk data reduction scripts, 1.0.0,  Zenodo,
  \dodoi{10.5281/zenodo.7986682}

\bibitem[{{Tobin} {et~al.}(2020){Tobin}, {Sheehan}, {Megeath},
  {D{\'\i}az-Rodr{\'\i}guez}, {Offner}, {Murillo}, {van 't Hoff}, {van
  Dishoeck}, {Osorio}, {Anglada}, {Furlan}, {Stutz}, {Reynolds}, {Karnath},
  {Fischer}, {Persson}, {Looney}, {Li}, {Stephens}, {Chandler}, {Cox},
  {Dunham}, {Tychoniec}, {Kama}, {Kratter}, {Kounkel}, {Mazur}, {Maud},
  {Patel}, {Perez}, {Sadavoy}, {Segura-Cox}, {Sharma}, {Stephenson}, {Watson},
  \& {Wyrowski}}]{Tobin:2020aa}
{Tobin}, J.~J., {Sheehan}, P.~D., {Megeath}, S.~T., {et~al.} 2020, \apj, 890,
  130, \dodoi{10.3847/1538-4357/ab6f64}

\bibitem[{{Tychoniec} {et~al.}(2018){Tychoniec}, {Tobin}, {Karska}, {Chandler},
  {Dunham}, {Harris}, {Kratter}, {Li}, {Looney}, {Melis}, {P{\'e}rez},
  {Sadavoy}, {Segura-Cox}, \& {van Dishoeck}}]{Tychoniec:2018ab}
{Tychoniec}, {\L}., {Tobin}, J.~J., {Karska}, A., {et~al.} 2018, \apjs, 238,
  19, \dodoi{10.3847/1538-4365/aaceae}

\bibitem[{{Ueda} {et~al.}(2020){Ueda}, {Kataoka}, \&
  {Tsukagoshi}}]{Ueda:2020aa}
{Ueda}, T., {Kataoka}, A., \& {Tsukagoshi}, T. 2020, \apj, 893, 125,
  \dodoi{10.3847/1538-4357/ab8223}

\bibitem[{{van der Walt} {et~al.}(2011){van der Walt}, {Colbert}, \&
  {Varoquaux}}]{van-der-Walt:2011aa}
{van der Walt}, S., {Colbert}, S.~C., \& {Varoquaux}, G. 2011, Computing in
  Science and Engineering, 13, 22, \dodoi{10.1109/MCSE.2011.37}

\bibitem[{{van't Hoff} {et~al.}(2020){van't Hoff}, {Harsono}, {Tobin},
  {Bosman}, {van Dishoeck}, {J{\o}rgensen}, {Miotello}, {Murillo}, \&
  {Walsh}}]{van-t-Hoff:2020aa}
{van't Hoff}, M. L.~R., {Harsono}, D., {Tobin}, J.~J., {et~al.} 2020, \apj,
  901, 166, \dodoi{10.3847/1538-4357/abb1a2}

\bibitem[{{Velusamy} {et~al.}(2014){Velusamy}, {Langer}, \&
  {Thompson}}]{Velusamy:2014aa}
{Velusamy}, T., {Langer}, W.~D., \& {Thompson}, T. 2014, \apj, 783, 6,
  \dodoi{10.1088/0004-637X/783/1/6}

\bibitem[{{Villenave} {et~al.}(2020){Villenave}, {M{\'e}nard}, {Dent},
  {Duch{\^e}ne}, {Stapelfeldt}, {Benisty}, {Boehler}, {van der Plas}, {Pinte},
  {Telkamp}, {Wolff}, {Flores}, {Lesur}, {Louvet}, {Riols}, {Dougados},
  {Williams}, \& {Padgett}}]{Villenave:2020aa}
{Villenave}, M., {M{\'e}nard}, F., {Dent}, W.~R.~F., {et~al.} 2020, \aap, 642,
  A164, \dodoi{10.1051/0004-6361/202038087}

\bibitem[{{Virtanen} {et~al.}(2020){Virtanen}, {Gommers}, {Oliphant},
  {Haberland}, {Reddy}, {Cournapeau}, {Burovski}, {Peterson}, {Weckesser},
  {Bright}, {van der Walt}, {Brett}, {Wilson}, {Jarrod Millman}, {Mayorov},
  {Nelson}, {Jones}, {Kern}, {Larson}, {Carey}, {Polat}, {Feng}, {Moore}, {Vand
  erPlas}, {Laxalde}, {Perktold}, {Cimrman}, {Henriksen}, {Quintero}, {Harris},
  {Archibald}, {Ribeiro}, {Pedregosa}, {van Mulbregt}, \&
  {Contributors}}]{Virtanen:2020aa}
{Virtanen}, P., {Gommers}, R., {Oliphant}, T.~E., {et~al.} 2020, Nature
  Methods, 17, 261, \dodoi{https://doi.org/10.1038/s41592-019-0686-2}

\bibitem[{{Voirin} {et~al.}(2018){Voirin}, {Manara}, \&
  {Prusti}}]{Voirin:2018aa}
{Voirin}, J., {Manara}, C.~F., \& {Prusti}, T. 2018, \aap, 610, A64,
  \dodoi{10.1051/0004-6361/201731153}

\bibitem[{{Wakelam} {et~al.}(2005){Wakelam}, {Ceccarelli}, {Castets},
  {Lefloch}, {Loinard}, {Faure}, {Schneider}, \& {Benayoun}}]{Wakelam:2005aa}
{Wakelam}, V., {Ceccarelli}, C., {Castets}, A., {et~al.} 2005, \aap, 437, 149,
  \dodoi{10.1051/0004-6361:20042566}

\bibitem[{{Weidenschilling}(1977)}]{Weidenschilling:1977ab}
{Weidenschilling}, S.~J. 1977, \mnras, 180, 57, \dodoi{10.1093/mnras/180.2.57}

\bibitem[{{Whitney} {et~al.}(2003){Whitney}, {Wood}, {Bjorkman}, \&
  {Wolff}}]{Whitney:2003aa}
{Whitney}, B.~A., {Wood}, K., {Bjorkman}, J.~E., \& {Wolff}, M.~J. 2003, \apj,
  591, 1049, \dodoi{10.1086/375415}

\bibitem[{{Williams} {et~al.}(2019){Williams}, {Cieza}, {Hales}, {Ansdell},
  {Ruiz-Rodriguez}, {Casassus}, {Perez}, \& {Zurlo}}]{Williams:2019aa}
{Williams}, J.~P., {Cieza}, L., {Hales}, A., {et~al.} 2019, \apjl, 875, L9,
  \dodoi{10.3847/2041-8213/ab1338}

\bibitem[{{Wilner} \& {Welch}(1994)}]{Wilner:1994aa}
{Wilner}, D.~J., \& {Welch}, W.~J. 1994, \apj, 427, 898, \dodoi{10.1086/174195}

\bibitem[{{Woitas} {et~al.}(2001){Woitas}, {Leinert}, \&
  {K{\"o}hler}}]{Woitas:2001aa}
{Woitas}, J., {Leinert}, C., \& {K{\"o}hler}, R. 2001, \aap, 376, 982,
  \dodoi{10.1051/0004-6361:20011034}

\bibitem[{{Wu} {et~al.}(2017){Wu}, {Sheehan}, {Males}, {Close}, {Morzinski},
  {Teske}, {Haug-Baltzell}, {Merchant}, \& {Lyons}}]{Wu:2017aa}
{Wu}, Y.-L., {Sheehan}, P.~D., {Males}, J.~R., {et~al.} 2017, \apj, 836, 223,
  \dodoi{10.3847/1538-4357/aa5b96}

\bibitem[{{Yen} {et~al.}(2019){Yen}, {Gu}, {Hirano}, {Koch}, {Lee}, {Liu}, \&
  {Takakuwa}}]{Yen:2019ac}
{Yen}, H.-W., {Gu}, P.-G., {Hirano}, N., {et~al.} 2019, \apj, 880, 69,
  \dodoi{10.3847/1538-4357/ab29f8}

\bibitem[{{Yen} {et~al.}(2013){Yen}, {Takakuwa}, {Ohashi}, \&
  {Ho}}]{Yen:2013aa}
{Yen}, H.-W., {Takakuwa}, S., {Ohashi}, N., \& {Ho}, P.~T.~P. 2013, \apj, 772,
  22, \dodoi{10.1088/0004-637X/772/1/22}

\bibitem[{{Yen} {et~al.}(2014){Yen}, {Takakuwa}, {Ohashi}, {Aikawa}, {Aso},
  {Koyamatsu}, {Machida}, {Saigo}, {Saito}, {Tomida}, \&
  {Tomisaka}}]{Yen:2014aa}
{Yen}, H.-W., {Takakuwa}, S., {Ohashi}, N., {et~al.} 2014, \apj, 793, 1,
  \dodoi{10.1088/0004-637X/793/1/1}

\bibitem[{{Zhang} {et~al.}(2015){Zhang}, {Blake}, \& {Bergin}}]{Zhang:2015aa}
{Zhang}, K., {Blake}, G.~A., \& {Bergin}, E.~A. 2015, \apjl, 806, L7,
  \dodoi{10.1088/2041-8205/806/1/L7}

\bibitem[{{Zhu} {et~al.}(2012){Zhu}, {Nelson}, {Dong}, {Espaillat}, \&
  {Hartmann}}]{Zhu:2012aa}
{Zhu}, Z., {Nelson}, R.~P., {Dong}, R., {Espaillat}, C., \& {Hartmann}, L.
  2012, \apj, 755, 6, \dodoi{10.1088/0004-637X/755/1/6}

\bibitem[{{Zhu} {et~al.}(2019){Zhu}, {Zhang}, {Jiang}, {Kataoka}, {Birnstiel},
  {Dullemond}, {Andrews}, {Huang}, {P{\'e}rez}, {Carpenter}, {Bai}, {Wilner},
  \& {Ricci}}]{Zhu:2019aa}
{Zhu}, Z., {Zhang}, S., {Jiang}, Y.-F., {et~al.} 2019, \apjl, 877, L18,
  \dodoi{10.3847/2041-8213/ab1f8c}

\bibitem[{{Zinnecker} {et~al.}(1999){Zinnecker}, {Krabbe}, {McCaughrean},
  {Stanke}, {Stecklum}, {Brandner}, {Padgett}, {Stapelfeldt}, \&
  {Yorke}}]{Zinnecker:1999aa}
{Zinnecker}, H., {Krabbe}, A., {McCaughrean}, M.~J., {et~al.} 1999, \aap, 352,
  L73

\bibitem[{{Zucker} {et~al.}(2020){Zucker}, {Speagle}, {Schlafly}, {Green},
  {Finkbeiner}, {Goodman}, \& {Alves}}]{Zucker:2020aa}
{Zucker}, C., {Speagle}, J.~S., {Schlafly}, E.~F., {et~al.} 2020, \aap, 633,
  A51, \dodoi{10.1051/0004-6361/201936145}

\bibitem[{{Zucker} {et~al.}(2019){Zucker}, {Speagle}, {Schlafly}, {Green},
  {Finkbeiner}, {Goodman}, \& {Alves}}]{Zucker:2019aa}
---. 2019, \apj, 879, 125, \dodoi{10.3847/1538-4357/ab2388}

\bibitem[{{Zurlo} {et~al.}(2020){Zurlo}, {Cieza}, {P{\'e}rez}, {Christiaens},
  {Williams}, {Guidi}, {C{\'a}novas}, {Casassus}, {Hales}, {Principe},
  {Ru{\'\i}z-Rodr{\'\i}guez}, \& {Fernandez-Figueroa}}]{Zurlo:2020aa}
{Zurlo}, A., {Cieza}, L.~A., {P{\'e}rez}, S., {et~al.} 2020, \mnras, 496, 5089,
  \dodoi{10.1093/mnras/staa1886}

\bibitem[{{Zurlo} {et~al.}(2021){Zurlo}, {Cieza}, {Ansdell}, {Christiaens},
  {P{\'e}rez}, {Lovell}, {Mesa}, {Williams}, {Gonzalez-Ruilova}, {Carraro},
  {Ru{\'\i}z-Rodr{\'\i}guez}, \& {Wyatt}}]{Zurlo:2021aa}
{Zurlo}, A., {Cieza}, L.~A., {Ansdell}, M., {et~al.} 2021, \mnras, 501, 2305,
  \dodoi{10.1093/mnras/staa3674}

\end{thebibliography}
\bibliographystyle{aasjournal}



\end{document}